\documentclass[prx,onecolumn,notitlepage,superscriptaddress]{revtex4-1}

\usepackage{graphicx,epsfig}% Include figure files
\usepackage{epstopdf}
\usepackage{bbm, amsmath, amssymb, mathrsfs, float, mathtools, cases, hyperref, color}
\usepackage{dsfont}
\usepackage{color}

\begin{document}

\newcommand{\id}{\mathds{1}}
\newcommand{\bs}{\boldsymbol}
\newcommand{\E}{\mathds{E}}
\newcommand{\beq}{\begin{equation}} 
\newcommand{\eeq}{\end{equation}}
\newcommand{\bem}{\begin{multline}}
\newcommand{\bes}{\begin{split}} \newcommand{\ees}{\end{split}} 
\newcommand{\bea}{\begin{eqnarray}} \newcommand{\eea}{\end{eqnarray}}
\newcommand{\dd}{\partial}
\newcommand{\sm}{\setminus}
\newcommand{\avg}[1]{\left\langle #1 \right\rangle}
\newcommand{\comment}[1]{\textcolor{red}{#1}}
\title{Coordination problems on networks revisited: statics and dynamics}

\author{Luca Dall'Asta}
\affiliation{Department of Applied Science and Technology DISAT, Politecnico di Torino, Corso Duca degli Abruzzi 24, 10129 Torino, Italy}
\affiliation{Collegio Carlo Alberto, P.za Arbarello 8, 10122 Torino, Italy}

\begin{abstract} 
Simple binary-state coordination models are widely used to study collective socio-economic phenomena such as the spread of innovations or the adoption of products on social networks. The common trait of these systems is the occurrence of large-scale coordination events taking place abruptly, in the form of a cascade process, as a consequence of small perturbations of an apparently stable state. The conditions for the occurrence of cascade instabilities have been largely analysed in the literature, however for the same  coordination models  no sufficient attention was given to the relation between structural properties of (Nash) equilibria and possible outcomes of dynamical equilibrium selection. Using methods from the statistical physics of disordered systems, the present work investigates both analytically and numerically, the statistical properties of such Nash equilibria on networks, focusing mostly on random graphs. We provide an accurate description of these properties, which  is then exploited to shed light on the mechanisms behind the onset of coordination/miscoordination on large networks. This is done studying the most common processes of dynamical equilibrium selection, such as best response, bounded-rational dynamics and learning processes. In particular, we show that well beyond the instability region, full coordination is still globally stochastically stable, however equilibrium selection processes with low stochasticity (e.g. best response) or strong memory effects (e.g. reinforcement learning) can be prevented from achieving full coordination by being trapped into a large (exponentially in number of agents) set of locally stable Nash equilibria at low/medium coordination (inefficient equilibria). These results should be useful to allow a better understanding of general coordination problems on complex networks.
\end{abstract}
\pacs{}
\keywords{game theory | Nash equilibria | coordination | message passing | random graphs}

\maketitle

\tableofcontents

\section{Introduction}\label{sec:intro}
Over the past two decades, the centrality acquired by distributed autonomous systems in the field of information technology, the advent of online social networks and their integration with platforms for digital commerce and entertainment, have stimulated a growing  interest in decision-making and game-theoretic problems in which rational (or bounded-rational) individuals interact on networks \cite{easley2010networks}. In particular, positive network externalities, i.e. the tendency of individuals to align their decisions to those of the neighbours, are identified as a major driving force behind large-scale cascading phenomena often observed in relation to the spread of innovations and products adoption on modern socio-economic and technological systems \cite{watts2002simple,centola2007complex,borge2013cascading}.  Starting from the seminal works by S. Morris \cite{morris2000contagion} and D. Watts \cite{watts2002simple}, several aspects of the relation between coordination cascades and the structure of the underlying interaction network have been explained, such as the existence of a critical threshold to trigger a coordination cascade \cite{morris2000contagion,watts2002simple,gleeson2007seed,lelarge2012diffusion} or the speed of convergence to the (fully) coordinated state \cite{montanari2009convergence,young2011dynamics,kreindler2014rapid}. Much is however still unclear about the statics and dynamics of coordination problems. In particular, according to the standard microscopic game-theoretic formulation of coordination problems on networks \cite{jackson2015games}, it is possible that a high multiplicity of (pure) Nash equilibria exists, with possibly very different structural and stability properties. Rather than elaborating Nash equilibrium refinements or alternative solution concepts in the framework of games of incomplete information \cite{galeotti2010network}, it is convenient to dwell further to investigate such a large variety of equilibria, whose richness is not fully represented by the mean-field methods commonly used to derive cascading conditions and calls for a more detailed statistical analysis. The present work represents a step forward in this direction. 

We employ methods from the statistical physics of disordered systems \cite{mezard2009information} to study, both analytically and numerically on random graphs and more general networks, the statistical properties of the set of (pure-strategy) Nash equilibria of coordination games on networks. The non-rigorous methods employed here are based on the assumption of the local tree-like structure of the interaction graph, which is correct for uncorrelated random graphs, even though most results are expected to be qualitatively valid also for more general network structures. 
The accurate knowledge of the equilibrium landscape is then used to shed light on the mechanisms triggering or preventing the onset of global coordination in best-response dynamics as well as in more general processes of dynamical equilibrium selection, such as those based on bounded rationality and reinforcement learning.

The paper is organized as follows. In Section II, the coordination model is formulated together with a brief description of its main properties. Section III recovers some known results on cascade processes on networks interpreting them as the final outcome of a monotonic binary dynamics for which it is possible to provide a description in terms of dynamic message passing. We will show that these processes can be used to identify the boundaries of the spectrum of equilibria when described in terms of the fraction of individuals playing one particular action. In addition to minimal and maximal equilibria, the same process also identifies pivotal equilibria which play an important role in cascade processes. In Section IV, the rich structure of the landscape of Nash equilibria will be explored by means of an improved message-passing technique  based on the cavity method from statistical mechanics. We will show that depending on the extent of the external disorder the system can undergo structural phase transitions which imply a complete reorganization of the equilibrium landscape. Furthermore, known results about network cascades will be reinterpreted in light of the structural properties of the equilibrium landscape. Section V is devoted to analyse, both numerically and analytically in some specific limit, the most common dynamical processes of equilibrium selection, such as best response, weighted best response, fictitious play and reinforcement learning. Some advances in understanding how the properties of the underlying network and the organization of the equilibrium landscape influence dynamical equilibrium selection are presented.  Finally, a discussion of major results and conclusions are presented in Section VI.

\section{A coordination model with heterogeneous payoffs}\label{sec:model}
Following the seminal work by Watts \cite{watts2002simple}, we consider a simple binary-state decision problem with positive externalities on a network $\mathcal{G}=(\mathcal{V},\mathcal{E})$ of agents, in which agents agree to coordinate on some action depending on the fraction of their neighbours choosing the same action. Agents intrinsic aversion to coordination is taken into account by including individual thresholds which have to be exceeded for the agent to align with her neighbours. More precisely, the action $x_i \in \{0,1\}$ is chosen by player $i\in V$  in order to maximise the individual utility function $u_i(x_i; \vec{x}_{\partial i})$, which depends on the actions  $\vec{x}_{\partial i} = \{ x_j | j \in \partial i\}$ of the neighbours of agent $i$ as follows 
\begin{equation}
 u_i(x_i; \vec{x}_{\partial i})   =  \theta_i k_i (1- x_i) + x_i\sum_{j\in \partial i} x_j, 
 \label{utility-function0}
\end{equation}
in which $k_i=|\partial i|$ is the degree of node  $i$ in the (undirected) network $\mathcal{G}$ and $\theta_i$ is her threshold value. Binary decision problems of this kind fall into a class generally known as {\em linear threshold models} \cite{granovetter1978threshold} because an agent $i$ plays action 1 only if a number $m  \geq  \theta_i k_i$ among her neighbours also plays 1, namely there is a threshold rule that depends linearly on the aggregate choice of the neighbours.  We are interested in the case in which the thresholds $\theta_i$ are time-independent ({\em quenched}) random variables, possibly different for each agent, but drawn from a common distribution $f(\theta)$. We also assume that individuals know the realisation $\vec{\theta} = (\theta_1,\dots,\theta_N)$  of threshold values, i.e. we are dealing with a {\em game of complete information} \cite{osborne1994course}.  The choice of such an idealised situation makes possible to analyse the structural equilibrium properties of the strategic interaction, which could be then related with those emerging from a possibly more realistic setting provided by games of incomplete information. 
The present decision problem is the generalisation to a  multi-agent system defined on a fixed network of a classical two-player bi-matrix normal-form coordination game of the stag-hunt type (see Appendix~\ref{app-game} for details). For a given realisation of thresholds $\{\theta_i\}_{i\in \mathcal{V}}$, the resulting game-theoretic problem of complete information, defined by the tuple $\Gamma=\left(\mathcal{G},\{0,1\}^{|\mathcal{V}|},\{u_i\}_{i\in \mathcal{V}},\{\theta_i\}_{i\in \mathcal{V}}\right)$, naturally admits Nash equilibria as rational solution concept. In particular, the game $\Gamma$ admits a multiplicity of {\em pure Nash equilibria}, which we indicate with the set $S_{\rm NE}$. These Nash equilibria can be obtained as the solutions $\vec{x}^{*}\in S_{\rm NE}$ of the fixed-point equations $x_i =b_i(\vec{x}_{\partial i}; \theta_i)$, $\forall i \in \mathcal{V}$, known as {\em best-response} relations, with reaction functions 
\begin{subequations}
\begin{align}
 b_i(\vec{x}_{\partial i}; \theta_i) & = {\arg\max}_{x \in \{0,1\}}  u_i(x; \vec{x}_{\partial i}) \\
&=  \Theta\left(  \sum_{j\in \partial i} x_j  - \theta_i k_i \right) \qquad \forall~i \in \mathcal{V},
\end{align}\label{best-response}
\end{subequations}
where $\Theta(x)=1$ if $x\geq 0$, otherwise $\Theta(x)=0$ (the choice of strict or weak inequality is not crucial as long as the threshold values are real).  
It seems natural to assume that threshold values are defined on the unit interval $[0,1]$, 
in which case the uniform profile $\vec{x}=\vec{1}=(1,1,\dots,1)$ is always a Nash equilibrium on any network and it is always the (Pareto) efficient one, namely the Nash equilibrium with the highest value of global utility $U = \sum_i u_i$. On the other hand, when thresholds are allowed to exceed the interval $[0,1]$, the model admits stubborn individuals having an intrinsic opinion that does not change due to peer effects. For instance, an agent with $\theta_i >1$ (resp. $\theta_i \leq 0$) always plays action $x_i=0$ (resp. $x_i=1$) independently of the actions of the neighbours.  

An important property of the present model is that the tuple $\Gamma =\left(\mathcal{G},\{0,1\}^{|\mathcal{V}|},\{u_i\}_{i\in \mathcal{V}},\{\theta_i\}_{i\in \mathcal{V}}\right)$ is a {\em potential game}, that is all pure Nash equilibria of the game are in one-to-one correspondence with the local maxima of a potential function given by
\begin{equation}\label{potential0}
V(\vec{x};\vec{\theta})  =   \sum_{(i,j)} x_i x_j  - \sum_i  \theta_i k_i x_i,  
\end{equation}
with $(i,j)\in \mathcal{E}$ and $i\in \mathcal{V}$
(see App.~\ref{app-game} for a derivation).  The existence of a potential function will be exploited in Sec.~\ref{sec-stochstability} to characterise pure Nash equilibria in terms of their stability with respect to small stochastic perturbations.

Although the methods and main results presented in this study are mostly independent of the details of the distribution of thresholds considered, yet a choice for $f(\theta)$ is necessary to illustrate them. The most common choice in the literature \cite{watts2002simple,gleeson2007seed,gai2010contagion}, which we shall adopt here as well, is a Gaussian distribution $\mathcal{N}(\mu,\sigma)$ of thresholds  with mean $\mu$ and standard deviation $\sigma$.

\section{Monotone decision processes}\label{sec:monotoneBR}
Before discussing in detail the static and dynamical properties of the coordination model, it is worth remarking that previous literature mostly focused on irreversible cascade processes that can lead to the efficient equilibrium from initial conditions containing just a small set of agents playing action $1$. A general condition for the existence of global cascades on random graphs can be derived \cite{watts2002simple,gleeson2007seed}, and it depends on the network structure, the distribution of thresholds and the initial seed size. In this Section, we recover such condition using a slightly different approach and discuss the relation with other relevant results present in the literature.
Suppose that all agents initially play action $0$ and they are iteratively called to decide whether they want to switch to $1$ or stay with action $0$. Without loss of generality, we can assume that this process occurs in discrete time, with parallel update. If some of the agents have negative threshold values, they spontaneously switch to action $1$ independently of the behaviour of their neighbours, this way acting as ``seeds" or initiators of a monotone dynamical process that leads towards more efficient equilibra. The monotonicity of this process, for any realisation $\vec{\theta}$ of the random variables, is easy to prove \cite{topkis1979equilibrium,vives1990nash,jackson2015games}, because agents that have already chosen action $1$ would never switch back to $0$ as a consequence of one or more neighbours switching to $1$ (see Appendix~\ref{app-game} for a discussion in terms of the property of increasing differences or supermodularity \cite{topkis2011supermodularity}). The fixed point of this process is by definition the Nash equilibrium with minimum number of agents playing action $1$, which here we call {\em minimal equilibrium} \cite{jackson2015games}. Since the number of agents playing action $1$ does not decrease in time, the dynamics corresponds to that of the permanent adoption model, discussed by several authors over the last decades, e.g. \cite{watts2002simple,gleeson2007seed,acemogludiffusion,lelarge2012diffusion,karsai2016local}. In particular, in the absence of agents with negative thresholds acting as seeds, the initial state $\vec{0}$ is the minimal Nash equilibrium. The stability of the minimal equilibrium with respect to an ``adoption cascade" can be analysed by considering the best-response process after monotone deviation in the action of a single agent, in which the strategy space is restricted to actions larger or equal to those played in the minimal equilibrium (see App.~\ref{app-game}). Because of the property of increasing differences, the response to monotone deviations is also a monotone process which ends into another Nash equilibrium.  The difference in the number of agents playing action $1$ between the initial (minimal) equilibrium and the one obtained at the end of the monotone process can be considered as a measure of distance between them.  The process can be averaged over different choices of the one-agent deviation. If the average distance obtained is $O(N)$ with $N=|\mathcal{V}|$, the minimal equilibrium can be considered unstable. It means that a small random perturbation can easily lead to Nash equilibria that are very different from the minimal one. Echenique \cite{echenique2007finding} showed that this monotone perturbation process can be applied iteratively in a gradually refined strategy space in order to find all Nash equilibria of a coordination game (the algorithm is briefly described for the present case in App.~\ref{app-game}). In the limit of large graphs ($N \to \infty$ limit), a similar result should hold for any random perturbation of finite size with respect to $N$. 

 A key remark for the statistical analysis of the average behaviour of this monotone adoption process over the distribution of $\vec{\theta}$ is that, after switching to $1$, the behaviour of an agent becomes independent of what might happen later,  making it possible to use the Principle of Deferred Decisions \cite{mitzenmacher2005probability}: the average over the random variable $\theta_i$ can be performed ``on the fly" at the time the choice of agent $i$ is analysed. Using the language of disordered systems, during the monotone adoption process, the quenched average over random variables is equivalent to an annealed one. In the following, a dynamic message-passing approach is employed to study the statistical properties of this monotone process and the resulting Nash equilibria. The method is exact on trees and usually provides good approximations on sparse uncorrelated random graphs in the large size limit.  

\subsection{Message-passing equations for minimal equilibria and  global cascade condition}\label{subsec:minimal}

On a tree-like graph, the monotone adoption process can be analysed by ideally starting from the leaves and progressively moving towards the bulk of the graph, defining a recursive probabilistic approach. This method was originally introduced in the statistical physics literature by D. Dhar and coworkers \cite{dhar1997zero,sabhapandit2000distribution,sabhapandit2002hysteresis} to compute the magnetisation curve along the hysteresis loop in the Random Field Ising Model (RFIM) on a Bethe lattice. The  same approach was recently reinterpreted in the dynamical framework of contagion processes \cite{ohta2010universal,karrer2010message,altarelli2013large,lokhov2014inferring,lokhov2015dynamic,paga2015contagion}. 
Given a tree graph $\mathcal{G}=(\mathcal{V},\mathcal{E})$, we consider the dynamic ``message" $h_{ij}(t)$ defined as the probability that agent $i$ plays strategy $x_i = 1$ at time $t$ given that the downstream agent $j$ was not touched by the best-response dynamics yet, i.e. agent $j$ still plays strategy $x_j = 0$.   
This probability can be computed as a function of similar quantities defined on the upstream nodes $\{k : k \in \partial i\sm j\}$ of the tree, 
\begin{subequations}
\begin{align} 
  {h}_{ij}(t+1) & =  \mathbb{E}_{\theta}\left[\sum_{\vec{x}_{\partial i}} \id\left[ 1 = b_i(\{\vec{x}_{\partial i\sm j}, 0\}; \theta_i) \right]  \prod_{k \in \partial i\sm j} \left[ x_k {h}_{ki}(t) + (1-x_k) (1-{h}_{ki}(t)) \right]\right] \\
& =   \sum_{\vec{x}_{\partial i}}  F_{1}\left(\frac{1}{|\partial i |} \sum_{k\in \partial i\sm j} x_k \right)     \prod_{k \in \partial i\sm j} \left[ x_k { h}_{ki}(t) + (1-x_k) (1-{ h}_{ki}(t)) \right], 
\end{align}\label{bp-minimal} 
\end{subequations}
for each directed edge $(i,j) \in \mathcal{E}$ and 
where $F_{1}(y)= \text{Prob}{\left[ \theta \le y \right]} = \int_{0}^{y}f(\theta) d\theta$. Notice that we could explicitly perform the average $ \mathbb{E}_{\theta}\left[\cdot\right]$ over the random variable $\theta_i$ only because of the previously mentioned property of monotonicity of the dynamics.
In the infinite time limit, the self-consistent equations for directed messages $\{h_{ij}(\infty)\}$ describe the statistical properties of the fixed-points  (absorbing states) of the adoption process. The equations for $\{h_{ij}\}$ refer to the ``lower edge" of the equilibrium landscape of the original coordination game, i.e. the (possibly not unique) minimal Nash equilibrium characterised by the minimum density of agents choosing action $1$. 
Although Eqs.~\eqref{bp-minimal} apply on a given graph instance, one can also analyse the behaviour on ensembles of (infinitely large) uncorrelated random graphs, that are completely specified by their degree distribution $p_k$. In this case, one can neglect the identity of the directed edge $(i,j)$ and consider the probability $h$ that, at the fixed point, an edge reaching an agent that plays action 0 comes from an agent playing instead action 1.
This is given by the solution of the self-consistent equation 
\beq\label{bp-lo}
h = \Phi_{\rm m}\left[h\right]  =    \sum_{k} \frac{k p_k}{\avg{k}} \sum_{m=0}^{k-1} B_{k-1,m}(h) F_{1}\left(m/k\right) 
 \eeq
 where $B_{k,m}(h) = \binom{k}{m} h^m (1-h)^{k-m}$ and $\avg{k}=\sum_k k p_k$ is the average degree. Similarly, the probability that a randomly chosen agent plays action $1$ is given by 
 \beq
 \label{bp-lo-tot} \rho_{\rm m}  = \sum_{k} p_k \sum_{m=0}^{k} B_{k,m}(h) F_{1}\left(m/k\right).
\eeq 
For large uncorrelated random graphs, the quantity $\rho_{\rm m}$ provides a good estimate (exact in the infinite size limit) of the lowest density of agents that could play action 1 in a pure-strategy Nash equilibrium.
Obviously, $h = 0$ is a solution  of \eqref{bp-lo} (and thus $\rho_{\rm m} =0$ from \eqref{bp-lo-tot}) as long as $F_1(0)=0$, i.e. there are no individuals acting as seeds of the adoption/contagion process. In addition to the minimal equilibrium, both Eq.~\eqref{bp-minimal} on single instances and Eq.~\eqref{bp-lo} at the ensemble level may admit other solutions,  corresponding to equilibria with a higher density of action $1$, which can be reached by means of monotone dynamics provided that a small perturbation of the minimal equilibrium has triggered a cascade phenomenon. Such equilibria were called {\em pivotal equilibria} \cite{morris2000contagion,lelarge2012diffusion}, because they are induced by a set of pivotal agents, whose degree is strictly less than the inverse of the corresponding threshold value. Pivotal agents are susceptible to small perturbations because they immediately switch to action $1$ as soon as one of the neighbours does. 
Assuming that the minimal Nash equilibrium corresponds to the $h=0$ solution of \eqref{bp-lo}, this is unstable under small perturbations if 
\beq\label{rho1-0low}
\Phi_{\rm m}'[0] =  \sum_{k}  p_k \frac{k(k-1)}{\avg{k}} F_{1}\left(\frac{1}{k}\right)> 1,
\eeq
where we used $F_1(0)=0$. The condition \eqref{rho1-0low}, originally derived by Watts \cite{watts2002simple}, admits a very intuitive interpretation in terms of (not necessarily homogeneous) percolation theory. 
Take as  a control parameter the (degree dependent) probability $F_1(1/k)$ that agents of degree $k$ playing $0$ would turn to $1$ by having at least one neighbour playing action $1$, then Eq.\eqref{rho1-0low} says that when this quantity is above the percolation threshold of the underlying graph, the $\vec{0}$ equilibrium is unstable to perturbations. As explained in Appendix~\ref{app-pivotal}, the same condition also determines the percolation properties of pivotal agents.  

\begin{figure}[t]
\begin{center}
\includegraphics[width=0.6\columnwidth]{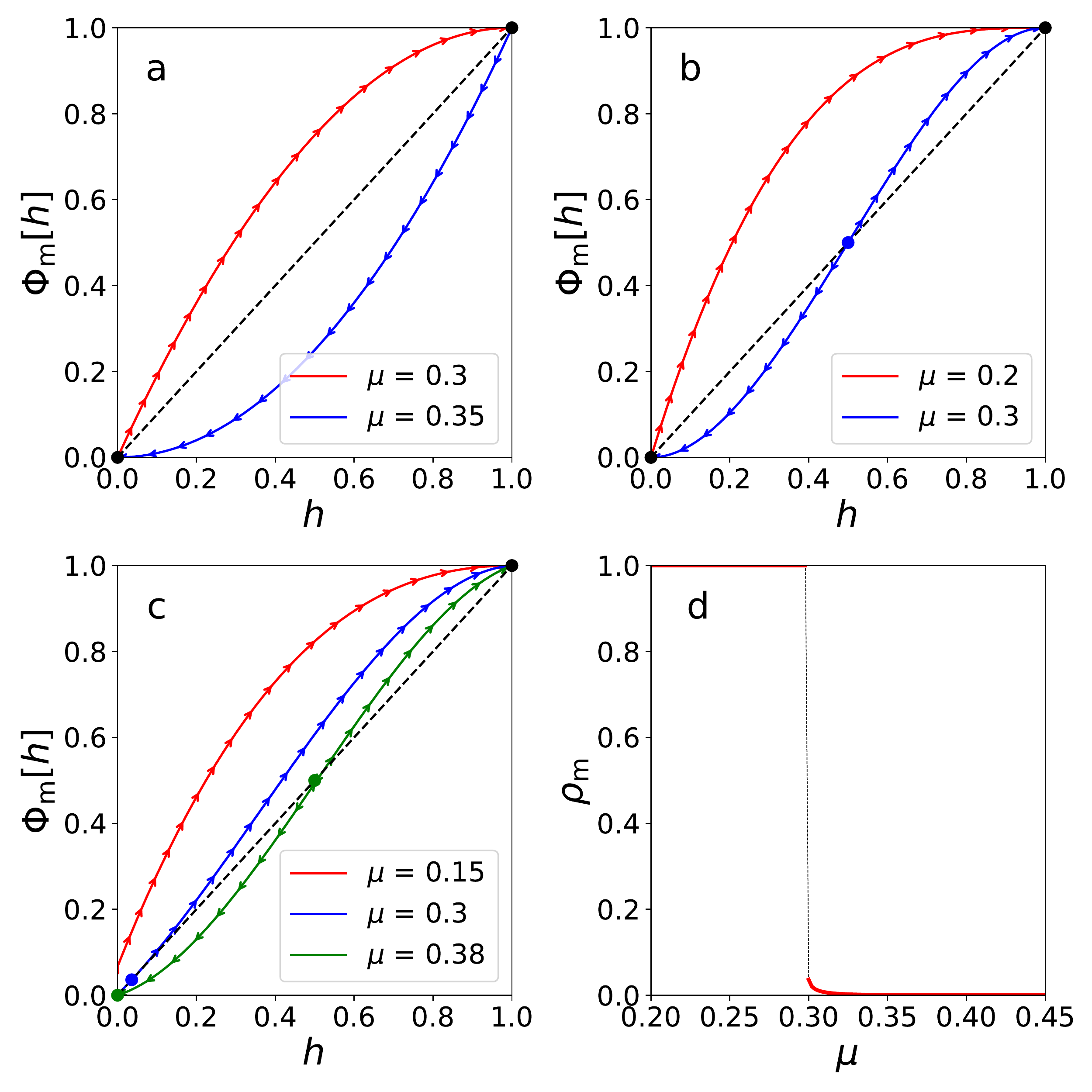}
\caption{Fixed points and flow lines of the Eq.\eqref{bp-lo} for the monotone adoption process with gaussian distribution $\mathcal{N}(\mu,\sigma)$  of threshold values on random regular graphs with degree (a) $K=3$ and (b) $K=4$, both for $\sigma =0$, and for (c) $K=4$ and $\sigma=0.1$. (d) Density $\rho_{\rm m}$ of agents playing action $1$ in the minimum stable equilibrium solution of the adoption process as function of $\mu$ for $K=4$ and $\sigma=0.1$ ($\mu_{\rm c,m}\approx 0.3$).}\label{fig:rho_lowREG}
\end{center}
\end{figure}

As an explanatory example of condition \eqref{rho1-0low}, in Fig.\ref{fig:rho_lowREG} we consider the model with gaussian thresholds, i.e. $\theta \sim \mathcal{N}(\mu,\sigma)$ introduced in Sec.\ref{sec:model}, on random regular graphs of degree $K$. For given values of $K$ and $\sigma$,  \eqref{rho1-0low} provides a threshold value $\mu_{\rm c,m}$ for the instability. For $K=3$ and $\sigma=0$, there are only two  solutions of Eq.~\eqref{bp-lo}, i.e. $h=0$ and $h=1$, which exchange stability at $\mu_{\rm c,m} = 1/K=1/3$ (see Fig.\ref{fig:rho_lowREG}a): for thresholds lower than $1/K$ an adoption cascade to the pivotal equilibrium $\vec{1}$ can be triggered by any small perturbation of the minimum equilibrium $\vec{0}$, because all nodes are pivotal ones. Figure~\ref{fig:rho_lowREG}b shows that for $K=4$ and $\sigma=0$ the instability occurs at $\mu_{\rm c,m} = 1/4$, with the difference that now both equilibria are (locally) stable for $\mu > \mu_{\rm c,m}$ and their basins of attraction are separated by an intermediate unstable fixed point. This result is reminiscent of the instability of mixed-strategy equilibria that occurs in coordination games such as the present one \cite{echenique2004mixed} (see also discussion in App.~\ref{app-game}).  A similar behaviour is observed for larger degree values. Note that the appearance of two distinct basins of attraction for Eq.~\eqref{bp-lo} is the signature of a structural change in the space of pure Nash equilibria that will be discussed in detail in the next sections.

For $\sigma>0$, fully homogeneous action profiles might not be equilibria anymore because seed individuals with negative thresholds exist for sufficiently low values of $\mu$, i.e. $F_1(0)>0$.  This  is illustrated in Fig.\ref{fig:rho_lowREG}c-\ref{fig:rho_lowREG}d for $K=4$ and $\sigma=0.1$. If $\mu > \mu_{\rm c,m}$ both the minimal and the pivotal stable fixed-point exist, the minimal one being different from $\rho_{\rm m}=0$ and approaching it for large values of $\mu$. The lower fixed point disappears abruptly at  $\mu_{\rm c,m}\approx 0.3$ with a discontinuous saddle-node bifurcation, so that the minimal equilibrium coincides with the pivotal one for $\mu < \mu_{\rm c,m}$ at sufficiently large values of $\sigma$. The value of $\mu_{\rm c,m}$ can be determined from a generalisation of the cascade condition \eqref{rho1-0low} derived by Gleeson and Cahalane \cite{gleeson2007seed} taking into account small but not vanishing density of initial adopters. Expanding the polynomial equation \eqref{bp-lo} to the second order in $\rho_{\rm m}$, the occurrence of global cascades corresponds either to the usual condition  $\Phi_{\rm m}'[0]>1$ or to the negative discriminant condition~\cite{gleeson2007seed}   
\begin{equation}\label{eq-gleeson}
(\Phi_{\rm m}'[0]-1)^2 - 2 \Phi_{\rm m}[0]\Phi_{\rm m}''[0] < 0.
\end{equation}
For values of $\sigma$ that are not too large, the discontinuous saddle-node bifurcation point can be located verifying that both conditions \eqref{rho1-0low} and \eqref{eq-gleeson} are violated.

\subsection{Properties of the maximal equilibria}\label{subsec:maximal}
By symmetry, a monotone best-response process, similar to the one considered in the previous subsection, can be used to investigate the properties of  {\em maximal equilibria}, i.e. equilibria with maximum density of agents playing action $1$ (as defined in App.~\ref{app-game}). Maximal equilibria are also the most (Pareto) efficient ones as they maximise the global utility $U$. The fully coordinated action profile $\vec{1}$ can either be a Nash equilibrium or spontaneously evolve towards a less coordinated profile (under best response) because of the existence of a fraction of agents with too high values of their thresholds. Applying on a tree-like graph the same construction we used for minimal equilibria, the relevant quantity to analyse is now the probability $H_{ij}(t)$ that the agent on node $i$ plays strategy $x_i = 0$ given that the agent in the downstream node $j$ still plays strategy $x_j = 1$. We obtain the equations 
\begin{subequations}
\begin{align}
 H_{ij}(t+1)  = & \mathbb{E}_{\theta}\left[\sum_{\vec{x}_{\partial i}} \id\left[ 0 = b_i(\{\vec{x}_{\partial i\sm j}, 1\}; \theta_i) \right]  \prod_{k \in \partial i\sm j} \left[ x_k (1-H_{ki}(t)) + (1-x_k) H_{ki}(t) \right]\right] \\
 = &  \sum_{\vec{x}_{\partial i}}  F_{0}\left( \frac{1}{|\partial i|}\left(1+ \sum_{k\in \partial i\sm j} x_k\right) \right)   \prod_{k \in \partial i\sm j} \left[ x_k (1-H_{ki}(t)) + (1-x_k) H_{ki}(t) \right]
\label{bp-maximal}
\end{align}
\end{subequations}
where $F_{0}(y) = 1- F_{1}(y)$.
For $t \to \infty$, the quantity $H_{ij} = H_{ij}(\infty)$ can be used to compute the density of agents playing action $0$ in the Nash equilibria reached by means of this process. 
Considering again the ensemble of uncorrelated random graphs with degree distribution $p_k$, we find the fixed-point equation
\beq\label{bp-up}
H = \Phi_{\rm M}\left[ H \right]  =   \sum_{k} \frac{k p_k}{\avg{k}} \sum_{m=0}^{k-1} B_{k-1,m}(1-H) F_{0}\left(\frac{1+m}{k}\right) 
 \eeq
 and the density of agents playing action $1$ is given by 
 \beq
 \label{bp-up-tot} \rho_{\rm M}  = \sum_{k} p_k \sum_{m=0}^{k} B_{k,m}(1-H) F_{1}\left(m/k\right).
\eeq 
The solution $H=0$, which means $\rho_{\rm M}=1$, corresponding to maximal Nash equilibria, exists as long as $F_{1}(1)=1$ and it becomes unstable if
 \beq\label{rho1-0up}
\Phi_{\rm M}\left[ 0\right] =  \sum_{k} p_k\frac{k(k-1)}{\avg{k}} F_{0}\left(\frac{k-1}{k}\right) > 1
 \eeq
 where we used $F_0(1)=0$. The condition \eqref{rho1-0up} describes another percolation-like phenomenon driven by a set of pivotal agents:  the maximal equilibrium $\vec{1}$  is not stable against perturbations if the fraction of nodes (of degree $k$) that would deviate to $0$ if at least one of their neighbours does is beyond the node percolation threshold. 
When $F_{1}(1)<1$, the instability of the maximal equilibrium (albeit different from $\vec{1}$) is provided, in addition to this first-order condition, by the second-order condition $(\Phi_{\rm M}'[0]-1)^2 - 2 \Phi_{\rm M}[0]\Phi_{\rm M}''[0] < 0$, analogue to \eqref{eq-gleeson}.

\begin{figure}[tb]
\begin{center}
\includegraphics[width=0.55\columnwidth]{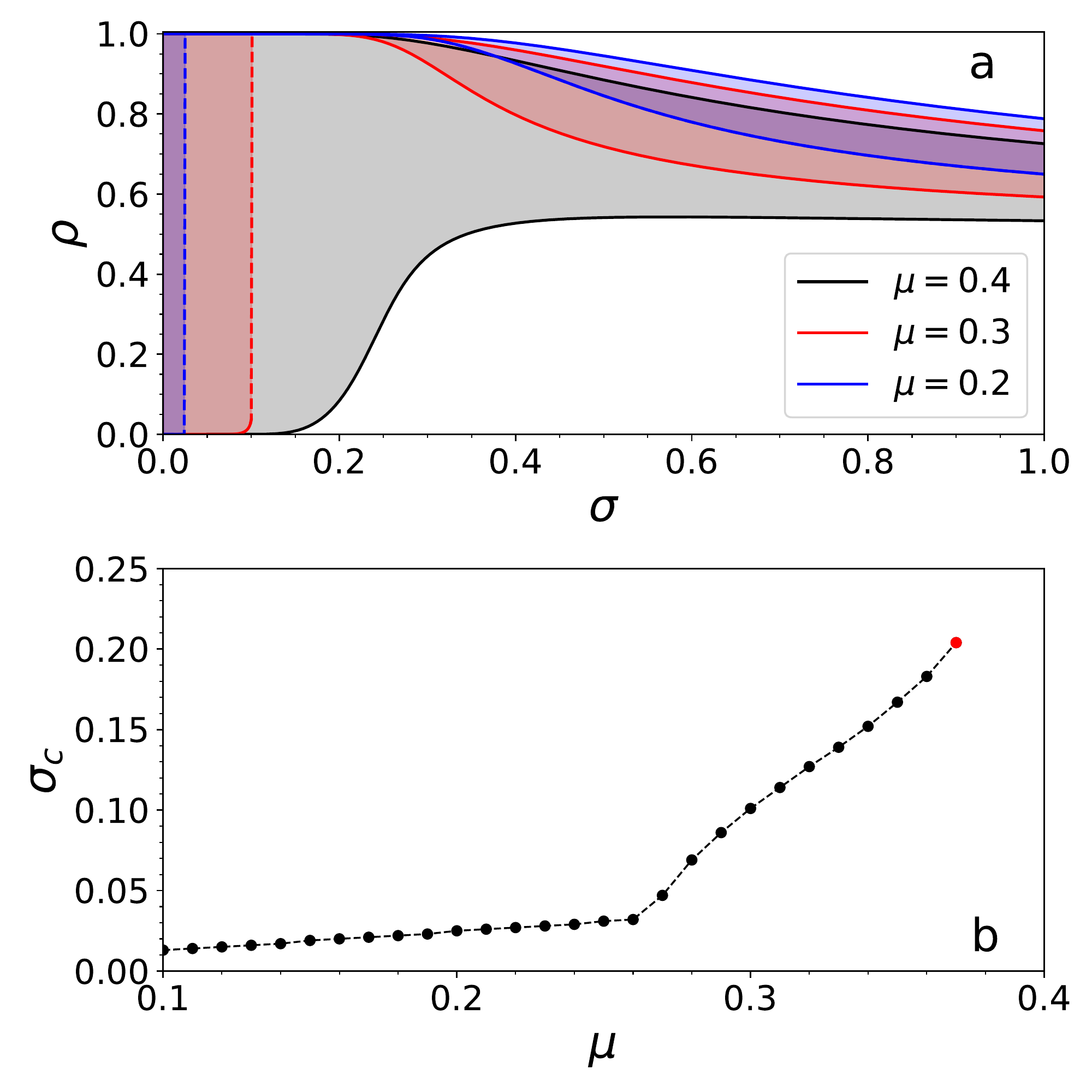}
\caption{(a) Density $\rho_{\rm m}$ and $\rho_{\rm M}$ of lower and upper edges of the equilibrium density spectrum, as function of the standard deviation $\sigma$ of the threshold distribution, obtained solving Eqs.~\eqref{bp-lo}-\eqref{bp-up} on  random regular graphs of degree $K=4$ and $\mu = 0.4, 0.3, 0.2$. (b) Critical value $\sigma_{\rm c}$ of threshold heterogeneity at which the instability in the lower edge (minimum equilibrium) takes place as function of the mean value $\mu$.}\label{fig:spectrumREG}
\end{center}
\end{figure}

The two monotone processes described in the present section define the existence and stability properties of the lower and upper edges of the spectrum (in terms of utility values) of equilibria for the coordination model $\left(\mathcal{G},\{0,1\}^{|\mathcal{V}|},\{u_i\}_{i\in \mathcal{V}},\{\theta_i\}_{i\in \mathcal{V}}\right)$ defined on random graphs with gaussian thresholds. Figure~\ref{fig:spectrumREG}a shows the lower and upper edges of the equilibrium density spectrum on random regular graphs of degree $K=4$ as function of the parameter $\sigma$ governing the heterogeneity of gaussian threshold values and for $\mu=0.2$ (blue), $0.3$ (red), $ 0.4$ (grey). The shaded regions cover the density intervals $[\rho_{\rm m},\rho_{\rm M}]$ for which it is possible to find Nash equilibria of the coordination model. This region always shrinks towards $\rho=0.5$ increasing $\sigma$; moreover, for sufficiently low values of $\mu$, the instability in the lower edge produces a discontinuous transition which corresponds to an overall rearrangement in the whole structure of the equilibrium landscape. Fig.~\ref{fig:spectrumREG}b shows that the corresponding critical line $\sigma_{\rm c}(\mu)$ as function of the other parameter $\mu$ (for $K=4$ in this example) terminates in a critical point (full red circle). Similar qualitative behaviour is observed for  random regular graphs with larger connectivity. The phenomenology of the monotone decision process is richer on general random  graphs as the properties of the minimal and maximal equilibria depend on the degree distribution $p_k$. For the sake of completeness, some results on non-regular random graphs are reported in App.~\ref{app-pivotal} and briefly resumed in Secs.~\ref{subsec:cavityGRG}.

A more detailed analysis of the properties of the space of equilibria is provided in the next Section by means of a purely static analysis using the cavity method.

\section{The equilibrium landscape}\label{sec:landscape}

Nash equilibria can be viewed as metastable configurations in discrete statistical mechanics models, in which each spin variable points in the direction of its local field, hence being stable against single-spin perturbations. The metastable states of this type are known as inherent structures in the theory of structural glasses \cite{sastry1998signatures} and kinetically-blocked configurations in jammed granular systems \cite{barrat2000edwards,biroli2001lattice}. Such systems are intrinsically out-of-equilibrium ones, therefore a long-standing debate concerns the proper use of thermodynamic approaches to investigate their physical properties \cite{baule2018edwards}. Approaches based on statistical mechanics were pioneered by S.F. Edwards \cite{edwards1989dislocations}, who conjectured that the physical properties of granular media should correspond to quasi-equilibrium steady states dominated by configurations maximizing the entropy (possibly plus constraints). This flat measure over the metastable states is essentially the same approach used in the present Section to describe the landscape of Nash equilibria for coordination games defined on graphs. Although this hypothesis has been criticized because it neglects completely the structure of the basins of attractions induced by system's dynamics (see e.g. \cite{eastham2006mechanism}), it proved extremely useful to understand general thermodynamic properties of amorphous systems \cite{baule2018edwards}. In agreement with the Edwards hypothesis, we use  statistical-mechanics methods to evaluate the flat-measure statistical properties of Nash equilibria, even though this approach has to be combined with more proper characterizations that could take into account the effects of dynamical equilibrium selection as well.  The cavity method was successfully used to study optimisation problems in finite-connectivity graphs \cite{mezard2002analytic,braunstein2005survey} and it was then employed to investigate the properties of Nash equilibria on networks for random games \cite{ramezanpour2011statistical}, public goods \cite{dall2009statistical,sun2016serving}, cooperation problems \cite{dall2012collaboration} and congestion games \cite{altarelli2015statics}. 

\subsection{Cavity method for Nash equilibria}\label{subsec:cavity}
The best-response conditions specified in Eq.~\eqref{best-response} can be viewed as a set of hard constraints between the binary variables defining the actions of the agents. Every solution of the set of constraints is a Nash equilibrium. For each realisation of the random variables $\vec{\theta}$, we can introduce a measure over the set of equilibria consistent with that realization by defining the partition function  
\begin{equation}\label{pf}
Z[\vec{\theta}] = \sum_{\vec{x}\in\{0,1\}^N} \prod_{i\in V} \id\left[x_i = b_i(\vec{x}_{\partial i}; \theta_i) \right] e^{-H(\vec{x})}.
\end{equation}
The energy-like term $H(\vec{x})$ is a function, composed of local interaction terms, that is used to bias the uniform Nash measure in order to analyse equilibria with some desired global or local property. In particular, we will consider $H(\vec{x}) = -\epsilon \sum_i x_i$, in which the Lagrange multiplier $\epsilon$ can be used to scan equilibria with a larger or smaller fraction of agents playing action $1$.

The cavity approach consists in assuming a (local) tree-like structure around each node $i$ and marginalise the probability measure associated with the partition function \eqref{pf} over the branches of the tree, obtaining a set of fixed-point equations for local probability marginals \cite{mezard2009information}. 
The best-response constraints involve a variable $x_i$ and all its neighbouring variables $\vec{x}_{\partial i}=\{ x_j| j\in \partial i\}$, as already shown for other game-theoretical problems \cite{dall2009statistical,ramezanpour2011statistical,dall2011optimal,altarelli2015statics,dall2012collaboration}. As a consequence,  the probability marginals defined on the directed edges $(i,j)$ of the graph and employed in the cavity approach have to depend on both variables $x_i$ and $x_j$. More precisely, we call $\eta_{ij}(x_i,x_j)$ the probability  that $i$ and $j$ take values $x_i$ and $x_j$ when all interactions involving $j$, but that with  $i$, are removed.  For a given realization $\vec{\theta}$ of the random variables, this probability can be computed as a function of similar quantities on the upstream nodes of the tree,
\begin{equation}\label{bp-instance}
  \eta_{ij}(x_i,x_j)   =\frac{1}{Z_{\rm c}} \sum_{\vec{x}_{\partial i\sm j}}\id\left[x_i = b_i(\vec{x}_{\partial i}; \theta_i) \right] e^{\epsilon x_i}  \prod_{k \in \partial i\sm j} \eta_{ki}(x_k,x_i).\\
\end{equation}  
The proportionality constant $Z_{\rm c}$ imposes the normalization condition $\sum_{x_i,x_j \in\{0,1\}} \eta_{ij}(x_i,x_j)=1$, hence leaving three independent probabilities for each directed edge.
Equations \eqref{bp-instance} are known as Belief Propagation (BP) equations and correspond to the cavity method in its replica-symmetric formulation on a given instance of the disorder (corresponding here both to the graph instance and the realization of threshold values).
The BP equations are exact on trees and usually provide very good approximations for the probability marginals on sparse graphs  \cite{mezard2009information}. 
From the fixed-point solution of the BP equations \eqref{bp-instance}, we can compute the Bethe approximation to the (negated) free-energy of the corresponding graphical model as 
\begin{equation}\label{bp-feps}
 Nf(\epsilon;\vec{\theta}\,)  =   \sum_{i\in V}f_i(\epsilon;\vec{\theta}\,) - \frac{1}{2}\sum_{(i,j)\in E} f_{ij}(\vec{\theta}\,)  
\end{equation}
with 
\begin{subequations}
\begin{align}
f_i(\epsilon;\vec{\theta}) & = -\ln \left\{\sum_{\vec{x}_{\partial i \cup i}} \id\left[x_i = b_i(\vec{x}_{\partial i}; \theta_i) \right]  e^{\epsilon x_i} \prod_{k \in \partial i} \eta_{ki}(x_k,x_i)  \right\},\\
f_{ij}(\vec{\theta}) & = - \ln\left\{ \sum_{ x_i,x_j} \eta_{ij}(x_i,x_j ) \eta_{ji}(x_j,x_i)\right\}.
\end{align}
\end{subequations}
The probability that agent $i$ plays action $1$ in the Nash equilibria weighted by the measure defined in Eq.~\eqref{pf} is estimated by means of the total marginals 
\begin{equation}\label{bp-rhoi}
\rho_i(\epsilon;\vec{\theta})  \propto   \sum_{\vec{x}_{\partial i \cup i}}  x_i \id\left[x_i = b_i(\vec{x}_{\partial i}; \theta_i) \right]  e^{\epsilon x_i}   \prod_{k \in \partial i} \eta_{ki}(x_k,x_i)
\end{equation}
and the average density of agents playing action $1$ in the equilibria is given by $\rho(\epsilon;\vec{\theta})= \sum_{i \in V} \rho_i(\epsilon;\vec{\theta}) /N$.  If we assume that for any density $\rho \in [0,1]$ of action $1$, the number of Nash equilibria could scale exponentially in $N$, in the large $N$ limit the partition function assumes the form 
\begin{equation}\label{bp-legendre}
e^{-N f(\epsilon;\vec{\theta})} = Z[\epsilon;\vec{\theta}]  \simeq \int d\rho e^{N s(\rho;\vec{\theta}) +  \epsilon N \rho(\vec{\theta})}
\end{equation}
 and relevant quantities such as the entropy $s(\rho;\vec{\theta})$ of Nash equilibria with a given density $\rho$ of agents playing action $1$ can be extracted by Legendre transform, performing the saddle-point integral. 
Using the same formalism, it is possible to weight Nash equilibria in terms of their (Hamming) distance from a reference configuration $x^*$ (which can be an equilibrium itself or not), investigating typical properties as well as large deviation properties. This is done by replacing the factor $e^{\epsilon x_i}$ with  $e^{\epsilon \left\{x_i^*(1-x_i)+(1-x_i^*)x_i\right\}}$ in the BP equations \eqref{bp-instance} and defining the intensive distance function $d(\epsilon;x^*,\vec{\theta})$ as follows 
\begin{equation}
d(\epsilon;x^*,\vec{\theta}) \propto \frac{1}{N}\sum_{i \in V} \sum_{\vec{x}_{\partial i \cup i}}   \left\{x_i^*(1-x_i)+(1-x_i^*)x_i\right\} \id\left[x_i = b_i(\vec{x}_{\partial i}; \theta_i) \right]  e^{\epsilon  \left\{x_i^*(1-x_i)+(1-x_i^*)x_i\right\}}   \prod_{k \in \partial i} \eta_{ki}(x_k,x_i)
\end{equation}
The entropy $s(d)=s(d;x^*,\vec{\theta})$ of Nash equilibria at some distance $d$ from a reference configuration $x^*$ can be computed by Legendre transform in analogy with $s(\rho;\vec{\theta})$.
 
 This is for a given realisation of the graph $\mathcal{G}$ and thresholds $\vec{\theta}$. The average value $f(\epsilon)=\mathbb{E}_{\theta}[f(\epsilon;\vec{\theta})]$ of the free energy over the quenched random variables $\vec{\theta}$ provides information on the Nash Equilibria with {\em typical} properties with respect to the distribution of  random variables, in the presence of the energetic bias $\epsilon$. From the average free-energy $f(\epsilon)$ and the corresponding average density $\rho(\epsilon)$, an approximation for the average entropy of the Nash equilibria can be computed straightforwardly as $s(\rho) = - f -\epsilon \rho$, which gives an estimate of the typical number $\mathcal{N}_{NE}(\rho)\simeq e^{N s(\rho)}$ of equilibria of a system with a given density $\rho$ of agents playing action $1$.

For convenience we can define on each directed edge a four-dimensional vector $\vec{\eta} = (\eta(0,0),\eta(1,0),\eta(0,1),\eta(1,1)) \in [0,1]^4$ with the normalization constraint $|\vec{\eta}|=\sum_{x_i,x_j \in\{0,1\}} \eta(x_i,x_j)=1$. Because of the random realization of the threshold values, to each directed edge $(i,j)$ we have to assign a distribution $P_{ij}[\vec{\eta}]$ of cavity marginals, which  satisfies a distributional (replica-symmetric) BP equation
\begin{equation}\label{bp-instanceP}
  P_{ij}[\vec{\eta}]   \propto \int \prod_{k\in \partial i\setminus j} \left[d\vec{\eta}_{ki} P_{ki}[\vec{\eta}_{ki}] \right] d\theta_i f(\theta_i) \delta \left(\vec{\eta} - \vec{\mathcal{F}}_{BP}[\{\vec{\eta}_{ki}\}_{k\in\partial i \setminus j}] \right).
\end{equation}
where $\vec{\mathcal{F}}_{BP}[\{\vec{\eta}_{ki}\}_{k\in\partial i \setminus j}]$ is a shorthand for the (vectorial version of the) r.h.s. of \eqref{bp-instance}.
The average free energy is then computed as
\begin{align}
\nonumber  f(\epsilon)  = & - \sum_{i\in V}   \int \prod_{k\in \partial i} d\vec{\eta}_{ki} P_{ki}[\vec{\eta}_{ki}] \mathbb{E}_{\theta}\left[ \ln \left\{\sum_{\vec{x}_{\partial i \cup i}} \id\left[x_i = b_i(\vec{x}_{\partial i}; \theta) \right]  e^{\epsilon x_i} \prod_{k\in\partial i} \eta_{ki}(x_k,x_i)  \right\}\right]  \\ 
  &  \quad + \frac{1}{2} \sum_{(i,j)\in E} \int d\vec{\eta}_{ij} d\vec{\eta}_{ji} P_{ij}[\vec{\eta}_{ij}] P_{ji}[\vec{\eta}_{ji}] \ln\left\{ \sum_{ x_i,x_j} \eta_{ij}(x_i,x_j) \eta_{ji}(x_j,x_i)\right\},\label{bp-fepsDIS}
\end{align}
while the probability that agent $i$ plays action $1$ in the set of Nash equilibria is  
\begin{equation}\label{bp-rhoiP}
\rho_i(\epsilon)  \propto  \int \prod_{k\in \partial i } \left[d\vec{\eta}_{ki} P_{ki}[\vec{\eta}_{ki}] \right] \mathbb{E}_{\theta}\left[   \sum_{\vec{x}_{\partial i \cup i}}  x_i \id\left[x_i = b_i(\vec{x}_{\partial i}; \theta) \right]  e^{\epsilon x_i}   \prod_{k \in \partial i} \eta_{ki}(x_k,x_i)\right].
\end{equation}

\begin{figure}[tb]
\begin{center}
\includegraphics[width=0.6\columnwidth]{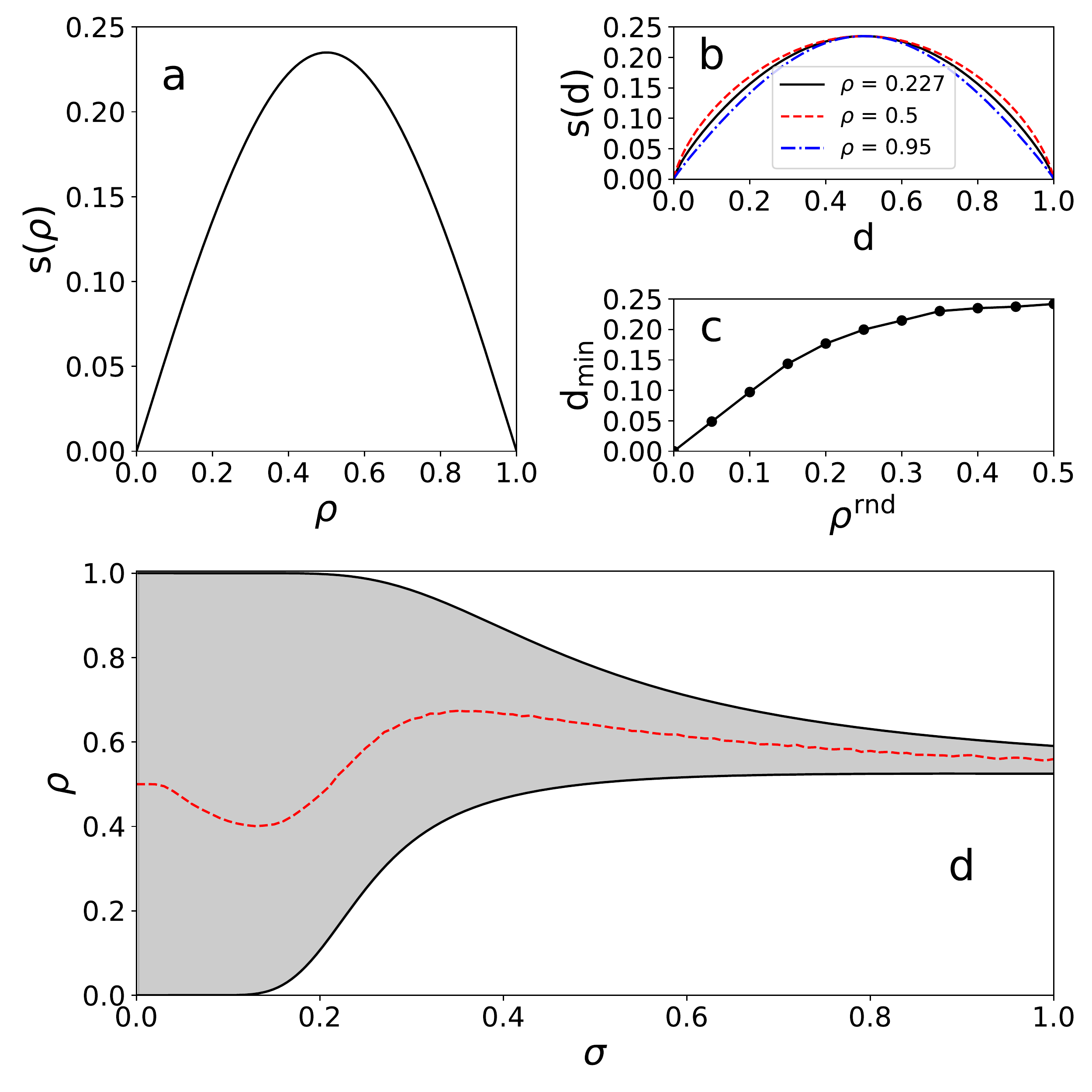}
\caption{Random regular graphs of degree $K=3$ and gaussian thresholds with mean $\mu=0.4$ and standard deviation $\sigma=0$: (a) Entropy $s$ of Nash equilibria as function of the corresponding density $\rho$ of agents playing action $1$; (b) Entropy $s$ of equilibria at a Hamming distance $d$ from a reference equilibrium, that is a typical one at density $\rho = 0.227$ (black solid line), $0.5$ (red dashed line), $0.95$ (blue dot-dashed line); (c) Minimum Hamming distance from a reference random profile at which Nash equilibria can be found as function of the density $\rho^{\rm rnd}$ of the latter. (d) Density spectrum of Nash equilibria as function $\sigma$; the dashed area is the density region in which equilibria can be found while the red dashed line is the density of most numerous equilibria.
}\label{fig:K3all}
\end{center}
\end{figure}

\subsection{Results for random regular graphs}\label{subsec-regularRG}
In the case of a random regular graph of degree $K$, all marginal distributions are equal, i.e. $P_{ij}[\vec{\eta}_{ij}]=P[\vec{\eta}]$, $\forall (i,j)\in E$, the distributional BP equations \eqref{bp-instanceP} simplify into 
\begin{equation}\label{bp-instancePregular}
  P[\vec{\eta}]   \propto \int \prod_{k=1}^{K-1} \left[d\vec{\eta}_{k} P[\vec{\eta}_{k}] \right] \mathbb{E}_\theta\left[ \delta \left(\vec{\eta} - \vec{\mathcal{F}}_{BP}[\{\vec{\eta}_{k}\}_{k=1}^{K-1}] \right)\right].
  \end{equation}
The distributional BP equations for ensembles of random graphs  can be obtained using a standard but more involved derivation by means of the replica method (see App.~\ref{sec:replica}). Equations \eqref{bp-instancePregular} can be solved by the following population dynamics approach (also known as density evolution) \cite{mezard2009information}. Consider a population $\{\vec{\eta}_i \}_{i=1}^M$ of $M$ vectors. For each vector $\vec{\eta}_i$, the four components are initially drawn from a uniform distribution in $[0,1]$ and the overall normalization $|\vec{\eta}_i|=1$ is imposed. Then,
\begin{enumerate}
\item Sample uniformly at random $K-1$ elements $\{\vec{\eta}_k\}_{k=1}^{K-1}$ from the population,
\item Draw a value of $\theta \sim f(\theta)$,
\item Compute the BP update $\vec{\eta}^{\rm new} = \vec{\mathcal{F}}_{BP}[\{\vec{\eta}_{k}\}_{k=1}^{K-1}]$,
\item Replace one element of the population (chosen uniformly at random) with $\vec{\eta}^{\rm new}$.  
\end{enumerate}
These steps have to be repeated until the empirical distribution over the population converges. More conveniently, one can focus on the convergence of some quantity of interest computed over the population of cavity marginals, such as the free energy $f$ and/or the density $\rho$. The population dynamics method is computationally demanding and requires large populations. A faster alternative method, which seems to provide comparable numerical results in the present case, is based on direct sampling, i.e. on the numerical calculation of the average free energy from a sample of free energy values obtained solving the BP equations for a large number of instances of the disorder $\vec{\theta}$ (on a single large instance of the random regular graph). The latter method can be very inaccurate in optimization problems, in which the static measure is usually dominated by rare events; on the contrary, it seems to work pretty well here as long as the replica symmetry holds. 

\begin{figure}[tb]
\begin{center}
\includegraphics[width=0.6\columnwidth]{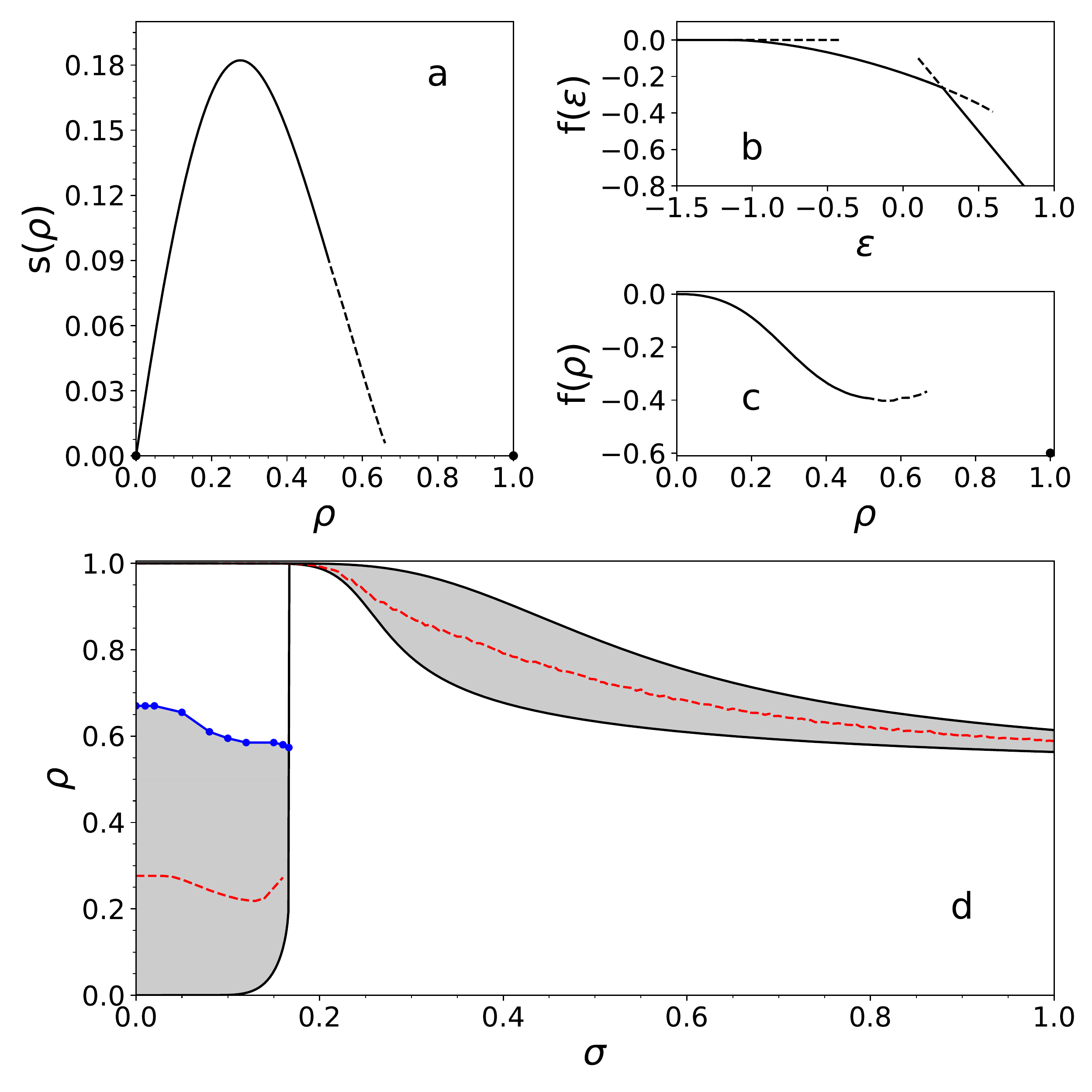}
\caption{Random regular graphs of degree $K=4$ and gaussian thresholds with mean $\mu=0.35$ and standard deviation $\sigma=0$: (a) Entropy $s$ of Nash equilibria as function of the corresponding density $\rho$ of agents playing action $1$; (b) Free-energy $f$ of equilibria as function of the control parameter $\epsilon$ conjugate to the density $\rho$; (c) Free-energy $f$ of equilibria as function of the density $\rho$. (d) Density spectrum of Nash equilibria as function $\sigma$; the dashed area is the density region in which equilibria can be found, with the upper boundary (blue) distinct from the isolated equilibrium at density $\rho=1$; the red dashed line is the density of most numerous equilibria.
}\label{fig:K4sigma}
\end{center}
\end{figure}

We used the belief propagation approach to explore the region of equilibria between minimum and maximum density values (located using the methods presented in Section~\ref{sec:monotoneBR}), where previous results suggest the existence of a number of Nash equilibria possibly scaling exponentially with the number of agents \cite{echenique2007finding,lucas2013multistable,lee2014simple,jackson2015games}. The simplest case under study is the one of  random regular graphs with homogeneous thresholds ($\sigma=0$), for which the set of pivotal agents is either the whole graph or the empty set, depending on the value of the thresholds \cite{lelarge2012diffusion}. It follows that, below $\mu_{\rm c,m}$, only the two trivial equilibria $\vec{0}$ and $\vec{1}$ exist, for any $K$. Above the percolation-like transition, two typical behaviours can be observed that are well represented by the cases $K=3$ and $K=4$. Figure~\ref{fig:K3all}a shows that for $K=3$ and $\mu =0.4 > \mu_{\rm c,m}$ the entropy of equilibria $s(\rho)$ has a continuous support for $\rho \in [0,1]$, meaning that one can find Nash equilibria with any fraction of agents playing action 1. At any density value $\rho$,  a Nash equilibrium is surrounded by many others with very similar density as demonstrated by the continuous curve obtained for the entropy $s(d)$ of equilibria at Hamming distance $d$ from an equilibrium taken as a reference point (Fig.~\ref{fig:K3all}b). Although their number is exponentially large in the size of the graph, Nash equilibria are very rare among possible action profiles and characterised by non-trivial correlations. This is also demonstrated by measuring the minimum distance of equilibria from a random action profile (not an equilibrium) with given density $\rho^{\rm rnd}$ taken as a reference point (Fig.~\ref{fig:K3all}c): the minimum distance is finite in general and vanishes only for $\rho \to0$, as expected. This result, which is observed also for larger values of $K$, has an implication on the dynamics, because it means that, {\em even in the absence of a cascade process, reaching a Nash equilibrium from a random profile requires a large number of local rearrangements}. 
Yet in terms of density $\rho$, a continuous set of equilibria exists for every $\sigma \geq 0$, even though the density interval covered by equilibria shrinks monotonically for larger values of $\sigma$ (see Fig.~\ref{fig:K3all}d) as  predicted also by \eqref{bp-lo}-\eqref{bp-lo-tot} and \eqref{bp-up}-\eqref{bp-up-tot}. The red dashed line indicates the location of the maximum of the entropy, which is the density corresponding to the most numerous equilibria. 

\begin{figure}[tb]
\begin{center}
\includegraphics[width=0.6\columnwidth]{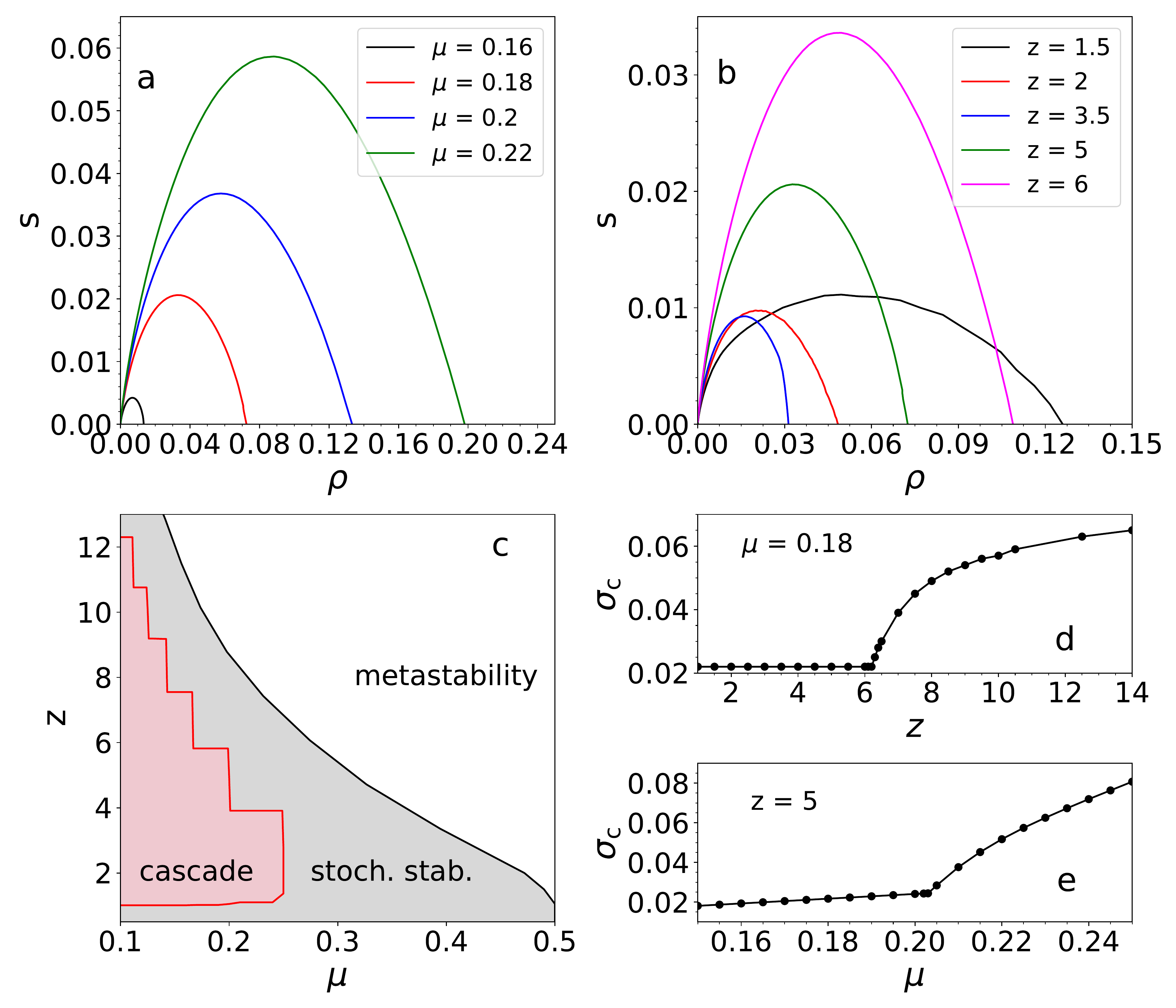}
\caption{Erd\H{o}s-R\'enyi (ER) random graphs of average degree $z$ and gaussian thresholds with mean $\mu$ and standard deviation $\sigma=0$: Entropy $s$ of Nash equilibria as function of the corresponding density $\rho$ of agents playing action $1$ for (a) $z=5$ and $\mu=0.16, 0.18, 0.2, 0.22$ and for (b) $\mu = 0.18$ and $z = 1.5, 2, 3.5, 5, 6$; (c) phase boundaries in the $(\mu,z)$-plane. (d-e) Critical value $\sigma_{\rm c}$ of heterogeneity at which the instability in the lower edge (minimum equilibrium) takes place as function of (d) $z$ for $\mu =0.18$ and (e) $\mu$ for $z=5$.}\label{fig:ER1}
\end{center}
\end{figure}

For $K=4$, a new non-trivial phenomenon emerges: the spectrum of Nash equilibria is not continuous with respect to $\rho$ in the whole support $[0,1]$, even in the absence of threshold heterogeneity ($\sigma=0$). Fig.~\ref{fig:K4sigma}a shows that for $\mu=0.35$, the fully coordinated equilibrium at $\rho=1$ is isolated as the entropic curve vanishes at much smaller density values. The dashed part of the curve represents equilibria that are ``thermodynamically'' unstable (see also the behaviour of the free-energy in Fig.~\ref{fig:K4sigma}b-\ref{fig:K4sigma}c) but, at least in principle, dynamically reachable by means of best-response and other rearrangement processes \footnote{In practice we found them using both Belief Propagation reinforcement and decimation processes, two popular techniques employed to find solution to combinatorial optimisation problems \cite{braunstein2007encoding,dall2008entropy,dall2009statistical}.}. No equilibria, except for $\vec{1}$, were found on finite graphs in the density region beyond the point where the entropy is predicted to vanish (within the replica-symmetric approximation). Rather strong correlations are observed in the high density region close to the zero entropy point, suggesting that a more accurate description of the properties of equilibria in this region could require going beyond the replica symmetric approach considered in the present work. The properties observed for $\mu=0.35$ are common to the whole region in which the minimum and maximum equilibria are locally stable under best-response dynamics. Remarkably, the density region in which equilibria with a non-trivial coexistence of actions $0$ and $1$ are stable seems to approximately correspond to the attraction basin of the $h=1$ solution of Eq.~\eqref{bp-lo} (compare with the corresponding curves in Fig.\ref{fig:rho_lowREG}). Notice that, for  random regular graphs of low degree and uniform threshold values, the replica-symmetric belief propagation approach does not require distributional equations, but rather BP equations of the type of \eqref{bp-instance}, therefore the mechanism underlying the onset of non-trivial equilibria can be analysed in detail. It turns out that the fixed points corresponding to non-trivial Nash equilibria (i.e. equilibria with density $\rho\in(0,1)$) emerge as function of the control parameter $\epsilon$ by means of a general combined mechanism involving a transcritical bifurcation and a saddle-node bifurcation in a reduced two-dimensional space, which depends quantitatively but not qualitatively on the values of degree and threshold (see App.~\ref{app-BPregular} for the details of the analysis). 

The effect of non-homogeneous thresholds is displayed in Fig.~\ref{fig:K4sigma}d (for $K=4$ and $\mu=0.35$), reporting the lower and upper edges of the spectrum of equilibria in which the grey area represents the regions in which Nash equilibria can be actually found. Again, the red dashed line indicates the location of the maximum of the entropy, that is the density corresponding to most numerous equilibria. Without explicitly reporting the results, we notice that the qualitative behaviour of $s(d)$ and $d_{\rm min}$ is the same as for $K=3$, although for $K\geq 4$ the range of variation of the Hamming distance is limited by the reduced density interval for which equilibria exist. 

\begin{figure}[tb]
\begin{center}
\includegraphics[width=0.45\columnwidth]{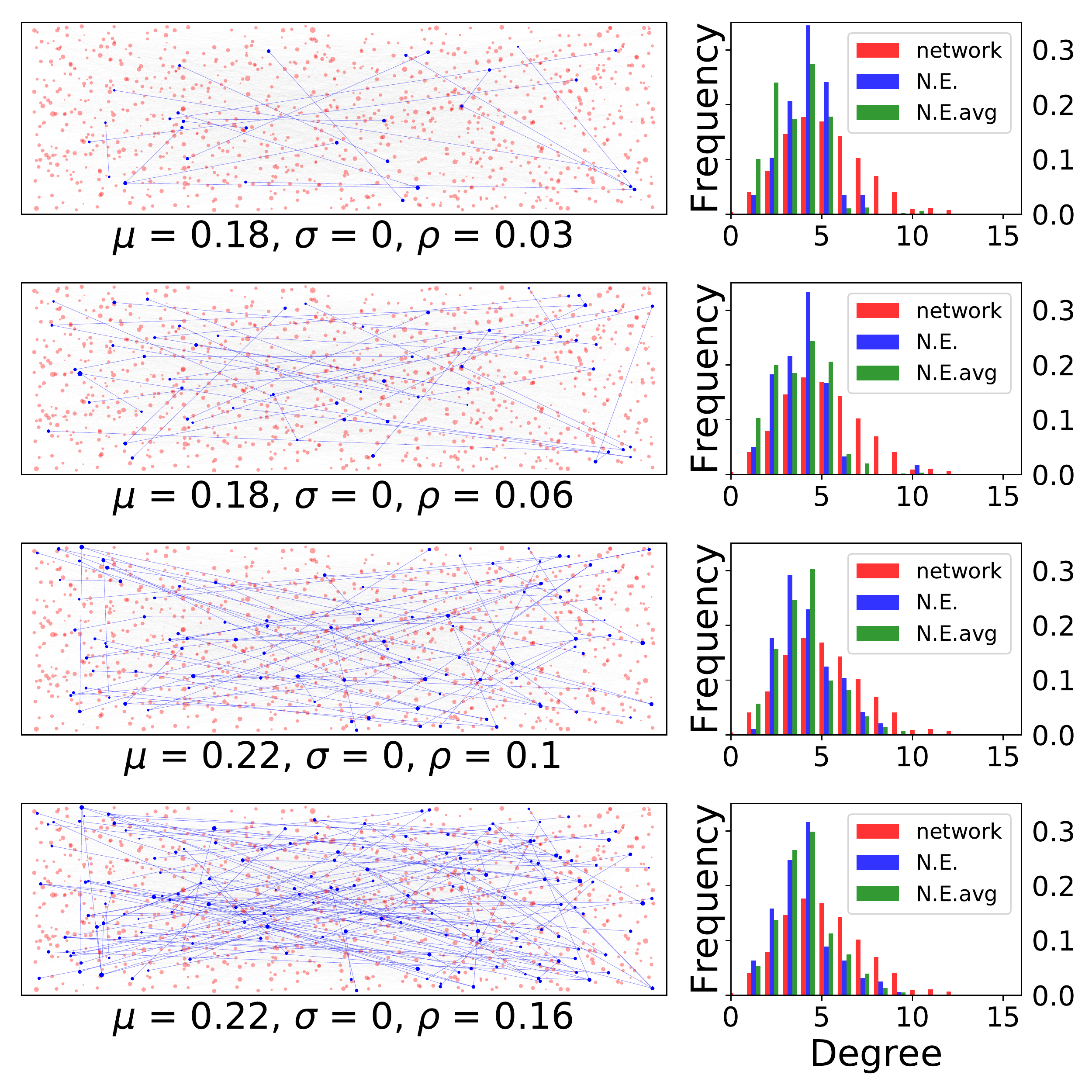} \quad
\includegraphics[width=0.45\columnwidth]{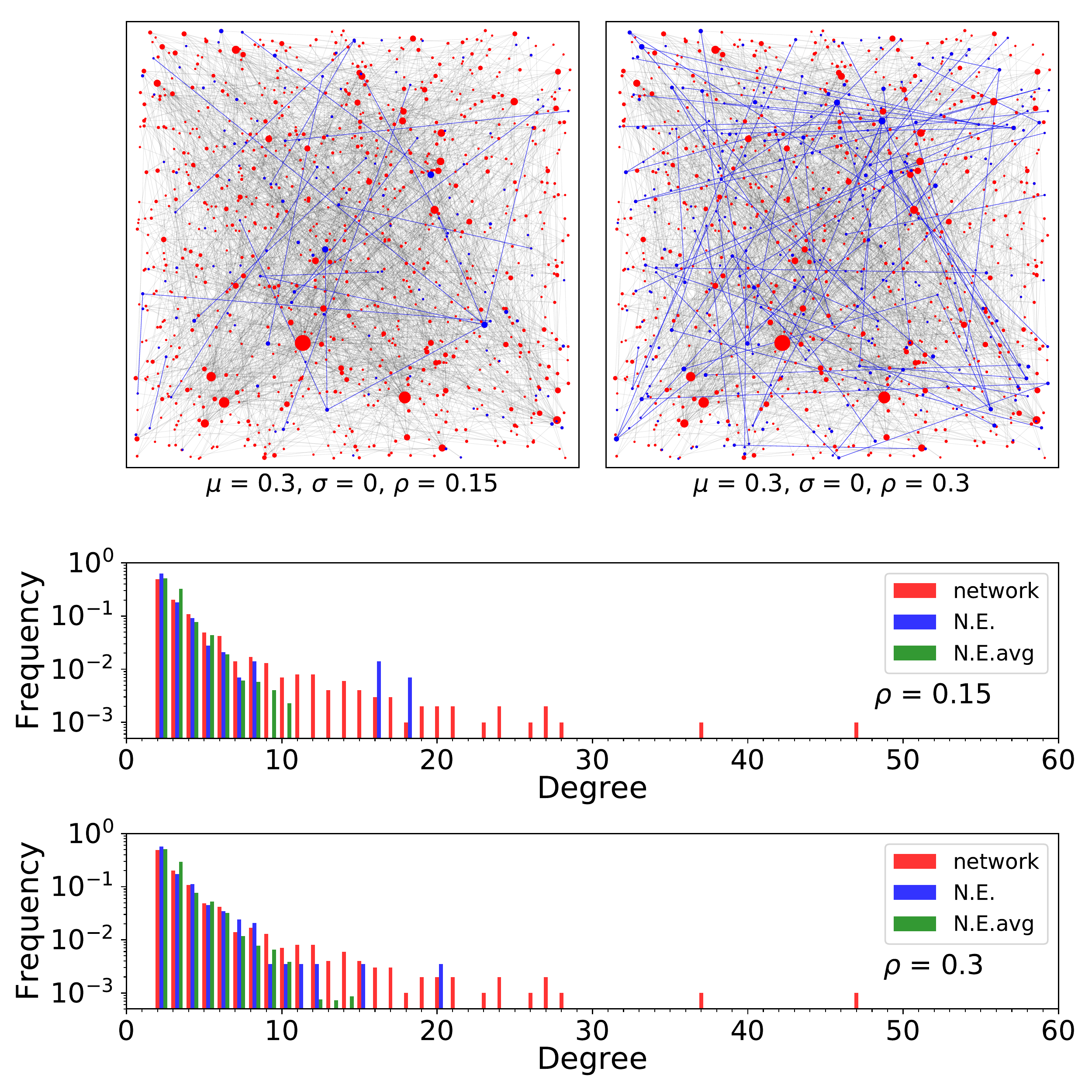}
\caption{(Left) Examples of Nash equilibria with different density of agents playing action $1$ for the model with gaussian thresholds (mean $\mu$ and standard deviation $\sigma$) on Erd\H{o}s-R\'enyi (ER) random graphs of average degree $z=5$. The plots in the left panels display the subgraphs induced by agents playing action $1$ in a Nash equilibrium. The histograms on the right panels represent the degree distribution of the network (red), the theoretical prediction of the experimental degree distribution in the subnetwork induced by agents playing action $1$ in a typical Nash equilibrium (blue) and the corresponding theoretical prediction obtained from the cavity method approach (green). The size of the nodes is proportional to their degree.\\
(Right) Examples of Nash equilibria with different density of agents playing action $1$ for the model with gaussian thresholds (mean $\mu=0.3$ and standard deviation $\sigma=0$) on Barab\`asi-Albert (BA) random network of $N=1000$ nodes and minimum degree $k_{\rm min}=2$. The top panels display the subgraphs induced by agents playing action $1$ in Nash equilibria with $\rho=0.3$ and $\rho=0.15$ . The histograms on the bottom panels represent the degree distribution of the network (red), the theoretical prediction of the experimental degree distribution in the subnetwork induced by agents playing action $1$ in a typical Nash equilibrium (blue) and the corresponding theoretical prediction obtained from the cavity method approach (green).  The size of the nodes is proportional to their degree.
}
\label{fig:ERBAdraws}
\end{center}
\end{figure}

\subsection{Results on non-regular random graphs}\label{subsec:cavityGRG}

Classical results on the cascade properties of the monotone decision process on Erd\H{o}s-R\'enyi (ER) random graphs in the case of uniform thresholds ($\sigma=0$), based on the role of pivotal agents \cite{morris2000contagion,watts2002simple,lelarge2012diffusion}, are discussed in App.~\ref{app-pivotal}. In brief, isolated agents always choose action $0$, while a non-trivial behaviour can be observed on the connected components of the graph. For average degree $z$ in the range $[z_{\rm m}(\mu),z_{\rm M}(\mu)]$, the minimal equilibrium at $\vec{0}$  becomes unstable with respect to perturbations on pivotal nodes, which trigger a cascade leading to a non-trivial pivotal equilibrium with a finite density of agents playing action $1$ (see also Fig.~\ref{fig:ERlelarge}a). The cascade phase boundary identified in this way \cite{watts2002simple} is displayed  in Fig.~\ref{fig:ER1}c (red line). A further result by M. Lelarge  \cite{lelarge2012diffusion}, reproduced in  App.~\ref{app-pivotal}, shows that the coexistence, in the same pivotal equilibrium, of an extensive connected components of agents playing action $0$ and one of agents playing action $1$ is possible only for very low values of the average degree, close to the lower boundary of the cascade region. Coexistence is however possible in other Nash equilibria, whose properties can be studied by means of the BP approach developed in Section \ref{subsec:cavity}. In particular, a large set of equilibria different from the pivotal ones but containing a finite density of nodes playing action $1$ is predicted  to exist almost everywhere in the $(\mu,z)$ plane, independently of the phase boundaries determined by the  cascade process. 
The entropy vs. density curves for different values of  $z$ and $\mu$ in Fig.~\ref{fig:ER1}a-\ref{fig:ER1}b unveil that no qualitative change is observed crossing the cascade boundaries: {\em a large number of equilibria at small density values exists below as well as above the cascade transition}. The main difference resides in their stability properties, because in the cascade region, such equilibria are not stable under small perturbations and best-response dynamics rapidly converge to the most efficient ones (see Section \ref{subsec:BRdynamics}).

A  difference in the structure of equilibria below and above the cascade condition is highlighted by Fig.~\ref{fig:ERBAdraws} (left), in which we report some examples of Nash equilibria at low density on ER random graphs, for $\mu$ values belonging to the two regions. The empirical degree distribution of nodes corresponding to agents playing action $1$ (blue) is compared with the theoretical prediction for the average behaviour obtained with the cavity method (green) and with the degree distribution of the underlying random graph (red). In low-density equilibria, action $1$ is played preferentially by low-degree nodes; this is particularly apparent in the region in which cascades take place, since  high-degree nodes should be avoided due to their crucial role in triggering cascades. The absence of high-degree nodes from the set of action-$1$ players in low-density (low-efficiency) equilibria is even more evident in very heterogeneous graphs (see right panels in Fig.~\ref{fig:ERBAdraws}). For heterogeneous random networks, a qualitative picture of the thermodynamic properties of equilibria is shown in Fig.~\ref{fig:BASFthermo}, in which it is clearly visible the density values at which low-efficiency equilibria become (thermodynamically, not dynamically) unstable with respect to the one of maximum coordination. A direct comparison with homogeneous random graphs with same size and average degree unveils that, in heterogenous graphs, the continuum part of the equilibrium spectrum is composed by a larger number of equilibria (larger entropy) and its density support is larger. The reason is purely combinatorial: heterogenous random graphs contain a proportionally larger fraction of low-degree nodes and the latter are those which preferentially play action 1 in these equilibria. A relevant question which will be addressed in the next section is about the dynamic stability of this huge set of equilibria also in relation to the same properties for homogenous random networks.

\begin{figure}[tb]
\begin{center}
\includegraphics[width=0.8\columnwidth]{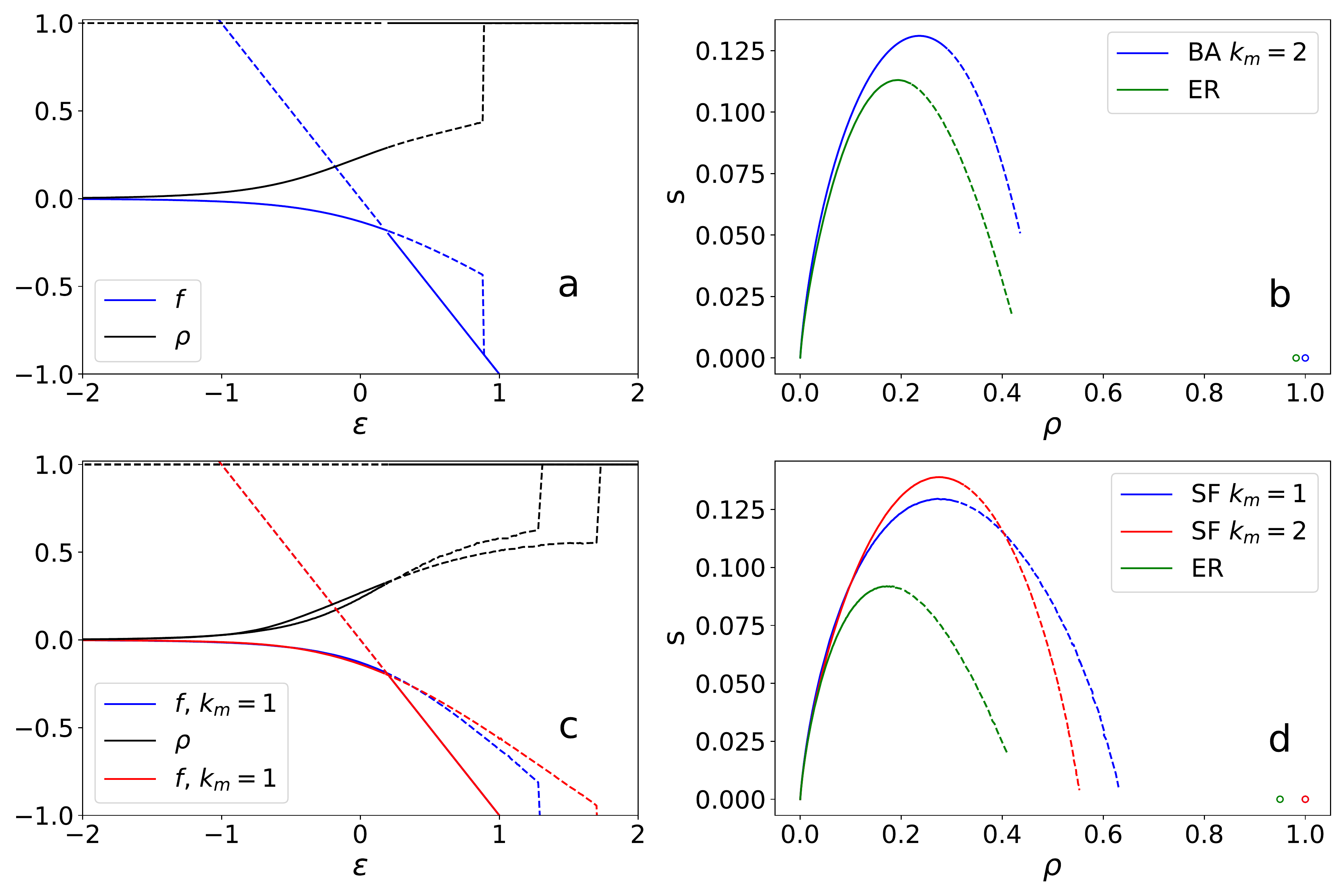}
\caption{Thermodynamic properties of equilibria in heterogeneous random networks for $\mu=0.3$ and $\sigma=0$: (a)-(c) Free energy $f$  (blue and red lines) and density $\rho$ (black lines) as function of the Lagrange multiplier $\epsilon$ and (b)-(d) entropy $s$ of Nash equilibria as function of the density $\rho$ (blue). Network considered are (a)-(b) Barab\`asi-Albert (BA) random network of $N=10^4$ nodes and minimum degree $k_{\rm min}=2$ and (c)-(d) scale-free random networks (SF) with power-law degree distribution with exponent $\gamma=2.8$, size $N \sim 10^4$ nodes and minimum degree $k_{\rm min}=1,2$. Dashed lines represent thermodynamically unstable branches (because the maximum-density equilibrium becomes stable). In (b)-(d), the entropy of equilibria (green) for homogeneous Erd\H{o}s-R\'enyi (ER) random graphs of the same size and average degree is also drawn for comparison.
 }\label{fig:BASFthermo}
\end{center}
\end{figure}

\section{Dynamical equilibrium selection}\label{sec:dynamics}
The existence of multiple equilibria in coordination games on networks raises a question about which of them can preferably be achieved through the various  dynamical self-organization processes in which agents can possibly take part. According to the static analysis performed in the previous Section, all pure Nash equilibria can be at least in principle achieved as fixed points of the deterministic best-response dynamics. As evidenced by the identified threshold condition for global contagion  \cite{morris2000contagion,watts2002simple,gleeson2007seed,lelarge2012diffusion}, Nash equilibria have instead rather different basins of attraction, whose properties are very difficult to study analytically and even computationally \cite{pangallo2019best}. We refer to App.~\ref{app-equilibrium-selection} for a brief review of most relevant theoretical and experimental results on dynamical equilibrium selection in coordination games, especially those defined on networks. In this Section, we employ both numerical simulations and approximated methods to study and compare representative processes of three major classes of dynamical equilibrium selection: best response, bounded-rational dynamics and learning processes. We mostly focus on random regular graphs and on a region of model parameters ($\mu<0.5$ and $\sigma$ small), in which the most efficient (payoff-dominant) equilibrium is also globally stochastically stable (risk-dominant), although both equilibria could be simultaneously locally stochastically stable. Metastability together with the entropic effects associated with the existence of many equilibria at low efficiency strongly affect the convergence properties of all types of dynamics under study.  Finally, we perform a comparative static analysis evaluating the effects on convergence properties of varying most relevant network properties, such as edge density, degree heterogeneity and clustering. We also briefly address the case in which the globally stochastically stable (risk-dominant) equilibrium differs from the payoff-dominant one, a situation which was largely investigated in the experimental literature.

\subsection{Best-response dynamics}\label{subsec:BRdynamics}
Myopic {\em best-response} (BR) dynamics is generally considered the preferential process of dynamical equilibrium selection in game theory and consists in the iterative adjustment of agents' actions in order to selfishly maximize individual payoffs at each step of the process  \cite{ellison1993learning}. We consider a random sequential update of individual actions, in which at each time interval $\Delta t$ a node $i$ is chosen uniformly at random to revise her action  by the rule
\beq\label{brDyn}
 x_i(t+\Delta t) = b_i(\vec{x}_{\partial i}(t); \theta_i),
\eeq
where $b_i(\cdot)$ is the best-response relation defined in \eqref{best-response}.
From any initial condition, when the best-response process converges to a fixed-point, the latter is a (possibly, mixed-strategy) Nash equilibrium. For potential games, the convergence of the random sequential process to a pure Nash equilibrium is always guaranteed (see e.g. \cite{monderer1996potential,nisan2007algorithmic}).  

\begin{figure}[tb]
\begin{center}
\includegraphics[width=0.6\columnwidth]{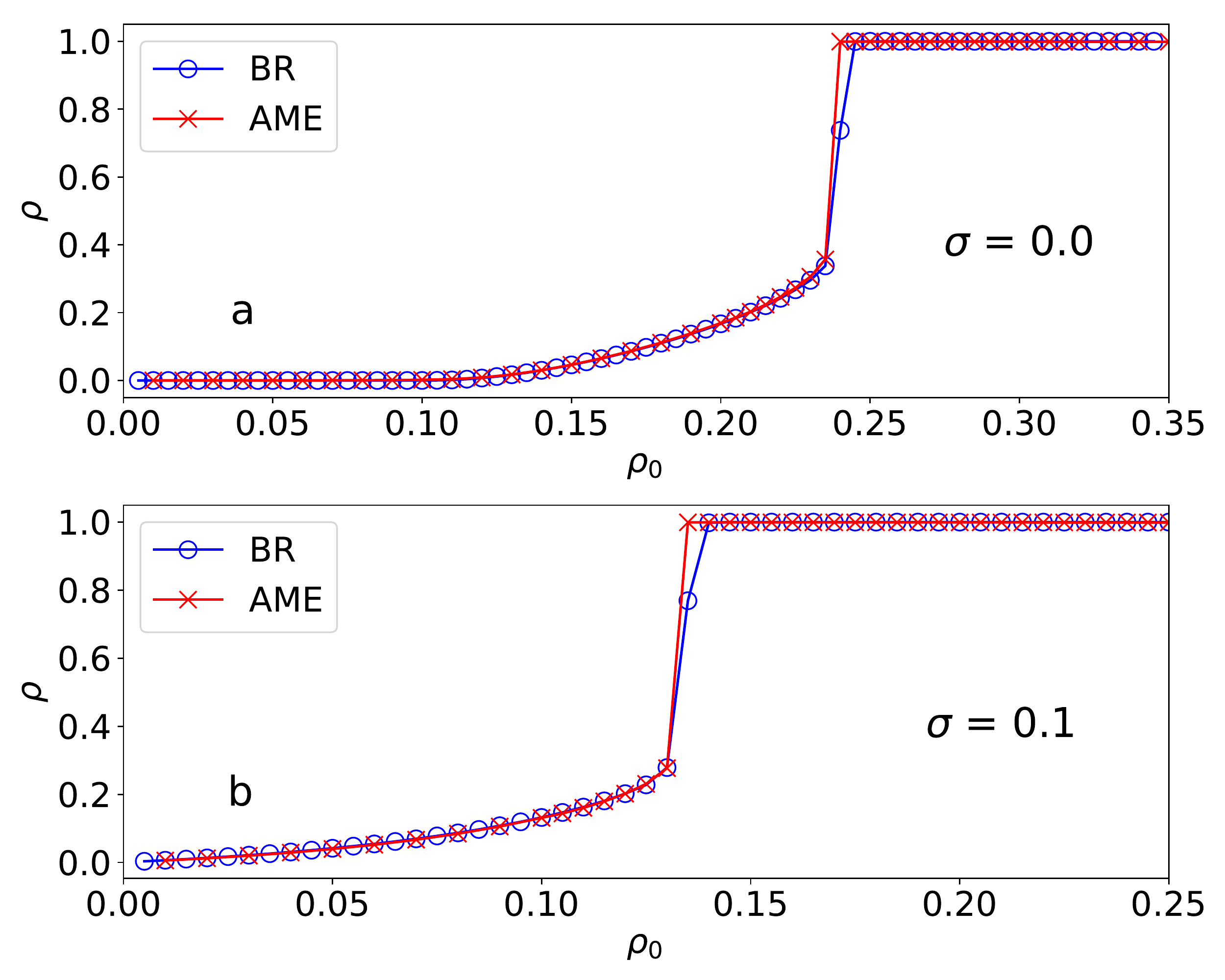}
\caption{Density $\rho$ of agents playing action $1$ in Nash equilibria reached by best-response (BR) dynamics as function of the initial density $\rho_0$ on random regular graphs of degree $K=4$ and disorder parameters (a) $\mu=0.35$, $\sigma=0$ and (b) $\mu=0.35$, $\sigma=0.1$. The prediction obtained from the approximate master equation (AME) method are in very good agreement with numerical simulations of BR dynamics on sufficiently large graphs ($N=5 \cdot 10^4$). The jump to the maximal equilibrium seems to occur at values of the initial density for which the low-coordination equilibria also becomes thermodynamically unstable.}\label{fig:BRame}
\end{center}
\end{figure}

When the initial condition of the BR process is just a small perturbation of the $\vec{0}$ profile, the dynamics converge to equilibria with statistical properties that are well described by the monotone decision process studied in Sec.~\ref{sec:monotoneBR}. However, best-response dynamics is not a  monotone process in general: when the initial condition is arbitrary, the density of agents playing action $1$ in the equilibrium eventually reached by the dynamics is not strictly larger than that of the initial configuration. It is thus interesting to investigate how much the initial conditions influence the final equilibrium and what is the minimum initial density for which complete coordination is reached as function of the parameters of the model. Numerical results obtained performing BR dynamics can be compared with results from a semi-analytic approximation scheme, called  {\em approximate master equation} (AME)  \cite{gleeson2011high,gleeson2013binary}. This method is widely used to study (markovian) binary-state dynamics on random graphs ensembles, where it turns out to be very accurate when the update rule of the binary variables defined on nodes only depends on the current state of the neighbouring variables. This condition is satisfied by the best-response dynamics considered here. The AME method gives a continuous-time evolution equation for the approximate marginal probability $q^x_{k}(m,t)$ that a node plays action $x$ and $m$ out of her $k$  neighbours play action $1$ at time $t$. The details of the method are described in App.~\ref{app-AME}. In order to average over the site disorder, the AME method is implemented by means of a population dynamics approach: we consider a large population of time-dependent marginals $\{q^{x}_{k_i,\theta_i}(m,t)\}_{i=1}^M$, each one characterized by a degree value $k_i$ and a threshold value $\theta_i$ drawn, respectively, from the distributions $p_k$ and $f(\theta)$.  Figure~\ref{fig:BRame} displays the stationary density $\rho$ as function of the initial density $\rho_0$ for ensembles of random regular graphs of degree $K$. It shows that for randomly chosen initial conditions with low density of agents playing action $1$, the best-response dynamics converges to a set of low-coordination Nash equilibria. The average density $\rho$ of agents playing action $1$ in the equilibria selected by best response increases monotonically increasing the density $\rho_0$ in the initial conditions. A sudden jump to the fully coordinated equilibrium ($\rho=1$) occurs beyond a threshold value $\rho_{\rm 0,c}^{\rm BR}$, that depends on the parameters of the coordination game (and on the properties of the underlying network). We remark that for all parameter values considered, even if not reported in the figures, {\em the discontinuity seems to take place approximately for the same values of $\rho$ for which we have the thermodynamic instability of partially-coordinated Nash equilibria in favour of the fully-coordinated one, i.e. discontinuity of free energy $f(\epsilon)$ as function of $\epsilon$ or when $f(\rho)$ changes convexity} (see eg, Fig.~\ref{fig:K4sigma}b-\ref{fig:K4sigma}c for $\sigma = 0$). Unfortunately, we have no deeper understanding of this phenomenon.   
 
 \begin{figure}[tb]
\begin{center}
\includegraphics[width=0.8\columnwidth]{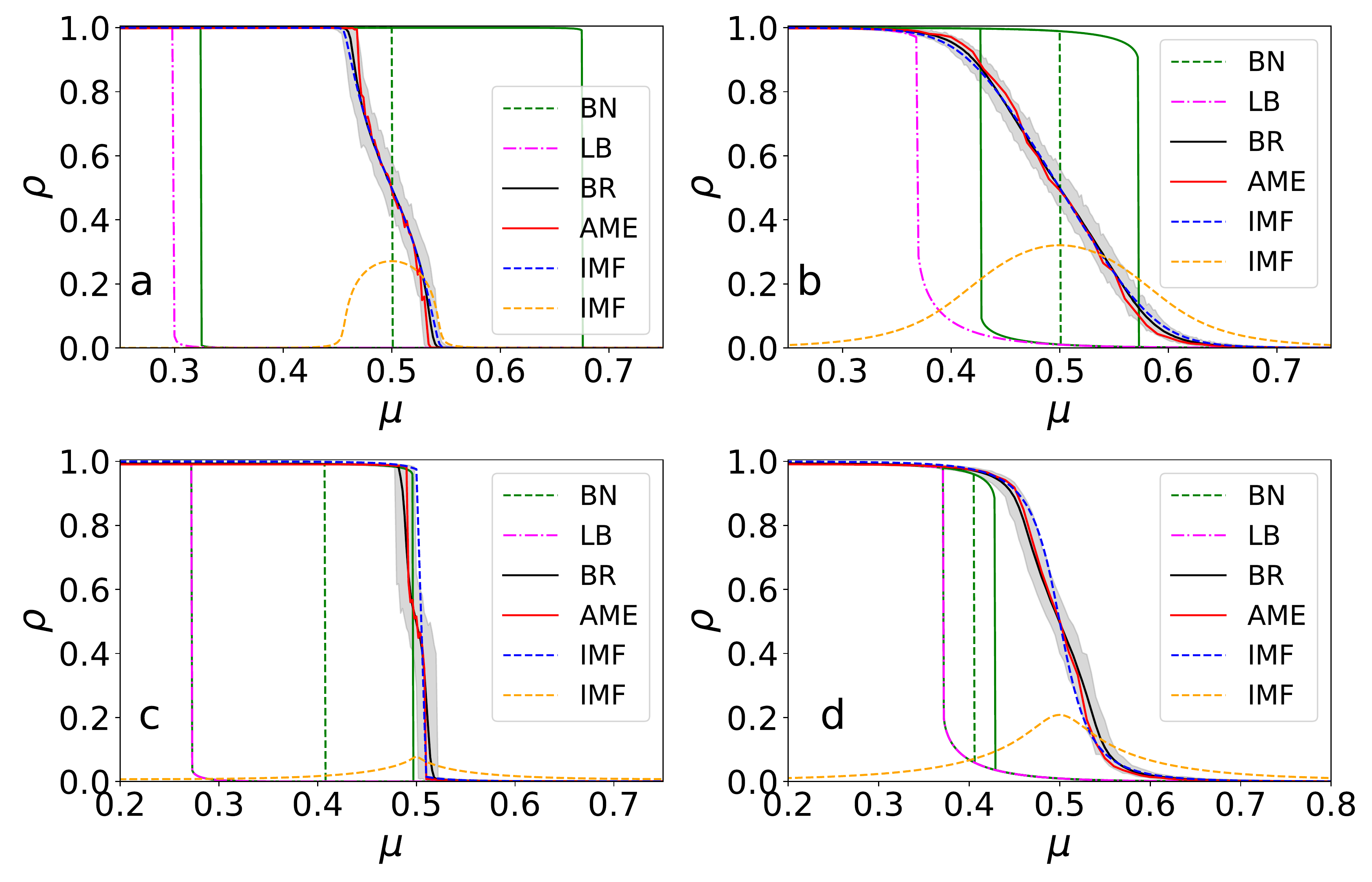}
\caption{Density $\rho$ of agents playing action $1$ obtained with best-response and other methods as function of $\mu$ for fixed values of $\sigma$ on (a) random regular  graphs of degree $K=4$ and $\sigma=0.1$,
(b) random regular  graphs of degree $K=4$ and $\sigma=0.2$, (c) Erd\H{o}s-R\`enyi random graphs of average degree $z=5$ and $\sigma=0.1$, (d) Erd\H{o}s-R\`enyi random graphs of average degree $z=5$ and $\sigma=0.2$. The fixed-point average density (solid black lines -- BR) of the best-response dynamics is compared with AME approximation of the BR dynamics (solid red lines -- AME) performed on a population of $10^5$ individuals, with an improved mean-field method (blue dashed lines -- IMF), density of the Bayes-Nash equilibria (green dashed lines -- BN) and the lower bound density (magenta dashed-dotted lines -- LB). The grey regions represent the minimum and maximum density observed over $10^4$ samples on graphs of $N=10^4$ nodes. Yellow dashed lines represent the fraction of edges that are expected to connect conditional dependent pairs of nodes as predicted by the improved mean-field (IMF) method. }\label{fig:BRcomparison}
\end{center}
\end{figure}

For the choice of parameters in Fig.~\ref{fig:BRame} and larger values of initial densities $\rho_0$, the selected equilibrium is always the maximum, Pareto-efficient one; however, the outcome strongly depends on the threshold distribution $f(\theta)$. A different reasonable assumption is that, in the initial state, agents choose independently and completely random among the available actions, i.e. $\rho_0 =0.5$. 
Figure~\ref{fig:BRcomparison} displays the behavior of best-response dynamics (solid black lines -- BR) on random regular  graphs (a-b) and Erd\H{o}s-R\`enyi  random graphs (c-d), when the initial density is $\rho_0 = 0.5$, as function of $\mu$ for fixed values $\sigma=0.1$ (a,c) and $\sigma=0.2$ (b,d). A direct comparison with other equilibrium concepts and approximation methods is provided. The properties of the equilibria selected by BR are obviously very different from those obtained with the monotone process from $\rho_0=0$ (magenta lines --  LB, in Fig.~\ref{fig:BRcomparison}) which corresponds to the lower bound (in terms of density) of the equilibrium spectrum.  A very striking, and a priori less predictable, difference is the one observed with Bayes-Nash (BN) equilibria (green lines -- BN, in Fig.~\ref{fig:BRcomparison}, solid lines are obtained following the metastable solutions, dashed lines mark the change of stability). Bayes-Nash equilibria are defined for a game of incomplete information, that is when the threshold and degree of an agent are considered as unknown random variables drawn from known distributions (see App.~\ref{app-BayesNash} for a definition). Since many real networks are partially unknown, these bayesian equilibria have played so far a central role in studies on network games \cite{jackson2007diffusion,galeotti2010network,jackson2015games}.  Bayes-Nash equilibria are found solving simple self-consistent mean-field equations, which  can be interpreted as the averaged version of the best-response relations in the presence of annealed disorder (see App.~\ref{app-BayesNash}).  In the case under study, regions of values of $\mu$ (and $\sigma$) exist for which coexistence of two solutions (i.e. of bayesian equilibria) is possible. However, {\em  the properties of Bayes-Nash equilibria seem to be unrelated to those of the pure Nash equilibria typically reached by BR in this regime, in particular for systems with non-uniform degrees and thresholds}.  Although BN equilibria are  based on a different game-theoretic formulation  with respect to pure Nash equilibria (incomplete information vs. complete information), this difference is remarkable and must be taken as an indication that the two formulations should not be freely interchanged in the description of coordination problems on networks.  Other naive mean-field approximations, such as \eqref{MFpk_rho}, provide equally unsatisfactory descriptions of the properties of pure Nash equilibria.

In search of a theoretical approach providing a more accurate description of the class of equilibria selected by BR starting from random initial conditions we again resort to the AME approach, valid for ensembles of uncorrelated random graphs.  The agreement between the results obtained by means of AME (solid red line -- AME, in Fig.~\ref{fig:BRcomparison}) and BR dynamics is pretty good, even though this method does not shed light on the underlying dynamical mechanisms of equilibrium selection at work. 
A qualitatively different approach that reaches a comparable level of agreement is based on the hypothesis that {\em best-response dynamics tends to get trapped into Nash equilibria in which a large fraction of agents is strongly conditioned on the neighbours}, because their best-response relation is only marginally satisfied, i.e. their choice would change if just one deviation in the neighbourhood occurs. A properly defined class of equilibria satisfying this hypothesis, which we call {\em marginal equilibria}, is defined in App.~\ref{app-marginal}, together with a mean-field approximation that explicitly distinguishes this marginal behaviour from those of agents that can safely play $0$ or $1$ (see equations \eqref{rho_IMF}-\eqref{rho_IMF2} in App.~\ref{app-marginal}). The results of this improved mean-field theory (blue dashed line -- IMF, in Figure~\ref{fig:BRcomparison}) are in rather good agreement with the typical outcomes of best-response dynamics. This method also makes possible to directly measure the fraction of connected conditional dependent pairs of nodes (yellow dashed lines -- IMF, in Fig.~\ref{fig:BRcomparison}). In the region where best response converges to non-trivial equilibria, this quantity is non-zero suggesting that standard mean-field methods,  neglecting conditional dependence, should be expected to fail to describe BR dynamics.

\subsection{Stochastic stability}\label{sec-stochstability}
A smoothed version of the best-response dynamics, addressing bounded rationality of agents, is the {\em logit update rule} \cite{mcfadden1973conditional,blume1993statistical},  which is a Markov chain in which at each time interval $\Delta t$ a node $i$ is chosen uniformly at random and revises her action with probability 
\beq\label{qbr}
W\left[ x_i(t+\Delta t)=x \right] = \frac{e^{\beta u_i(x; \vec{x}_{\partial i})}}{\sum_{x'\in\{0,1\}} e^{\beta u_i(x' ; \vec{x}_{\partial i})}},
\eeq
where $\beta$ is a parameter controlling the level of rationality of the agents. In the $\beta \to +\infty$ limit, the probability measure of the Markov chain concentrates on a subset of the binary configurations  called {\em stochastically stable} states  \cite{young1993evolution,kandori1993learning,kandori1995evolution}. In potential games, stochastically stable states coincide with the set of Nash equilibria corresponding to the global maximum of the potential function \cite{blume1993statistical}. For the present coordination model, the equilibrium selection of stochastically stable equilibria under the logit update rule reduces to a static analysis of the potential $V(\vec{x};\vec{\theta})$ in \eqref{potential0}  and of the properties of its global maxima (see also App.~\ref{app-game}). For any given realization $\vec{\theta}$ of the disorder, this can be done efficiently employing a zero-temperature cavity method applied on an auxiliary optimisation problem. This method, derived in App.~\ref{app-stoch}, is based on a set of self-consistent max-sum equations \eqref{eq:ms-cavity} for ``cavity fields'' $y_{ij}\in [0,1]$ defined on the directed edges $(i,j) \in \mathcal{E}$  and representing the relative preference of the agent $i$ to play action $1$ in the absence of node $j$ inside a configuration realising the maximum value $V_{\rm max}$ of the potential function. Using a population dynamics approach, it is then possible to perform the average over the threshold distribution $f(\theta)$ and study the statistical properties of  stochastically stable equilibria as function of the parameters of the coordination model.
Results on the stochastic stability of Nash equilibria in the coordination game defined on random regular graphs of degree $K=4$ are reported in Figure~\ref{fig:stochstab}a: lines represent theoretical predictions from the max-sum equations, whereas symbols stand for the outcome of direct numerical simulations of the dynamics with logit update rule. The latter have been performed by means of an annealing schedule in which $\beta$ is gradually increased from $\beta_{\rm min}=0$ to $\beta_{\rm max} > 10$ in order to practically ensure the convergence to the asymptotic measure (which formally occurs only for $\beta=\infty$).  When thresholds are weakly disordered ($\sigma < \sigma_{\rm c}$), the maximum potential value $V_{\rm max}$ has a discontinuity in the derivative at $\mu_{\rm c,s}=0.5$, marking a change in the stochastically stable state, which passes from the maximal equilibrium to the minimal one (resp. $\vec{1}$ and $\vec{0}$ for the choice of parameters done). A phenomenon of {\em metastability} is also observed, with the former global maximum of the potential that survives as a local maximum up to a spinodal point $\mu_{\rm sp,s}$ (dashed lines in Fig.~\ref{fig:stochstab}a). The transition disappears in the strong disorder region, after the structural transition in the space of Nash equilibria ($\sigma > \sigma_{\rm c}$). Notice that for $\sigma = 0$ the spinodal point $\mu_{\rm sp,s}$ coincides (at least numerically) with the critical point $\mu_{\rm c,m}$ marking the end of the cascade region, whereas in the presence of weak disorder (e.g. $\sigma >0.1$) we  find $\mu_{\rm sp,s} > \mu_{\rm c,m}$, meaning that {\em there is an intermediate region in which inefficient equilibria are robust with respect to small perturbations, preventing cascade phenomena, but they are not (even locally) stochastically stable}. Below $\mu_{\rm c,m}$, instead, $V_{\rm typ}(\rho)$ grows with $\rho$ for any $\nu\geq 0$, meaning that a small deviation from the minimal equilibrium is likely to be amplified by any dynamical process (not only by means of the logit dynamics in the $\beta \to \infty$ limit).  In fact, in this region, a non negligible fraction of agents has negative thresholds, thus acting as seeds for the propagation of action $1$ throughout the system. 

\begin{figure}[tb]
\begin{center}
\includegraphics[width=0.8\columnwidth]{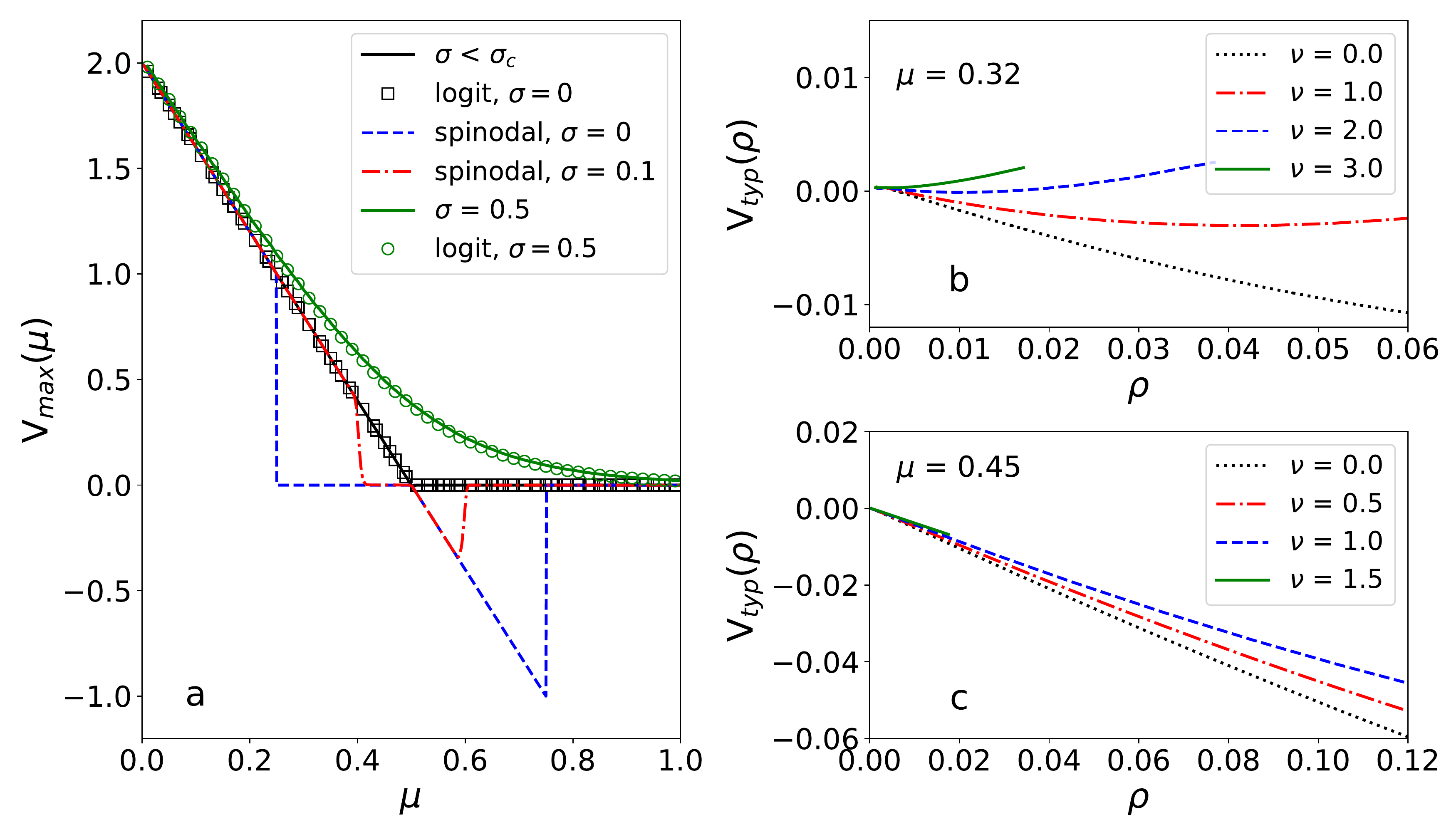}
\caption{(a) Maximum value $V_{\rm max}$ of the potential function on random regular graphs of degree $K=4$ as function of the threshold mean $\mu$ for $\sigma=0$ (black) and $\sigma = 0.5$ (green). Results obtained with the max-sum algorithm (solid lines) are compared with those obtained from numerical simulations of the logit update rule \eqref{qbr} using an annealing schedule from $\beta_{\rm min}=0$ to $\beta_{\rm max} > 10$ with uniform random initial conditions (symbols). For $\sigma < \sigma_{\rm c}$, the maximum of the potential changes continuously, but with discontinuous derivative, at $\mu_{\rm s,c}=0.5$. Metastable local maxima of the potential, for $\sigma=0$ (blue dashed line) and $\sigma =0.1$ (red dot-dashed line), survive up to a spinodal point $\mu_{\rm sp, s}$. (b-c) Average typical potential $V_{\rm typ}(\rho)$ as function of the density $\rho$ in random regular graphs of degree $K=4$, $\sigma=0.1$, (b) $\mu=0.32$ and (c) $\mu = 0.45$, for different values of the parameter $\nu\geq 0$. For $\mu<\mu_{\rm sp,s}$, it is possible to bias the set of equilibria favouring higher potential ones (by increasing $\nu$) so that the minimal equilibrium stops to represent a local maximum of the typical potential; in the metastability region, instead, $V_{\rm typ}(\rho)$ always decreases from zero in the proximity of the minimal equilibrium for any $\nu\geq0$.}\label{fig:stochstab}
\end{center}
\end{figure}

In addition to the value of the maximum potential, we also computed a different quantity that we called {\em typical potential} $V_{\rm typ}(\rho)$, defined as the average potential value computed on typical Nash equilibria at a given density value $\rho$. This average value is not much informative because, in the presence of many different equilibria for the same value of density $\rho$, one could expect that they also correspond to  rather different values of the potential function. This heterogeneity can be exploited to investigate the mechanism with which logit dynamics explores the equilibrium landscape and to understand the origin of the observed metastability. To this end, we extend the analysis performed in Sec.~\ref{sec:landscape}, appropriately modifying the energy function $H(\vec{x})$ in \eqref{pf} as follows 
\begin{equation}\label{eq-HV}
H(\vec{x}) =  \sum_i \left(\nu\theta_i k_i  - \epsilon\right) x_i - \nu \sum_{(i,j)} x_i x_j.
\end{equation}
Since the potential is the sum of local terms, one can still employ the BP approach in order to approximately evaluate the associated probability distribution and other statistical properties (e.g. local marginals) as function of the density $\rho$ and potential $V$,  respectively tuned by means of the Lagrange multipliers $\epsilon$ and $\nu$. A detailed derivation of the modified BP equations is reported in App.~\ref{BPVtyp}. 
In relation with the metastability phenomenon observed in Fig.~\ref{fig:stochstab}a for random regular graphs, we used this method to investigate the equilibrium landscape in the neighbourhood of the minimal equilibrium, when the maximal one is stochastically stable ($\mu<0.5$ for $\sigma=0.1$). Figures ~\ref{fig:stochstab}b-\ref{fig:stochstab}c display the average values of the typical potential $V_{\rm typ}$ computed on Nash equilibria with density close to $0$, changing the parameter $\nu$ in order to favor equilibria corresponding to larger/smaller values of the potential. These values are obtained from the solutions of the BP equations \eqref{bp-instanceV}, valid for a fixed instance $\vec{\theta}$ of the disorder. For simplicity, the average over the threshold distribution $f(\theta)$ is here performed by sampling, i.e. averaging the corresponding values of $\rho$ and $V_{\rm typ}$ (computed as in \eqref{Vtyp}) over many instances of the random variables $\vec{\theta}$.
First consider the case in which the minimal equilibrium is not expected to be a local maximum of the potential, e.g. the case $\mu=0.32$ in Fig.~\ref{fig:stochstab}b. For $\nu=0$, the average typical potential $V_{\rm typ}(\rho)$ decreases when the density departs from zero, meaning that most equilibria at low density are less stochastically stable than the minimal one. On the other hand, the behavior changes when, by increasing $\nu$, equilibria with possibly larger values of the potential are weighted more in the analysis. The green curve for $\nu=3.0$ demonstrates that it is in fact possible to increase the value of the typical potential increasing continuously the density from zero by properly selecting Nash equilibria, meaning that {\em even though the majority of equilibria at small density are less stochastically stable with respect to the minimal equilibrium, the latter is not a local maximum of the potential}. It follows that the logit update rule should always eventually escape from the minimal equilibrium, even in large systems, a behaviour actually observed in numerical simulations.   
On the contrary, in the metastable region (e.g. for $\mu=0.45$ in Fig.~\ref{fig:stochstab}c), the average typical potential always decreases as the density continuously grows from zero (at least for all accessible values of the parameter $\nu$), suggesting that the minimal equilibrium is a real local maximum of the potential function.  

\subsection{Learning dynamics}\label{sec-reinforcement}
Both best response and smoothed best response dynamics are myopic processes in which agents only care about their current payoffs and actions. Here we partially relax this setting, considering learning rules in which the agents use information obtained from the analysis of their own past play and  of that of other agents, in order to  modify and possibly improve their choices. On the other hand, agents are still myopic, that is they do not elaborate time-dependent strategies based on the calculation of some discounted future expected payoff but  merely repeatedly play one-shot actions $x(t) \in \{0,1\}$ at discrete times $t$.  Two different and widely-known prototypes of learning processes belonging to this class are Fictitious Play and Reinforcement Learning \cite{fudenberg2009learning,fudenberg1998theory}.

{\em Fictitious Play} (FP) was proposed as an algorithm for solving zero-sum games \cite{brown1951iterative}, but it is probably the simplest process of myopic learning, in which agents at each stage $t$ of a repeated game best reply to their beliefs about the strategies of their opponents. Initially, agents have some prior information on the strategy used by the others. During the discrete-time (e.g. $\Delta t=1$) synchronous dynamics, the only information that agents can use to update their beliefs is the observation of the actions played by neighbours in the stage games. Let us define a weight $\kappa_{j}^{x}(t)=\kappa_{j}^{x}(0)+n_j^{x}(t)$ for $t\geq 0$ in which  $n_{j}^{x}(t)$ is the number of times that agent $j$ plays action $x$ up to time $t\geq 1$,  while $\kappa_{j}^{x}(0)$ is some initial information on the propensity of  agent $j$ to play action $x$. According to the FP dynamic rule, the mixed strategy $\pi_j$ of agent $j$ at time $t$ can be estimated as 
\begin{equation}\label{fpBelief}
 \hat{\pi}_j(t)  =  \frac{\kappa_{j}^{1}(t)}{\sum_{x\in \{0,1\}} \kappa_{j}^{x}(t)}   = \begin{cases} 
  \frac{\kappa_{j}^{1}(0)}{\sum_x \kappa_{j}^{x}(0)} & \text{for } t=0 \\
  \\
  \left(1- \frac{1}{t + \sum_x \kappa_{j}^{x}(0)}\right) \hat{\pi}_j(t-1)+ \frac{1}{t + \sum_x \kappa_{j}^{x}(0)}\id\left[x_j(t-1) =1\right] &  \text{for } t>0. 
 \end{cases}
\end{equation}
For zero initial weights ($\kappa_{j}^{x}(0)=0$ for $x\in \{0,1\}$), the belief update rule \eqref{fpBelief} coincides with the empirical frequency distribution $\frac{1}{t}\sum_{s=1}^t \id\left[x_j(s) =x\right]$.
In the deterministic FP dynamics, agents form beliefs about others' strategies, but they play pure actions: at time $t$, agent $i$ best replies to the empirical distribution of neighbours' play, i.e.
\begin{equation}\label{brFictitious}
x_i(t) = {\arg\max}_{x \in \{0,1\}} \mathbb{U}_i\left[ x| \{\hat{\pi}_j(t)\}_{j \in \partial i}\right],
\end{equation}
where 
\begin{subequations}
\begin{align}
 \mathbb{U}_i\left[ x | \{\hat{\pi}_j(t)\}_{j \in \partial i}\right]  & = \sum_{\vec{x}_{\partial i}}  u_i\left(x; \vec{x}_{\partial i}\right) \prod_{j \in \partial i} \left[ x_j \hat{\pi}_j(t) + (1-x_j)(1-\hat{\pi}_j(t)) \right]\\
  & = \sum_{\vec{x}_{\partial i}}  \left(x_i \sum_{j \in \partial i} x_j + (1-x_i)\theta_i k_i \right) \prod_{j \in \partial i} \left[ x_j \hat{\pi}_j(t) + (1-x_j)(1-\hat{\pi}_j(t)) \right] \\
 & = x_i \sum_{j \in \partial i} \hat{\pi}_j(t) + (1-x_i)\theta_i k_i 
\end{align}
\end{subequations}
 is the expected utility of agent $i$ when agents randomize their actions from the partial mixed strategy profile given by the set of beliefs $\{\hat{\pi}_j(t)\}_{j\in \partial i}$. 
 
 In this formulation of FP dynamics, agents consider a factorized form for the partial belief profile of their neighbours, as it happens for the product measure defined on mixed strategies. This is not the case in general: for instance, when agents have access to additional information provided by third parties, such as randomization devices, correlation between agents' strategies could exist \cite{fudenberg1998theory}. Suppose now that the individual mixed strategies $\{\pi_{j}\}_{j\in \partial i}$ of the neighbours of agent $i$ are time-independent, then the pure actions played over time by her neighbours $j \in \partial i$ are i.i.d. random variables drawn from binomial distributions with unknown parameters. Under the stationarity assumption, the empirical frequency distributions of past play give precisely the maximum likelihood estimators of such parameters. In the presence of initial weights, the belief update rule \eqref{fpBelief} can be given a Bayesian interpretation \cite{fudenberg1998theory,shoham2008multiagent}. Suppose that, at any time $t$ in a stationary environment, the prior distributions for parameters $\pi_j$ are conjugate Beta distributions with parameters $\kappa_{j}^{x}(t)$, it turns out that the expected mixed strategy of agents $j$ (posterior mean) is given by  $\hat{\pi}_j(t) = \kappa_{j}^{1}(t)/\sum_x \kappa_{j}^{x}(t)$, which is exactly the fictitious play update rule. 
In practice, every agent updates her strategy using fictitious play, therefore strategies are not actually stationary during the dynamics. Stationarity is usually a reasonable approximation in the long run, but when this is not the case, beliefs do not tell much about real strategies.

If the empirical distribution of strategies converges during the (deterministic) FP dynamics, then it converges to a (possibly mixed) Nash equilibrium of the stage game. There are several results on the convergence of fictitious play in games of strategic complements (see e.g. \cite{berger2009convergence}), and it is believed that FP dynamics always converge in such games \cite{krishna1992learning}. The game being a potential game is a sufficient condition for the empirical distribution of strategies to converge \cite{monderer1996fictitious}. 
Suppose that a pure action profile $\vec{x}^\ast$ is an absorbing state of fictitious play dynamics, i.e. there is a time $t^\ast$ such that  $\forall t\geq t^\ast$ agents play action profile $\vec{x}(t) = \vec{x}^\ast$. If a pure-strategy profile is an absorbing state of fictitious play, then it is also a pure Nash equilibrium of the game (the converse is also true). Echenique and Edlin \cite{echenique2004mixed}  demonstrated that mixed strategy equilibria are unstable for a large class of dynamics, including fictitious play, therefore we expect FP dynamics to converge to absorbing states corresponding to pure Nash equilibria. These results suggest that, in the present coordination game defined on graphs, FP dynamics should converge to pure Nash equilibria, which ones, depending on the parameters and on the initial priors on agents' beliefs, is not apparent a priori.

We slightly relax the  FP update rule by considering a smoothed version in which \eqref{brFictitious} is replaced by a stochastic choice of pure actions $x_i$ from a mixed strategy $\pi_i(t)$ of agent $i$ at time $t$ defined, e.g. using a logit function with inverse temperature $\beta$, as
\begin{equation}\label{brFictitious-stoch}
\pi_i(t) = \frac{e^{\beta \mathbb{U}_i\left[x_i =1 | \{\hat{\pi}_j(t)\}_{j \in \partial i}\right]}}{\sum_{x\in X} e^{\beta \mathbb{U}_i\left[ x| \{\hat{\pi}_j(t)\}_{j \in \partial i}\right]}}.
\end{equation}
Taking the continuous-time limit, after a logarithmic rescaling of time $\tau = \log(1+t)$, the dynamics of beliefs can be approximated by 
\begin{align}\label{eq-fpcont_pihat}
 \frac{d}{d\tau}\hat{\pi}_i(\tau) & \approx \omega_i(\tau)\left(\pi_i(\tau) - \hat{\pi}_i(\tau) \right), 
\end{align}
with $\omega_i(\tau) =1- e^{-\tau}\sum_x \kappa_{i}^{x}(0)$, suggesting that mixed strategies and beliefs are asymptotically equal and they satisfy a sort of 
 average best response relation, whose solutions are known as {\em Nash distributions} \cite{fudenberg1998theory}. Nash distributions were originally introduced as a solution concept for games with randomly perturbed payoffs, such that they could converge back to Nash equilibria when the perturbation is sent to zero. Here instead they can be interpreted as a result of the learning process. 
The speed of convergence to the fixed-point Nash distributions depends on the pre-factors $\omega_i$, i.e. on the initial values of the beliefs, which are  chosen as follows: a random fraction $\rho_0$ of the agents have $\kappa_{j}^{1}(0)=\epsilon> 0$ and $\kappa_{j}^{0}(0)=0$, whereas the opposite holds for the remaining nodes. As the parameter $\epsilon$ tunes the memory of the learning process, when multiple solutions of the fixed point relations exist, dynamics with different values of $\epsilon$ can have different asymptotic results. For large $\epsilon$, the FP dynamics evolves very slowly, hardly forgetting the initial conditions and getting more easily trapped into Nash equilibria characterized by a lower level of efficiency. 
Figure~\ref{fig:fpK4numerics} displays some examples of long run behaviour of stochastic FP trajectories ($\beta=10$ and $\epsilon=10^{-4},1,10^4$) for random regular graphs of degree $K=4$ and $\sigma= 0, 0.1$. In panels \ref{fig:fpK4numerics}a-\ref{fig:fpK4numerics}b, the initial conditions are random profiles at very low density ($\rho_0\sim 10^{-3}$) of agents playing action $1$. The curves in log-log plot clearly show that the qualitative behavior of the dynamics is similar to the one already observed for best response, in which the low-efficiency equilibria become locally attractive when $\mu$ grows beyond a threshold. The latter depends on the parameter $\epsilon$, the memory being larger for larger values of $\epsilon$. Similar curves obtained at fixed $\mu=0.35$ and varying the initial density $\rho_0$ are shown in panels \ref{fig:fpK4numerics}c-\ref{fig:fpK4numerics}d. As already observed for the best response dynamics, there is a threshold value of initial density $\rho_0$ above that the time-dependent density rapidly converges to 1, otherwise remaining trapped into low-efficiency equilibria. The threshold value, however, depends on $\epsilon$. We obtained qualitatively similar results on more general random networks.

\begin{figure}[tb]
\begin{center}
\includegraphics[width=0.95\columnwidth]{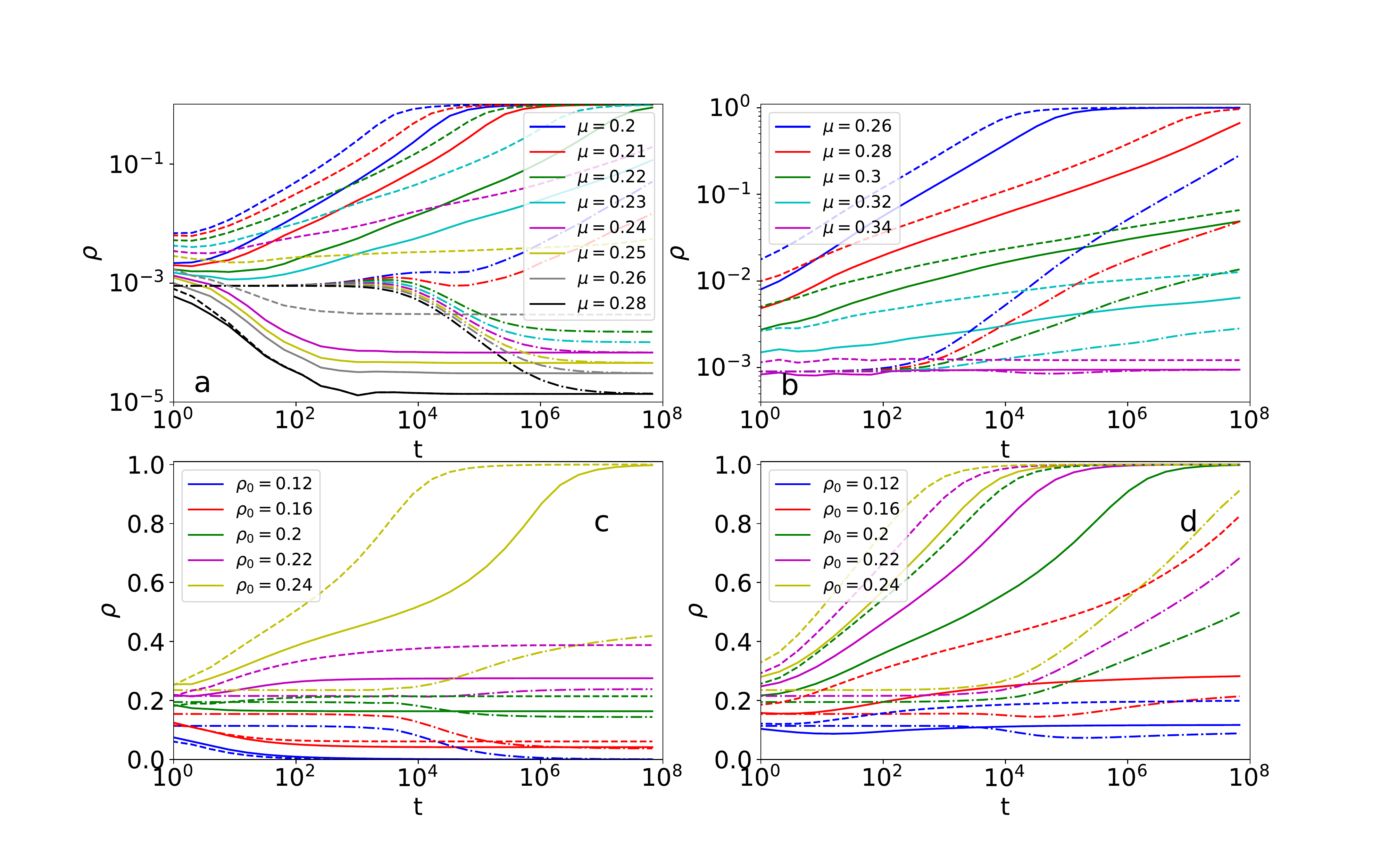}
\caption{(a)-(b) Density $\rho(t)$ of agents playing action $1$ as function of time $t$ during fictitious play dynamics on a random regular graph of size $N=10^4$ nodes, degree $K=4$ and thresholds with variable mean $\mu$ and standard deviation (a) $\sigma =0$, and (b) $\sigma =0.1$ and $\epsilon=1$ (full lines), $\epsilon =10^{-4}$ (dashed lines) and $\epsilon =10^{4}$ (dot-dashed lines). Initial density is $\rho_0=10^{-3}$. (c)-(d) Density $\rho(t)$ of agents playing action $1$ as function of time $t$ during fictitious play dynamics for various initial density values $\rho_0$ on a random regular graph of size $N=10^4$ nodes, degree $K=4$ and thresholds with mean $\mu=0.35$ and standard deviation (c) $\sigma =0$, and (d) $\sigma =0.1$ and $\epsilon=1$ (full lines), $\epsilon =10^{-4}$ (dashed lines) and $\epsilon =10^{4}$ (dot-dashed lines). }\label{fig:fpK4numerics}
\end{center}
\end{figure}

\begin{figure}[tb]
\begin{center}
\includegraphics[width=0.9\columnwidth]{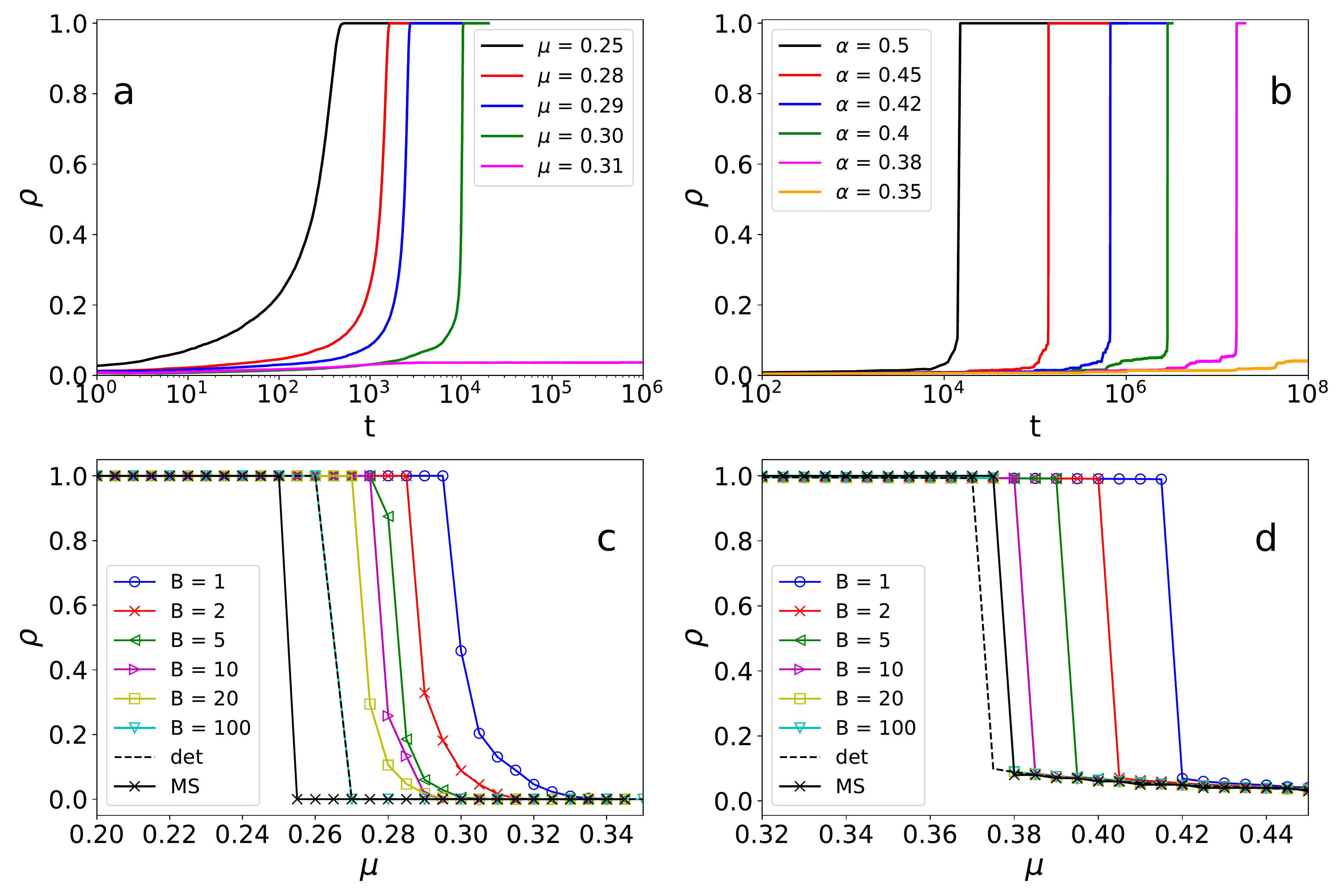}
\caption{(a-b) Density $\rho(t)$ of agents playing action $1$ as function of time $t$ during for reinforcement learning on a  random regular graph of size $N=10^4$ nodes and degree $K=4$ with $\beta = 10$ and $B=1$: (a) $\alpha=0.01$ for $\sigma=0.1$ and varying  $\mu$; (b) $\mu = 0.35$ and $\sigma =0.1$ varying $\alpha$. Initial conditions are random strategy profiles such that $\rho(t=0)<0.01$. (c-d) The long-time stationary density $\rho$ of agents playing action $1$ as function of the mean threshold value $\mu$ (and for $\sigma=0$) on a random regular graph of size $N=10^4$ nodes and degree $K=4$ obtained by means of reinforcement learning with $\beta=1$ for various values of $B$ and with (c) $\alpha=0.01$, (d) $\alpha = 0.5$. Initial conditions are random strategy profiles such that $\rho(t=0)<0.01$. For comparison, results obtained running the deterministic version of the learning dynamics (Sato-Crutchfield equations \eqref{SCeq}) or evaluating its internal fixed points by means of the max-sum (MS) equations (see \eqref{eq:maxsumSC} in App.\ref{app-MSsato}) are also reported.
}\label{fig:orlK4numerics}
\end{center}
\end{figure}

 {\em Reinforcement Learning} is based on an opposite principle with respect to Fictitious Play: agents do not directly care about  strategies of the others but keep track of a discounted cumulative payoff obtained during the dynamics and take decisions in the attempt of maximizing it \cite{bush1955stochastic,sutton2018reinforcement}. Following previous literature in the field \cite{galla2009intrinsic,kianercy2012dynamics,galla2013complex,realpe2012fixation,sanders2018prevalence}, we consider a particular case of Q-learning \cite{watkins1992q} and Experience Weighted Attraction (EWA) rules \cite{camerer1999experience,ho2007self}. Each agent $i$ has a time-dependent mixed-strategy profile $\pi_i(t)$, defined as the individual probability of playing action $x_i=1$ at time $t$; each discrete time step $t$, agents are involved in $B$ sequential fictitious stage games ($\{x_i^b\}_{i=1,\dots,N}^{b=1,\dots, B}$) with the neighbours, in which  pure strategies are sampled according to $\pi_i(t)$. Single-stage games are used to update a score function, called {\em attraction} and defined as 
\begin{equation}\label{RLrule2}
A^{x}_i(t+1)  = (1-\alpha) A^{x}_i(t)+ \frac{1}{B}\sum_{b=1}^{B} u_i\left(x^b=x; \vec{x}_{\partial i}^b\right),
\end{equation}
where $u_i(x^b; \vec{x}_{\partial i}^b)$ is the payoff that agent $i$ gets from playing action $x^b$ at stage $b$ when the neighbours play partial action profile $\vec{x}_{\partial i}^b$.
The mixed strategies are then updated by means of a logit rule 
\begin{equation}\label{RLrule1}
\pi_i(t) = \frac{e^{\beta A^{1}_i(t)}}{\sum_{x \in X} e^{\beta A^{x}_i(t)}}.
\end{equation}
For $B=1$, the dynamics is equivalent to {\em online learning}, whereas in the limit $B\to \infty$ the empirical average of the payoffs computed over $B$ sequential stage games should reproduce the expected payoff obtained when the neighbouring agents follow their mixed strategies $\vec{\pi}_{\partial i}(t)$. The corresponding deterministic reinforcement learning rule obtained in the limit $B\to \infty$ consists of Eq.~\eqref{RLrule1} with
\begin{equation}\label{RLrule3}
A^{x}_i(t+1)  = (1-\alpha) A^{x}_i(t) +  \mathbb{U}_i\left[ x | \vec{\pi}_{\partial i}(t)\right].
\end{equation}

Besides the batch size $B$, the learning process is influenced by other two parameters: the discount factor $\alpha \in (0,1]$, describing memory loss during the learning process, and the intensity of choice $\beta \geq 0$. 
We first analyze the learning dynamics in the large $\beta$ limit, in which agents tend to play pure strategies corresponding to the actions with highest attraction. In this regime, only the action ranking is relevant for polarizing mixed strategies, not the relative size of the corresponding attractions. It follows that the global long-time behaviour of the learning dynamics does not depend much on the size $B$ of the batch when $\beta$ is large, therefore we only report numerical results for online learning ($B=1$) on a random regular graph of size $N=10^4$ and degree $K=4$.
For different values of $\mu$ (at fixed $\sigma=0.1$ and $\alpha =0.01$), Figure~\ref{fig:orlK4numerics}a shows the time-dependent behavior of the average density $\rho(t)$ of agents playing action $1$ from initial conditions in which a very small fraction of agents does (less than 1\%). The dynamics converges to the Pareto efficient equilibrium only for $\mu < \mu_{\rm c, rl}(\sigma,\alpha)$, with $\mu_{\rm c, rl}(\sigma,\alpha) \approx \mu_{\rm c,m}(\sigma)$ for $\alpha \to 0$, reproducing the threshold phenomenon already observed for the best-response dynamics and for cascade processes. Moreover, the time required to reach $\rho=1$ grows when decreasing the memory-loss parameter $\alpha$ at fixed value of $\mu$ (Fig.~\ref{fig:orlK4numerics}b). In the notable case $\alpha = 1$ agents do not remember past actions and the learning dynamics becomes equivalent to  smoothed best-response. Since the most efficient equilibrium is also risk-dominant for $\mu<0.5$, we expect online learning to reach the maximum density in this whole region for $\alpha=1$.

Decreasing $\beta$, the batch size $B$ starts playing a role in the long-time behaviour of the reinforcement learning dynamics. {\em Learning processes with small batch sizes more easily escape towards the efficient equilibrium with respect to those for large values of $B$.} There is a combined effect with the existence of low-density equilibria: for weak disorder ($\sigma \ll \sigma_c$), such equilibria are rather stable with respect to perturbations and for large $B$ the stochasticity of the process is too weak to enable large fluctuations towards the efficient equilibrium. When $B$ is reduced, the region of values of the parameter $\mu$ for which the efficient equilibrium is reached is considerably broadened. Figures~\ref{fig:orlK4numerics}c-\ref{fig:orlK4numerics}d show such a phenomenon for $\beta=1$ and $\alpha=0.01, 0.5$ on a random regular graph of size $N=10^4$, degree $K=4$ and $\sigma=0$. The effect disappears as $\sigma$ is increased towards $\sigma_c$.

\begin{figure}[tb]
\begin{center}
\includegraphics[width=0.9\columnwidth]{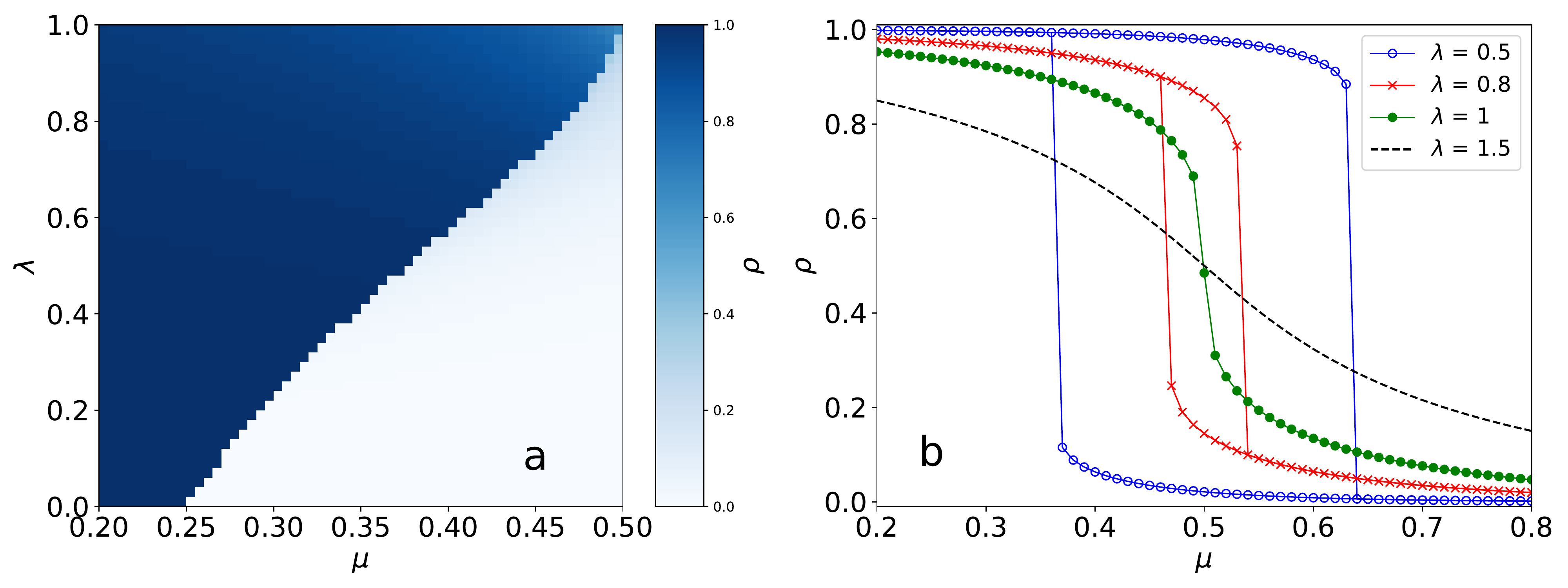} 
\caption{Analysis of the internal fixed points of the Sato-Crutchfield equations for continuous-time reinforcement learning in the case of random regular graphs of degree $K=4$ with homogeneous Gaussian thresholds ($\sigma=0$): (a) density $\rho$ of agents playing action $1$ as function of $\lambda$ and $\mu$; (b) hysteresis phenomenon in the density profile $\rho$ as function of $\mu$ for various values of $\lambda$. Results are obtained using the max-sum equations (see \eqref{eq:maxsumSC} in App.\ref{app-MSsato}).}\label{fig:rlK4theory}
\end{center}
\end{figure}

In the continuous time limit (or in the limit in which $\beta$ is small) after time rescaling $t \to \beta t$, the deterministic cumulative reinforcement learning rule can be mapped on the following set of continuous-time equations \cite{kianercy2012dynamics,galla2013complex},  
\begin{subequations}
\begin{align}
 \dot{\pi}_i  &=  \pi_i \left[ \beta\left( \sum_{j \in\partial i} \pi_j - (1-\pi_i) k_i \theta_i  - \pi_i \sum_{j\in \partial i} \pi_j \right)  - \alpha \left(\log{\pi_i} - \pi_i \log{\pi_i} - (1-\pi_i)\log{(1-\pi_i)}  \right) \right]  \\
   & =  \beta \pi_i (1-\pi_i) \left[ \left( \sum_{j \in\partial i} \pi_j -  k_i \theta_i  \right)  - \lambda\left(\log{\frac{\pi_i}{1-\pi_i}}  \right) \right], \label{SCeq}
\end{align} 
\end{subequations}
known as the {\em Sato-Crutchfield (SC) equations} \cite{sato2002chaos,sato2003coupled,galla2009intrinsic} (see App.~\ref{app-MSsato} for a derivation). The fixed points of SC equations only depend on the ratio $\lambda = \alpha/\beta \geq 0$.
While leaving unchanged the long-time behaviour of the deterministic learning dynamics, SC equations have the advantage of simplifying the analytical treatment. In particular, small noise approximations \cite{galla2009intrinsic,bladon2010evolutionary,galla2011cycles,realpe2012fixation} and large deviation theory \cite{nicole2017stochastic,nicole2018dynamical} have been applied in games with few players and generating functional path-integral techniques were used in the case of a fully-connected group of agents \cite{galla2013complex,sanders2018prevalence}. A generalization of such techniques to large sparse graphs is both theoretically  and computationally  challenging and will not be pursued here. We adopt a different approach by analyzing directly the fixed points of the Sato-Crutchfield equations, or more precisely the internal ones, corresponding to mixed-strategy Nash equilibria.  For $\lambda =0$, the Sato-Crutchfield equations coincide with multi-population replicator equations, which in the present context can be interpreted as evolutionary equations for ``populations of ideas'' \cite{galla2013complex,nicole2017stochastic}. It is known that the replicator flow generally converges to  mixed-strategy Nash equilibria. Internal equilibria are however unstable for coordination games, meaning that the dynamics for $\lambda =0$ tend to select pure Nash equilibria, reproducing the results of best-response dynamics (with random sequential update rule). For $\lambda > 0$, the single-agent entropic term $ s(\pi_i)= - \pi_i \log{\pi_i} - (1-\pi_i)\log{(1-\pi_i)}$ destabilizes pure-strategy equilibria, enforcing instead a dynamical trajectory that starts in the interior of the simplex and eventually converges to the existing interior fixed points, i.e. to the non-trivial zeros of the r.h.s. of Eq.~\eqref{SCeq}. Notice that the vertices of the strategy hypercube are still fixed points of the SC equations, but they cannot be selected in the dynamics for $\lambda >0$ just described. In order to study the mixed-strategy fixed points appearing when $\lambda > 0$, we focus on the generalized potential function
\begin{equation}
G(\vec{\pi})  = \sum_{(i,j)} \pi_i \pi_j + \sum_i \left[\lambda s(\pi_i) - k_i \theta_i \pi_i  \right]
\end{equation}
and the associated  exponential probability measure $P(\vec{\pi})\propto e^{\Gamma G(\vec{\pi})}$ with inverse temperature $\Gamma$. The zero-temperature cavity method with continuous variables (see the derivation of the corresponding max-sum equations \eqref{eq:maxsumSC} in App.\ref{app-MSsato}) is then used to extract information about the statistical properties of the maxima of $G$ that are the (internal) fixed points of the dynamics for $\lambda \neq 0$. For random regular graphs of degree $K=4$ with  homogeneous threshold values ($\sigma=0$), the results are displayed in Fig.~\ref{fig:rlK4theory}a  as function of $\mu$ and $\lambda$. The overall scenario is in agreement with what already described for the discrete deterministic reinforcement learning dynamics: for $\lambda \to 0$, the two symmetric equilibria with all agents playing action $0$ or $1$ are selected by the dynamics and their relative stability changes abruptly at a critical value $\approx \mu_{\rm c,m}(\sigma)$; at larger values of $\lambda$, two branches of internal equilibria exist and exchange stability with a hysteresis phenomenon (see both panels of Fig.~\ref{fig:rlK4theory}). Finally for $\lambda > 1$, i.e. for memory-loss stronger than the intensity of choice, the behaviour of equilibrium density becomes continuous in $\mu$.

\subsection{Comparative Statics: emergence of coordination on general random networks}\label{sec-comparison-networks}
Based on the theoretical understanding of the structural organization of equilibra and on results of numerical simulations, we briefly investigate the role played by most commonly considered structural network properties in determining dynamical selection of equilibria.  We focus on values of parameters for which the efficient (maximum-density) equilibrium is also globally stochastically stable, whereas the inefficient (minimum-density) equilibrium is only locally stochastically stable. Because of that, systems prepared in sufficiently low-density action profiles are expected to initially converge, under a bounded-rational dynamics, such as weighted best response, towards the minimum one. Because of the mean-field character of most networks, the large fluctuations necessary to escape the metastable state and ultimately converge to the maximum equilibrium can be very rare. It mainly depends on the density $\rho_0$ of agents playing action 1 in the initial conditions. As a result, the asymptotic properties of the dynamics do change, either continuously or abruptly, when  $\rho_0$ is increased.  These convergence properties depend on the internal organization of the continuous spectrum of equilibria at low density, therefore indirectly on the topological properties of the underlying interaction graph. We take into account four major network properties: average degree, degree heterogeneity, clustering, locality of interactions. 

 \begin{figure}[tb]
\begin{center}
\includegraphics[width=0.8\columnwidth]{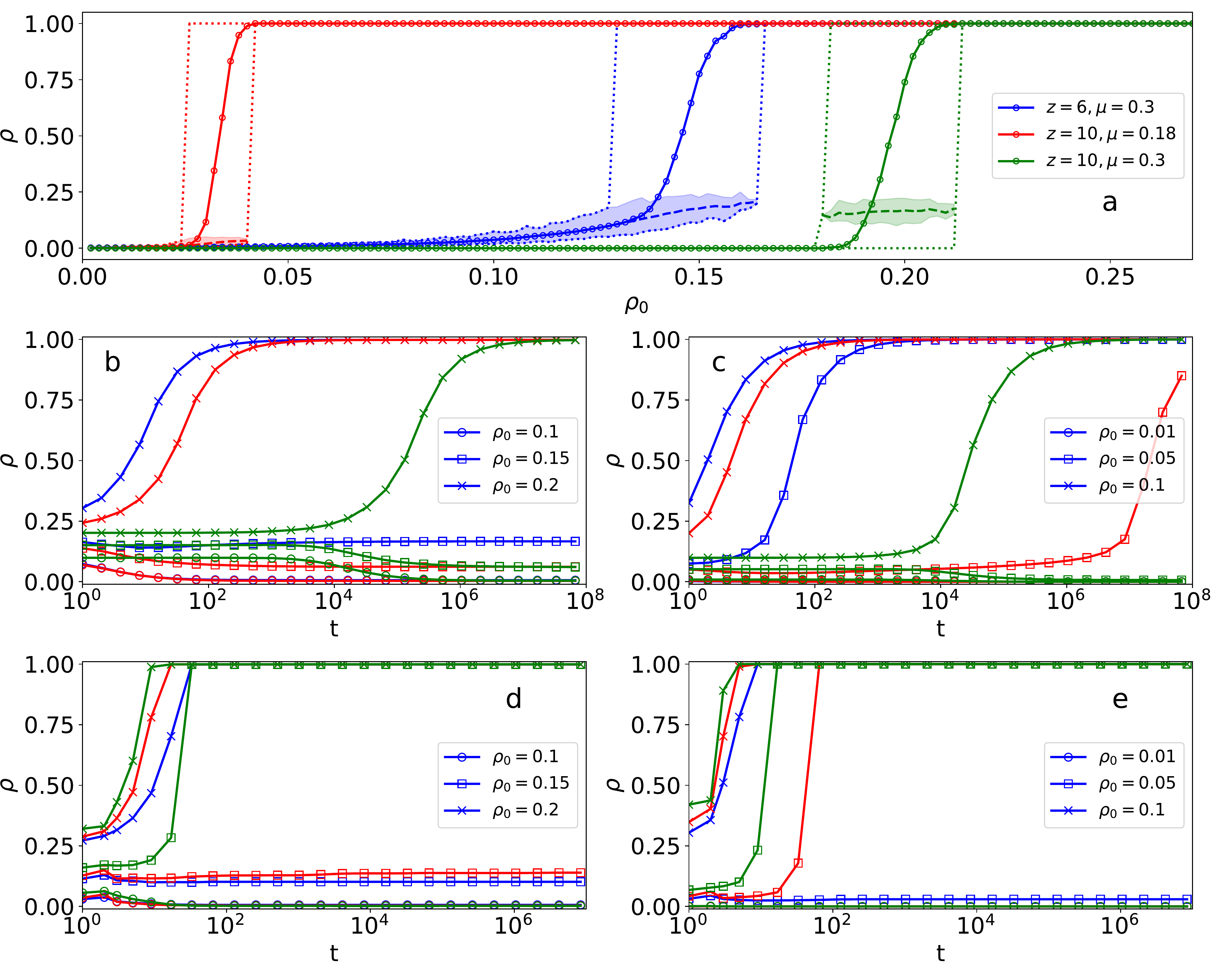}
\caption{(a) Equilibrium density $\rho$ reached by means of best response dynamics  as function of the initial density $\rho_0$ of agents playing action $1$ on Erd\H{o}s-R\`enyi  (ER) random graphs with $N=10^4$ nodes for $z=6$, $\mu=0.3$ and $\sigma=0$ (blue), $z=10$, $\mu=0.3$ and $\sigma=0$ (green), $z=10$, $\mu=0.18$ and $\sigma=0$ (red). Averages are taken over 1000 instances of the dynamics. Quantities reported are the  average density (full line), the minimum and maximum densities (dotted lines), the average density obtained when the maximum and minimum equilibria are excluded from the sample (dashed line), density values at which non-trivial equilibria are found dynamically (light shaded area). (b-c) Density $\rho(t)$ of agents playing action $1$ as function of time during the (deterministic) Fictitious Play dynamics with $\epsilon = 10^{-4}$ (blue), $\epsilon = 1$ (red) and $\epsilon = 10^{4}$ (green) for: (b) $z=6$ and $\mu = 0.3$ and initial density $\rho_0 = 0.1,0.15, 0.2$ and (c) $z=10$ and $\mu = 0.18$ and initial density $\rho_0 = 0.01,0.05, 0.1$. (d-e) Density $\rho(t)$ of agents playing action $1$ as function of time during stochastic Reinforcement Learning dynamics with $B= 10$ and $\alpha = 0.1$ (blue), $\alpha  = 0.5$ (red) and $\alpha  = 0.9$ (green) for: (d) $z=6$ and $\mu = 0.3$ and initial density $\rho_0 = 0.1,0.15, 0.2$ and (e) $z=10$ and $\mu = 0.18$ and initial density $\rho_0 = 0.01,0.05, 0.1$ }\label{fig:meta_ER}
\end{center}
\end{figure}

We have already observed in Fig.~\ref{fig:ER1}c that, apart from very small values of $z$, increasing the average degree determines a shift of the cascade instability towards lower values of $\mu$, that is the minimum equilibrium becomes increasingly (locally) stable. It follows that it is necessary to go to larger values of the initial density $\rho_0$ of agents playing action 1 in order to trigger the convergence towards the maximum equilibrium, which is also globally stochastically stable. This is shown by results of best response dynamics for Erd\H{o}s-R\`enyi random graphs with $z=6$ (blue) and $z=10$ (green) when $\mu=0.3$ in  Fig.~\ref{fig:meta_ER}a. Alternatively,  we have considered the scenario in which the product $\mu z$ is kept constant when $z$ is increased. Figure~\ref{fig:meta_ER}a displays results for $z=6, \mu=0.3$ and $z=10, \mu=0.18$ which give the same average threshold $\mu z = 1.8$ (in practice, on average one agent prefers to play action 1 if at least two of her neighbours also do the same). In this case, the support of low-efficiency equilibria narrows when the average degree is increased (compare blue and red curves and the entropy behaviour shown in the inset), suggesting that it becomes easier to escape from the low-efficiency region towards the payoff-dominant equilibrium. This is reflected in the results from other processes of dynamical equilibrium selection, such as fictitious play and reinforcement learning (see panels b-e in Fig.~\ref{fig:meta_ER}). We conclude that increasing network density has contrasting effects depending whether the individual thresholds are kept constant either intensively or extensively. 

In relation to the role played by degree heterogeneity, we have already shown in Fig.~\ref{fig:BASFthermo} that the density support of the equilibrium spectrum is larger for heterogeneous random networks (e.g. with power-law degree distribution) than for homogeneous networks (Erd\H{o}s-R\`enyi random graphs) with same average degree. Results for the asymptotic density reached by means of the best response dynamics in Figure~\ref{fig:FPRL_BASF}a-\ref{fig:FPRL_BASF}b confirm this intuition. Another effect of degree heterogeneity is that of widening  the interval of initial density values $\rho_0$ for the occurrence of the transition towards the most efficient (maximum) equilibrium. The precise location and properties of the transition are influenced by the minimum and maximum degree appearing in the graph, as they determine the extremal threshold values for coordination in the population of agents. The memory parameters of the dynamical rules (e.g. $\epsilon$ for FP in Fig.~\ref{fig:FPRL_BASF}c-\ref{fig:FPRL_BASF}d and $\alpha$ for RL in Fig.~\ref{fig:FPRL_BASF}e-\ref{fig:FPRL_BASF}f) mostly affect the convergence speed without changing the qualitative features of the asymptotic behaviour. We conclude that degree heterogeneity does not univocally promote the onset of efficient coordination although high-degree nodes can actually trigger its emergence; the outcome of dynamical equilibrium selection is less predictable on heterogenous networks compared to homogeneous ones. 

 \begin{figure}[tb]
\begin{center}
\includegraphics[width=0.8\columnwidth]{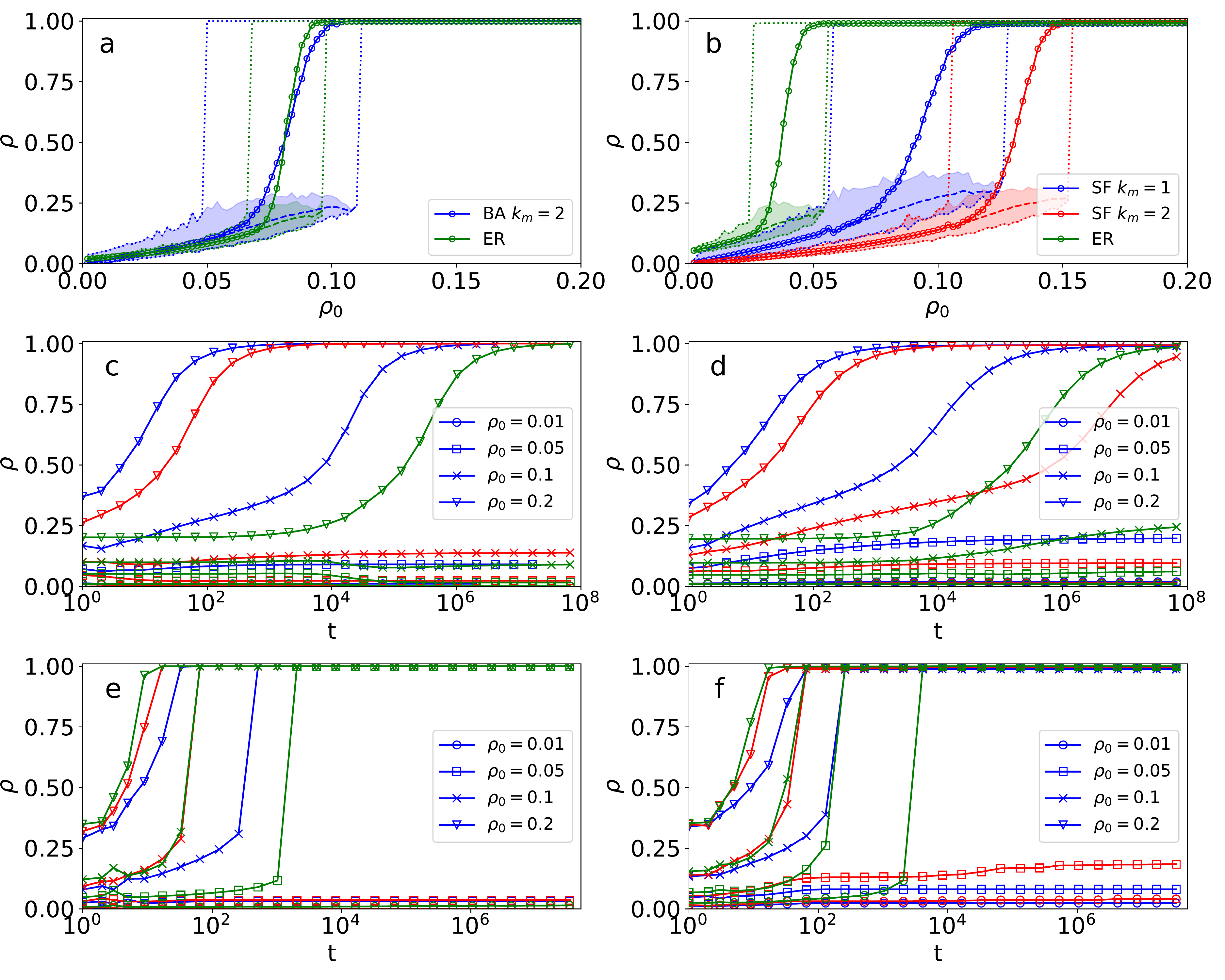}
\caption{Equilibrium selection on (a,c,e) Barab\`asi-Albert (BA) random network of $N=10^4$ nodes and minimum degree $k_{\rm m}=2$ and (b,d,f) scale-free random networks (SF) with power-law degree distribution $p(k)\sim k^{-\gamma}$ with exponent $\gamma=2.8$, size $N \sim 10^4$ nodes and minimum degree $k_{\rm min}=1,2$; both with  $\mu=0.3$ and $\sigma=0$. (a-b) Equilibrium density $\rho$ reached by means of best response dynamics  as function of the initial density $\rho_0$ of agents playing action $1$. Averages are taken over 1000 instances of the dynamics. Quantities reported are the  average density (full line), the minimum and maximum densities (dotted lines) and  the average density obtained when the maximum and minimum equilibria are excluded from the sample (dashed line), density values at which non-trivial equilibria are found dynamically (light shaded area). As a comparison, results for Erd\H{o}s-R\`enyi random graphs (green) with same size and average degree, respectively of (a) BA network with $k_{\rm m}=2$ and (b) SF network with $k_{\rm m}=1$, are also displayed.  
(c-d) Density $\rho(t)$ of agents playing action $1$ on (c) BA network and (d) SF network (with $k_{\rm m}=1$) as function of time during the (deterministic) Fictitious Play dynamics with $\epsilon = 10^{-4}$ (blue), $\epsilon = 1$ (red) and $\epsilon = 10^{4}$ (green) and with initial density $\rho_0 = 0.01,0.05,0.1, 0.2$. (e-f) Density $\rho(t)$ of agents playing action $1$ on (e) BA network and (f) SF network (with $k_{\rm m}=1$) as function of time during stochastic Reinforcement Learning dynamics with $B= 10$ and $\alpha = 0.1$ (blue), $\alpha  = 0.5$ (red) and $\alpha  = 0.9$ (green) and with initial density $\rho_0 = 0.01,0.05,0.1, 0.2$.}\label{fig:FPRL_BASF}
\end{center}
\end{figure}

Figure~\ref{fig:BR_other} displays results for dynamical selection processes on Watts-Strogatz (WS) random networks (a) with constant rewiring probability $p$ and variable average degree $\langle k\rangle$, (b) with constant $\langle k\rangle$ and variable $p$. At very low randomness (e.g. $p=0.01$) and relatively small degree (e.g. $\langle k \rangle=4$), the WS network structure is effectively one-dimensional and the coordination model admits a continuum of equilibria for all possible values of density $\rho$ (however, BP fails in correctly describing this limit because it neglects local correlations due to short loops). The dynamical equilibrium selection rules under study converge to a subset of these equilibria depending on the initial density $\rho_0$. Increasing the average degree $z$, the system becomes more globally connected; similarly, increasing the rewiring probability $p$ the coordination model  develops long-range interactions while remaining sparse (small-world property). In both cases, the locality of interactions is lost in favour of a mean-field character. The structure of the Nash equilibrium landscape changes, with the maximum-density equilibrium becoming more and more attractive for the dynamics. An abrupt transition in the values of the average equilibrium density $\rho$ reached during the dynamics appears for sufficiently large values of rearrangement probability $p$ (or sufficiently large average degree $\langle k\rangle$). Figures \ref{fig:BR_other}a-\ref{fig:BR_other}b show results for the best response dynamics, but similar behavior is observed also with other dynamic rules. In order to decouple the effect of clustering from the locality of interactions we considered best response dynamics on a random clustered graphs with  $\langle k\rangle\approx 4$ and variable clustering in Fig.~\ref{fig:BR_other}c. Increasing the clustering coefficient (i.e. increasing the number of triangles in the graph), the support of the low-density equilibria is reduced. This is because in the presence of many triangles, agents playing action 1 more easily induce coordination in their neighbourhood and the convergence to the maximum equilibrium is more likely to be activated. Finally, Fig.~\ref{fig:BR_other}d displays results of best response dynamics on networks formed by a ring of cliques of variable size. Increasing the size of the cliques makes the $\rho$ vs. $\rho_0$ curve increasingly steep although the equilibrium landscape always appears as a continuum of equilibria at any density value (as for any one-dimensional model). In the limit in which the network is formed by a unique clique, the asymptotic density develops a discontinuous transition from 0 to 1 at $\rho \approx \mu$, which is the well-known behavior in fully-connected graphs.

 \begin{figure}[tb]
\begin{center}
\includegraphics[width=0.8\columnwidth]{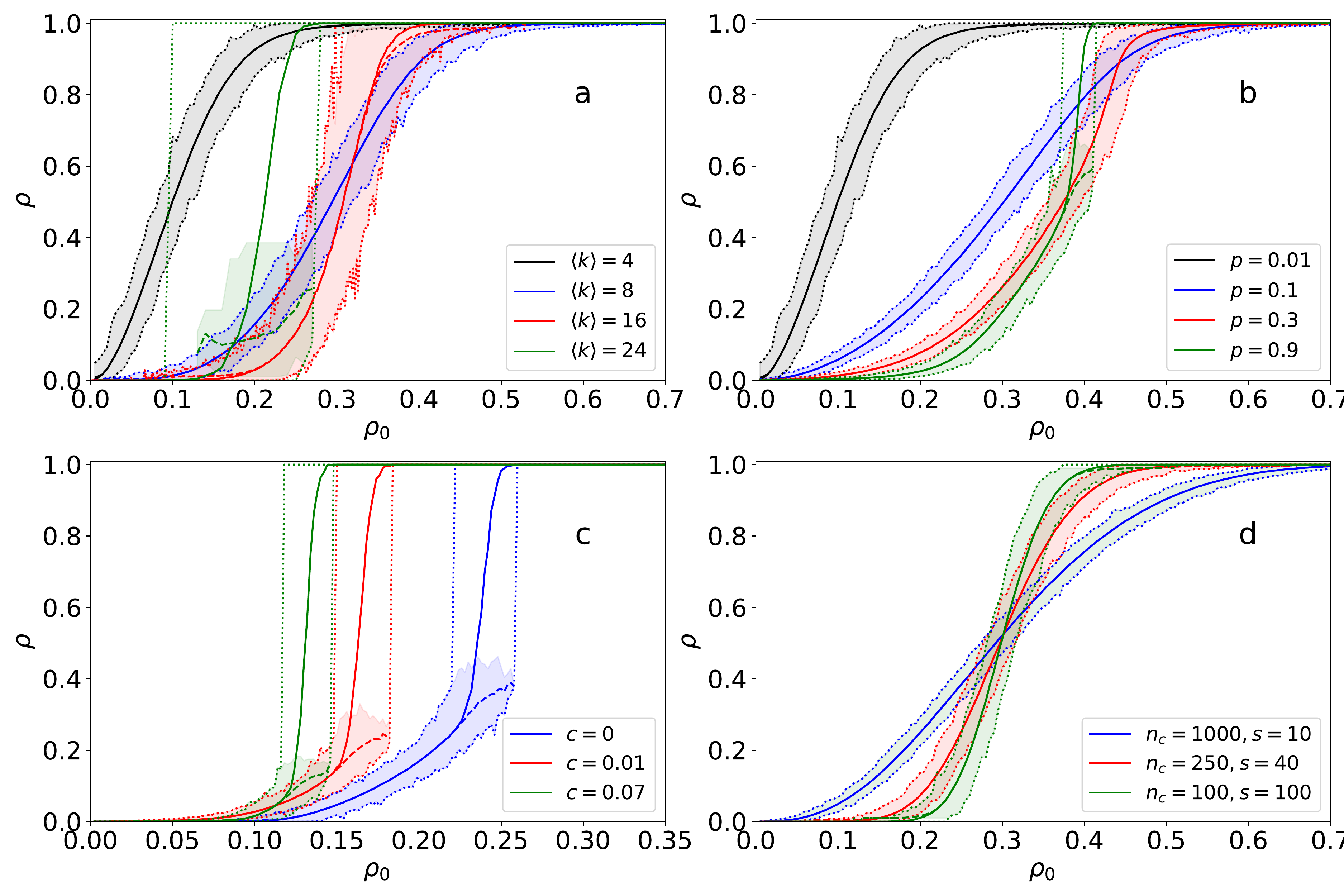}
\caption{Equilibrium density $\rho$ achieved by means of best response dynamics as function of the initial density $\rho_0$ for: (a) Watts-Strogatz random graphs with $N=10^4$ nodes, rewiring probability $p=0.01$ and average degree $\langle k \rangle =6, 8, 16, 20$, (b) Watts-Strogatz random graphs with $N=10^4$ nodes, average degree $\langle k \rangle =4$, and rewiring probability $p=0.01, 0.1, 0.5, 0.9$, (c) random clustered graphs with $N=10^3$ nodes, average degree $\langle k \rangle \approx 4$ and clustering coefficient $c = 0, 0.01, 0.055$, (d) ring of cliques with $N=10^4$ nodes partitioned as $n_c= 1000,250, 100$ cliques of equal size, respectively  $s=10,40,100$.
 Averages are taken over 1000 instances of the dynamics, with parameters (a-b) $\mu=0.45, \sigma=0$ and (c-d) $\mu=0.3, \sigma=0$. Quantities reported are the average density (full line), the minimum and maximum densities (dotted lines) and the average density obtained when the maximum and minimum equilibria are excluded from the sample (dashed line). }\label{fig:BR_other}
\end{center}
\end{figure}

\subsection{Payoff dominance vs. risk dominance}
In order to verify the relation between our theoretical and numerical understanding of dynamical equilibrium selection in coordination games and the results already present in the literature (which are briefly discussed in App.~\ref{app-equilibrium-selection}), we also analyse the case in which the efficient equilibrium is (at most) locally stochastically stable, whereas the risk-dominant equilibrium corresponds to the minimum one. In this regime it was theoretically predicted and experimentally verified that (see references in App.~\ref{app-equilibrium-selection}): (1) larger  connectivity density and clustering promote convergence to the efficient equilibrium, (2) degree heterogeneity often speeds up coordination to efficient outcomes but in general increases unpredictability, (3) the presence of local interactions and low-dimensional structures favour instead risk-dominant outcomes.  

 \begin{figure}[tb]
\begin{center}
\includegraphics[width=0.9\columnwidth]{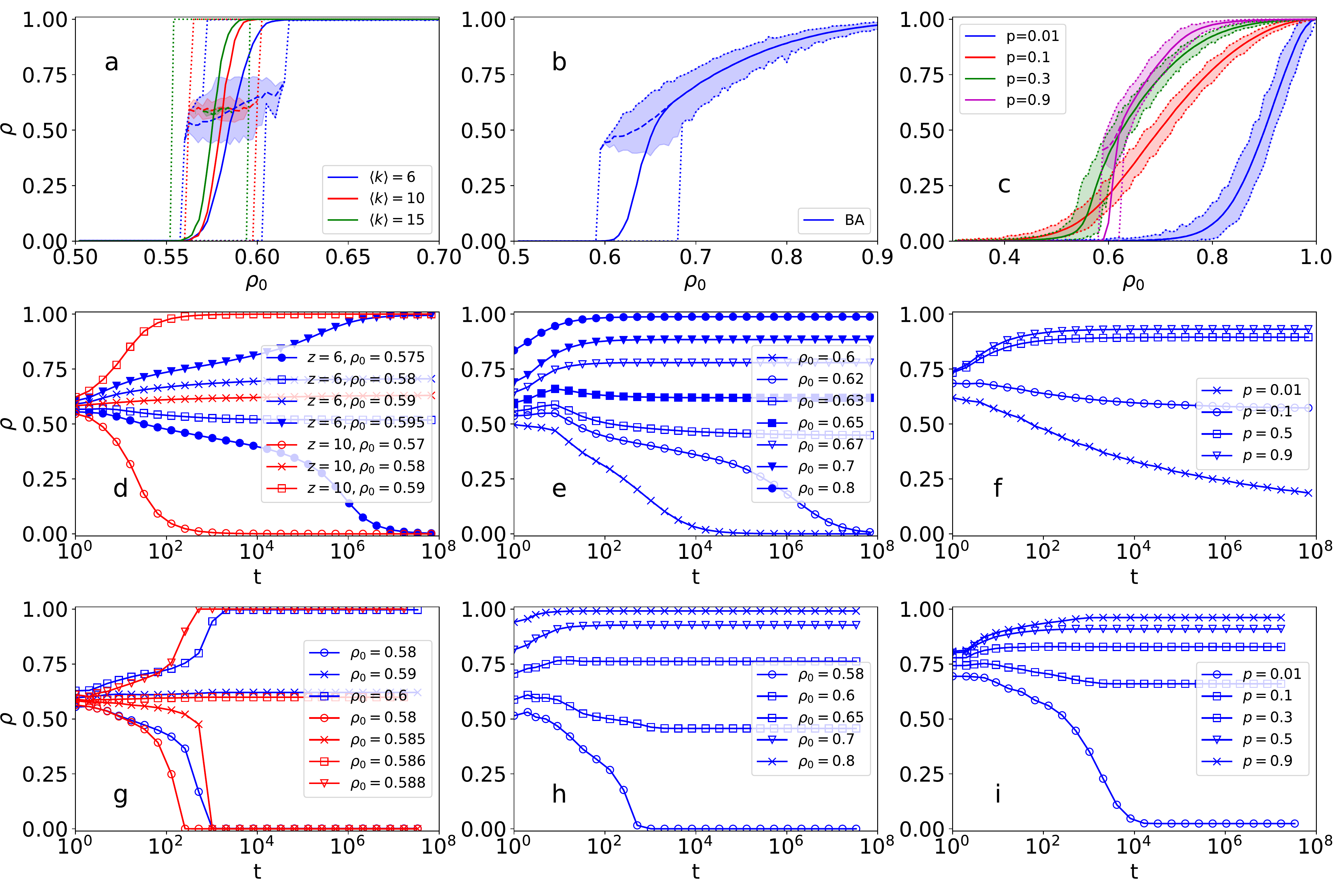}
\caption{Equilibrium selection when the minimum equilibrium is stochastically stable on (a,d,g) Erd\H{o}s-R\`enyi random graphs of $N=10^4$ nodes and variable average degree $z$,  (b,e,h) Barab\`asi-Albert (BA) random networks with minimum degree $k_{\rm min}=2$ and (c,f,i) Watts-Strogatz (WS) random graph with average degree $\langle k \rangle =4$ and variable  rewiring probability $p$. In all networks, individual thresholds are homogeneous, i.e.  Gaussian with mean  $\mu=0.55$ and $\sigma=0$. (a-c) Equilibrium density $\rho$ reached by means of best response dynamics  as function of the initial density $\rho_0$ of agents playing action $1$. Averages are taken over 1000 instances of the dynamics. Quantities reported are the average density (full line), the minimum and maximum densities (dotted lines) and  the average density obtained when the maximum and minimum equilibria are excluded from the sample (dashed line), density values at which non-trivial equilibria are found dynamically (light shaded area). 
(d-f) Density $\rho(t)$ of agents playing action $1$ as function of time during the (deterministic) Fictitious Play dynamics with $\epsilon = 10^{-4}$. (g-i) Density $\rho(t)$ of agents playing action $1$ as function of time during stochastic Reinforcement Learning dynamics with $\beta=10$, $B= 10$ and $\alpha = 0.01$ and with various initial density values $\rho_0$. In ER graphs (d,g), $\langle k \rangle =6$ (blue) and $\langle k \rangle =6$ (red) for various initial density values $\rho_0$, in BA networks (e,h) several values of $\rho_0$, in WS networks (f,i), same initial value $\rho_0=0.7$ and several values of rewiring probability $p$.}
\label{fig:LITER}
\end{center}
\end{figure}
Figure~\ref{fig:LITER} shows results obtained for ER random graphs, BA power-law random networks and WS small-world networks in the regime in which the minimum equilibrium is also globally stochastically stable and the maximum, efficient one, is at most locally stochastically stable. Clearly, if the fraction of players initially oriented to play action 1 is small, any dynamics rapidly converges to the minimum equilibrium. If instead $\rho_0$ is sufficiently large, i.e. at least half of the population, then some non-trivial effects can be observed. The increase of the average degree alone does not substantially alter the basins of attraction of the two stochastically stable states, but it thins the region where the dynamics can be trapped away from them, accelerating the convergence. Degree heterogeneity instead expands the region in which non-trivial high-density equilibria can be reached. It is not surprising that a larger variety of outcomes is observed in experiments performed on such network structures. Finally, the tendency to converge towards efficient equilibria increases with the rewiring probability $p$ in WS, suggesting that this is promoted by the introduction of long-range connections.  We conclude that our numerical tests confirm the main results already reported in the literature on the convergence to efficient or risk-dominant equilibria in coordination games on networks (when the two are distinct) and suggest a novel paradigm for their comprehension in light of the underlying structural properties of the equilibrium landscape.

\section{Conclusions}

In this paper we have conducted a detailed analysis of the static and dynamic properties of Nash equilibria in coordination games defined on random networks. By analyzing the class of pivotal equilibria, previous works pointed out the existence of cascade phenomena in which global coordination is suddenly activated beyond a critical threshold value, which depends on the parameters of the model. Here we integrated the already existing results on coordination models providing a statistical characterization of the properties of a large number of asymmetric equilibria, which even if not directly involved in the cascade processes still contribute to shape the equilibrium landscape in which such phenomena take place. The picture is clear if we adopt as a natural classification for Nash equilibria the one based on the density of agents playing the higher action, i.e. the level of efficiency. First, we have shown that the edges of the density-based equilibrium spectrum (i.e. minimum and maximum equilibria) can be explicitly computed through recursive equations analogous to those given to define the pivotal equilibria and cascade phenomena. This is a consequence of a monotonicity property of the best-response relation starting from the minimum/maximum action profiles in problems with supermodular game-theoretic interactions.  Furthermore, employing the cavity method, the statistical properties of the spectrum of Nash equilibria was studied in detail in the case of uncorrelated random graphs. Generally, for weak disorder on the individual (Gaussian) thresholds, the spectrum presents a continuous band of equilibria at densities ranging from zero to a finite value plus an isolated (Pareto) efficient equilibrium at density 1. The amplitude of the continuous band does not change much as the disorder in the thresholds increases up to a critical value $\sigma_{\rm c}$, beyond which a structural re-organization of the Nash equilibria takes place, which then forms a unique continuous band at high density (but lower than 1). This transition is the equivalent of the well-known transition between weak-disorder and strong-disorder regimes in the Random Field Ising model. The existence of this transition and the value of critical disorder $\sigma_{\rm c}$ at which it takes place was shown to depend on the underlying interaction graph. 

A natural question concerns the relation between the boundary of the cascade region and the structure of the equilibrium landscape. The key parameter is the mean value $\mu$ of individual thresholds. We found that there is no qualitative change in the structure of the equilibrium landscape by just crossing the cascade boundaries
increasing $\mu$, because a large number of low-efficiency equilibria exist below as well as above the cascade transition. 
What changes is their stability with respect to simple dynamic rules of equilibrium selection. Different dynamical processes of equilibrium selection have been studied. 
In the case of best-response dynamics, whenever a continuum spectrum of equilibria exists above the minimum equilibrium, initialising the dynamics from a low density profile favours the convergence to low-density equilibria, even though the process is not monotone. There exists a critical density of agents initially playing action 1 which triggers the instability towards the most efficient equilibrium. The instability is apparently unrelated to the already observed discontinuity due to the cascade process. It is likely that at density values higher than the critical one, the existing Nash equilibria become too rigidly correlated to be reached by a local rearrangement process.
A second result on equilibrium selection is obtained when relaxing the dynamics in order to admit bounded-rational moves: the efficient equilibrium was shown to be generally stochastically stable beyond the boundary of the cascade region, i.e. even though best response does not converge to the efficient equilibrium, a weighted best response still does. We distinguish two situations by increasing the mean threshold $\mu$ from the cascade boundary. Initially only the maximum equilibrium is stochastically stable and even though the majority of equilibria at small density are less stochastically stable with respect to the minimal equilibrium, the latter is not a local maximum of the potential. In other words, convergence to the maximum equilibrium can be prevented mostly for entropic reasons, as there are many low-efficiency equilibria in which myopic rational dynamics can easily get stuck. By further increasing $\mu$, instead, the minimum equilibrium becomes metastable and the large number of inefficient equilibria form a sort of basin of attraction around it. At this point, even bounded-rational dynamic rules can be trapped in low-efficiency equilibria. The stability and metastability properties of the maximum and minimum equilibria can finally exchange by further increasing $\mu$, the latter being the typical situation of most experiments about equilibrium selection performed in the laboratory: the maximum equilibrium is payoff-dominant and the minimum equilibrium is risk-dominant.

These different regimes are crucial to understand the asymptotic behavior of dynamical learning rules in which there is a sort of exploration/exploitation tradeoff represented by the interplay of stochasticity, utility maximization and memory effects. When the memory of the initial moves is strong, low-efficiency equilibria can become asymptotically robust even if the maximum equilibrium is the only stochastically stable state. Results qualitatively  similar to the best-response dynamics are recovered. On the other hand, higher stochasticity favours the local exploration of the equilibrium landscape and, consequently, the ultimate convergence to the stochastically stable equilibrium. 

Concerning the interesting experimental regime in which payoff-dominant and risk-dominant equilibria differ, we obtained numerical results which confirm the behaviour already observed and reported in the literature. Convergence to the efficient equilibrium is easier on networks with a more connected and clustered structure, while local interactions promote convergence to the risk-dominant equilibrium. We advance the explanation that while higher clustering and global interactions change the density support of the equilibrium landscape, higher connectivity produces a shrinkage of the internal part of the spectrum of equilibria, with the by-product of producing faster convergence to the extremes (minimum and maximum equilibria). Moreover, we found that inefficient equilibria preferentially involve low-degree nodes, meaning that increasing degree heterogeneity turns out to favour  the establishment of efficient equilibria (although high variability in the outcomes has to be expected).

In conclusion, the cascade properties only provide a limited understanding of the very rich phenomenology shown by coordination games defined on networks. The results presented in this paper give a new and deeper interpretation to these models and shed light on the relation between static structural properties of Nash equilibria and their fate under dynamic processes of equilibrium selection. Several aspects deserve further investigation. From a technical point of view, the message passing methods employed in this work rely on the assumption of local tree-like structure of networks, which is known to be correct for most random graphs, but unlikely to occur in practice. Even though some of the results were shown numerically to be qualitatively correct also in clustered or structured networks, a more specific analysis should be developed, for instance employing message-passing methods recently introduced in order to account for the presence of short loops \cite{cantwell2019message}. From a modelling point of view, instead, the class of coordination problems studied here is still far from realistic models, in which thresholds are not necessarily linearly related with the (in-)degree \cite{backstrom2006group} and the decision rule can include cooperative effects \cite{centola2007complex}.

\acknowledgments
The author acknowledges funding from Italian national PRIN project 2015592CTH.

\newpage
\appendix

\section{Game-theoretical formulation}\label{app-game}
A simple and common game-theoretic formulation of a coordination problem is the normal-form game composed of two players and two actions $\{A,B\}$ and payoffs given by a bi-matrix of the type shown in Tab.~\ref{tab1}-a, where $a>c$ and $d>b$. In this way, the two symmetric action profiles $(A,A)$ and $(B,B)$ are the only strict pure Nash equilibria. In addition, there is a mixed-strategy Nash equilibrium in which agents play $A$ with probability $q = (d-b)/(a-c-b+d)$ and $B$ with probability $1-q$. The mixed strategy equilibrium is however unstable to small perturbations (e.g. by Cournot-Nash tot\^onnement). We also assume $a>d$, implying that $(A,A)$ is the {\em Pareto efficient} equilibrium, namely no other outcome makes at least one player strictly better off without decreasing the payoff of the other player. When $c < d$, the coordination of players on some action is always better than mis-coordination, while for $c \geq d$, the payoff matrix represents a cooperation dilemma commonly known as {\em stag-hunt} game after the famous metaphor proposed by J.J. Rousseau. The latter condition, however, does not affect the properties of the equilibria \cite{cooper1999coordination}.  Another important concept for games with more than one Nash equilibrium is {\em risk dominance}, which is an attempt of establishing which pure Nash equilibrium is more robust against non-rational play from other agents. Following the definition given by Harsanyi and Selten \cite{harsanyi1988general}, a strict pure Nash equilibrium of a two-person game is risk-dominant if the equilibrium action of each player is a best response against any mixed strategy that assigns at least probability $1/2$  to her opponent equilibrium action. In order to understand that, consider a Markov chain in which an agent plays best-response with probability $q$ and, with probability $1-q$, she plays a randomly chosen strategy. In the $2\times 2$ game in Tab.~\ref{tab1}-a, the expected payoff of a player choosing action $A$ is $\E[u(A)]=q a + (1-q) b$, while for action $B$ it is $\E[u(B)]=q c + (1-q) d$. For both players, $\E[u(A)] > \E[u(B)]$ when $q> (d-b)/ (a - c - b + d )$. It follows that the equilibrium $(A,A)$ is less risky than $(B,B)$ if the smallest probability $q^*= (d-b)/ (a - c - b + d )$ for which $(A,A)$ is not worse than $(B,B)$ is smaller than $1/2$, i.e. if $d- b  < a - c$. If instead $d- b  > a - c$, the profile $(B,B)$ is the risk-dominant equilibrium.
\begin{table}[b]
 \begin{center}
a)  \begin{tabular}{|c|c|c|}
  \hline
  & $A$  & $B$ \\
   \hline
$A$ &  $a,a$ & $b, c$ \\ 
\hline
$B$ & $c,b$ & $d,d$ \\ 
    \hline
  \end{tabular} \hspace{2cm}
    b) 
   \begin{tabular}{|c|c|c|}
  \hline
  & $A$  & $B$ \\
   \hline
$A$ &  $1,1$ & $0, \theta_j$ \\ 
\hline
$B$ & $\theta_i,0$ & $\theta_i,\theta_j$ \\ 
    \hline
  \end{tabular}
\caption{(a) Payoff bi-matrix for a two-player coordination game, with $a>c$ and $d>b$ and $a> d$. (b) Payoff bi-matrix for the linear threshold model with heterogeneous thresholds.}\label{tab1}
\end{center}
\end{table}

When the agents are nodes of a fixed (exogeneous) network $\mathcal{G}=(\mathcal{V},\mathcal{E})$, the normal-form game in Tab.\ref{tab1}-a can be generalised into a game in which agents simultaneously play with all their neighbours through the same action choice \cite{jackson2010social,jackson2015games}. For convenience we map the actions $\{A,B\}$ onto the binary choice $x_i \in \{0,1\}$ of player $i$, that is selected maximising the individual utility function $u_i(x_i; \vec{x}_{\partial i})$, the latter  depending linearly on the sum of the actions of her neighbours $\vec{x}_{\partial i} = \{ x_j | j \in \partial i\}$, i.e.
\begin{subequations}
\begin{align}
 u_i(x_i; \vec{x}_{\partial i})   =&    \sum_{j\in \partial i} \left[ a x_i x_j  + b x_i(1-x_j)  + c (1-x_i) x_j  + d (1-x_i)(1-x_j) \right] \\
    = &  \left(d + (b -d) x_i\right)k_i    + \left((a -c +d -b )x_i + c -d \right) \sum_{j\in \partial i} x_j   \label{utility-function}
\end{align}
\end{subequations}
where $k_i=|\partial i|$ is the degree of agent $i$ in the (undirected) network  $\mathcal{G}$. Coordination models of this kind are also known as {\em linear threshold models} \cite{granovetter1978threshold} because an agent $i$ plays action 1 if a number $m  \geq  \theta k_i = \frac{d-b}{a -c + d -b} k_i$ among her neighbours play 1. Without loss of generality, we can set $a=1$, $b= 0$ and $d=c=\theta \in [0,1)$. We are interested in the case in which the payoff parameters, or equivalently the thresholds $\theta_i$, are time-independent ({\em quenched}) random variables as shown in Tab.\ref{tab1}-b, possibly different for each agent, but drawn from a given common distribution $f(\theta)$. For threshold models with heterogeneous thresholds, the utility functions read
\begin{subequations}
\begin{align}
 u_i(x_i; \vec{x}_{\partial i})  & =  \sum_{j\in \partial i} \left[ x_i x_j   + \theta_i (1-x_i) x_j  + \theta_i (1-x_i)(1-x_j) \right] \\
 & = \theta_i k_i (1- x_i) + x_i\sum_{j\in \partial i} x_j.  \label{utility-function1}
\end{align}
\end{subequations}
The Nash equilibria of the game defined by the utility function \eqref{utility-function1} with heterogeneous thresholds $\{\theta_i\}$ are in one-to-one correspondence with the solutions of the best-response relations \eqref{best-response}. For randomly drawn real-valued thresholds, one should expect all pure Nash equilibria to be strict, i.e. agents are never indifferent among outcomes; on the contrary, for some specific values of the threshold values utility ties are possible and agent's indifference has to be considered (here we assumed to arbitrarily break ties in case of zero utility by preferring action $1$). 
The assumption that the threshold value is linear in the degree of the agent (i.e. $\theta_i k_i$) is not necessary to define a coordination model and could be relaxed. For instance, Twitter data for the 2009 protest recruitment in Spain \cite{gonzalez2011dynamics} are consistent with a linear dependence of thresholds with the in-degree of individuals. On the other hand, Backstrom et al. \cite{backstrom2006group} have shown that a member's likelihood of joining a group on the on-line blogging and social networking site LiveJournal depends roughly logarithmically on the number of friends. This behaviour is consistent with a different class of threshold functions based on the rule of ``diminishing returns'' (see also \cite{kleinberg2007cascading}). Finally, there is an extensive economic literature on binary choice models with heterogeneous idiosyncratic preferences, that go under the name of {\em random utility models} (RUM) \cite{luce2012individual,manski1977structure,bouchaud2013crises,gordon2009discrete}. In this literature the most common choices for the preference distribution are gaussian and logistic, though other distributions, such as skewed lambda~\cite{kandler2009innovation} have been proposed. Empirical data seem to suggest that individuals have diverse thresholds for adopting opinions and taking decisions but the nature of the distribution depends strongly on the system under study.

In the economic literature, the coordination game as defined by the tuple $\Gamma=\left(\mathcal{G},\{0,1\}^{|\mathcal{V}|},\{u_i\}_{i\in \mathcal{V}},\{\theta_i\}_{i\in \mathcal{V}}\right)$ belongs to the important class of {\em  strategic complements} \cite{bulow1985multimarket}, because choices of neighbouring agents are complementary and they mutually reinforce each other. This is mathematically expressed by the property of {\em increasing differences} or, more generally, by the concept of {\em supermodularity} \cite{topkis2011supermodularity}. A {\em partially ordered set} is a set $X$ on which there is a binary order relation ``$\geq$" that is reflexive, antisymmetric, and transitive. Defining the order relation ``$\geq_i$" for each agent $i$ using the natural order between actions $0$ and $1$, the action space $X_i=\{0,1\}$ is a partially ordered set for each agent $i$. Moreover, the pair $(X_i,\geq_i)$ is trivially a complete lattice because any non empty subset of $X_i$ admits a supremum $\sup_i\{x_i \in X_i \}=1$ and an infimum $\inf_i \{x_i \in X_i \}=0$ in $X_i$. The space of action profiles $\mathcal{X}=\{0,1 \}^{|\mathcal{V}|}$ is also a complete lattice if one defines the partial order ``$\geq$" in such a way that $x \geq x'$ if and only if $x_i \geq x_i'$ $\forall i$ and for any subset $S\subset \mathcal{X}$ one defines the infimum and supremum  as the elements given by $\inf(X)=\left\{\inf_i \{x_i | x\in X \} \right\}_{i\in X}$ and $\sup(X)=\left\{\sup_i \{x_i | x\in X \} \right\}_{i\in X}$ \cite{jackson2015games}.\\
 A game $\Gamma$ is of strategic complements if it exhibits the property of  increasing differences: for all $i$ such that $x_i \geq x_i'$ and $\vec{x}_{\partial i} \geq \vec{x}_{\partial i}'$, then 
\begin{align}
u_i\left(x_i;\vec{x}_{\partial i} \right) - u_i\left(x_i';\vec{x}_{\partial i} \right) \geq u_i\left(x_i;\vec{x}_{\partial i}' \right) - u_i\left(x_i';\vec{x}_{\partial i}' \right).
\end{align} 

It is possible to prove that the set of pure Nash equilibria of a game of strategic complements with finite strategy space forms a complete lattice, i.e. it admits a partial order and every non-empty subset has a supremum and an infimum in the set \cite{topkis1979equilibrium,zhou1994set}. It means that games of strategic complements admit a {\em maximal equilibrium} and a {\em minimal equilibrium} that correspond to the extrema of the lattice structure. Moreover, pure Nash equilibria in such games are Pareto-rankable, that is they can be classified in terms of the achieved value of  the global utility function $U = \sum_i u_i$. Based on Pareto ranking, the best (resp. worst) equilibrium, i.e. the one achieving the maximum (resp. mimimum) global utility, corresponds to the maximal (resp. minimal) pure Nash equilibrium. The best equilibrium is also called the {\em (Pareto) efficient} one.  As long as the uniform profile $\vec{1}=(1,1,\dots,1)$ is a Nash equilibrium, it is always the efficient (or maximal) one. Strategic complements on networks also admit mixed strategy Nash equilibria, but their interest is very limited because they are unstable under a broad class of learning rule and perturbations \cite{echenique2004mixed}. The lattice structure of pure Nash equilibria implies that there is a rather simple algorithm to find the maximal and the minimal equilibria, as described in Section \ref{sec:monotoneBR}: starting from the maximal (resp. minimal) available action profile, i.e. the profile $\vec{1}$ (resp. $\vec{0}$), it is sufficient to iterate best-response dynamics until convergence, resulting in a monotone process that approaches the maximum (resp. minimal) equilibrium \cite{topkis1979equilibrium,vives1990nash,jackson2015games}. Echenique \cite{echenique2007finding} modified this process in order to devise an algorithm able to find all pure Nash equilibria in games of strategic complements. The algorithm is based on the idea of repeating the best-response iteration after an equilibrium is found on a reduced state space and force a monotone sequence of iterations that find additional equilibria in a constructive way. Let us call $b(\vec{x}; \vec{\theta}) = \prod_{i=1}^{|\mathcal{V}|} b_i(\vec{x}_{\partial i}; \theta_i)$ the (synchronous) best-response update for the overall action profile. Since we broke ties due to utility indifference of the agents, the pure Nash equilibria satisfy the condition $\vec{x} = b(\vec{x}; \vec{\theta})$. We also know that performing best-response iteration on the profile $\vec{0}$ brings to the minimal equilibrium $\vec{x}^{\rm m}$. The first phase of the algorithm consists in discovering potential equilibrium profiles. This is done as follows (for a formal definition see Algorithm 2 in \cite{echenique2007finding}). From the minimal equilibrium, one simply repeats the best-response dynamics at most $N$ times, each one starting from a minimal monotone perturbation of  $\vec{x}^{\rm m}$. Suppose agent $1$ is restricted to choose action larger than the one adopted in the minimal equilibrium, i.e. $\vec{x} = (x^{\rm m}_1+1,x^{\rm m}_2, \dots, x^{\rm m}_N)$ with $N=|\mathcal{V}|$. If $\vec{x}\in \{0,1\}^N$, because of monotonicity, the best-response dynamics will produce the minimum equilibrium larger or equal to $\vec{x}$. This is marked as a potential equilibrium. Then the algorithm moves to the following starting point $\vec{x} = (x^{\rm m}_1,x^{\rm m}_2+1, \dots, x^{\rm m}_N)$ and, if $\vec{x}\in \{0,1\}^N$, the best-response dynamics will produce another potential equilibrium. This is done starting from all allowed monotone perturbations of $\vec{x}^{\rm m}$ (at most $N$ if $\vec{x}^{\rm m}=\vec{0}$). Then, the algorithm moves to analyze the first potential equilibrium discovered and repeats the same procedure based on minimal monotone perturbations and best-response dynamics. At each step of the process one of the potential equilibria is analyzed and novel potential equilibria can be discovered. The algorithm stops when all discovered potential equilibria, including the maximal one $\vec{x}^M$ have been analyzed. Finally, all potential equilibria have to be checked against deviations towards the original equilibria from which they have been discovered (i.e. it is sufficient to check with respect to the action on which perturbation occurred). Even though the algorithm is exponential in the number of agents in the worst case, it tries to exploit monotonicity of the influence structure between agents often resulting in a much more efficient process (see also \cite{rodriguez2012structure}). In the present case, however, the cardinality of the set of pure Nash equilibria is itself in general exponential in $N$.

While the efficiency properties of the $2\times 2$ coordination game can be easily generalized to the game under study, the concept of risk-dominance has no trivial generalisation to multi-agent games. Among the proposed generalisations, {\em $p$-dominance} is one of the most appealing. Following Morris et al. \cite{morris1995p}, a strategy is $p$-dominant if it is best response whenever it is played by at least fraction $p$ of agents. So they defined a $(p_1, \dots, p_n)$-dominant equilibrium as the one in which the action of agent $i$ is the best response to any distribution that assigns a weight of at least $p_i$ to the equilibrium profile. In this way, one could generalise the concept of risk-dominance to arbitrary multi-player games, defining it as the selection of $1/2$-dominant equilibria \cite{ellison2000basins}. These are also called {\em globally risk dominant} equilibria \cite{maruta1997relationship}. 
More recently, other refinements of the Nash equilibrium concept, which more suitable for multi-player games defined on networks were introduced, such as the {\em ordinal generalized risk-dominance} \cite{peski2010generalized} with a clear graph-theoretic interpretation.
The main justification behind the introduction of the concept of risk-dominance in $2\times 2$ games is that it should identify the equilibrium with larger basins of attraction or, in other words, the one that is more secure because it is more robust against stochastic perturbations. Along the same lines,  the concept of {\em stochastically stable} equilibrium was introduced as the stable long-run outcome of stochastic evolutionary learning dynamics \cite{foster1990stochastic,young1993evolution,kandori1993learning,blume1993statistical,kandori1995evolution}. Stochastically stable equilibria correspond to the set of strategy profiles (pure or mixed ones) with positive  limiting distributions as the random mutation rate or noise strength of the evolutionary model tends to zero.  The {\em least resistance method} introduced by Young \cite{young1993evolution} to identify stochastically stable states is based on the construction of the minimum-cost spanning trees on the graph associated with the transition matrix \cite{chu1965shortest,edmonds1967optimum,tarjan1977finding,freidlin1998random}. Even though the algorithm for the minimum-cost spanning tree is polynomial, the number of vertices of the transition graph generally scales exponentially with the system size (e.g. it is $O(2^N)$ in the present case), making this computational approach unfeasible for large multi-player games. A more conceptual criticism to stochastic stability is that if perturbations were allowed to be state-dependent in an arbitrary fashion, any state could be made stochastically stable just by adjusting the rates at which perturbations vanish  \cite{bergin1996evolution}. However, when the game admits a potential function \cite{monderer1996potential}, one can reasonably define the stochastically stable states as those identified by the limiting distribution associated with {\em logit update rule} \cite{mcfadden1973conditional,blume1993statistical}. The coordination game defined by utility function  \eqref{utility-function} on a graph $\mathcal{G}$
is in fact a {\em potential game}, that is it admits the potential function
\begin{equation}\label{potentialfunx}
 V(\vec{x})  =   (a-c+d-b)\sum_{(i,j)} x_i x_j  - (d-b) \sum_i  k_i x_i.
\end{equation}
The generalization to heterogeneous parameters is straightforward when the interaction is symmetric, in particular when $a_i-c_i+d_i-b_i$ is a constant independent of $i$. Threshold models with heterogenous thresholds $\vec{\theta}$ defined in Table \ref{tab1}-b is such that $a_i-c_i+d_i-b_i=1-\theta_i+\theta_i= 1$, therefore they are potential games with potential function 
\begin{equation}\label{potentialfunction}
V(\vec{x};\vec{\theta})  =   \sum_{(i,j)} x_i x_j  - \sum_i  \theta_i k_i x_i.  
\end{equation}
The change in the potential function as consequence of action revision $x_i \to x_i'$ by a single agent $i$ is equal to the corresponding individual utility change for agent $i$, i.e.  given two profiles $\vec{x}$ and $\vec{x}'$ differing only in the action played by agent $i$, it holds  that 
\begin{subequations}
\begin{align}
   V(\vec{x}';\vec{\theta})  - V(\vec{x};\vec{\theta})  & =   \left[\sum_{j\in\partial i}x_j -k_i \theta_i \right](x_i'-x_i) \\
   &=  u_i(x_i',\vec{x}_{\partial i})-u_i(x_i,\vec{x}_{\partial i}).
\end{align}
\end{subequations}
It follows that under best-response dynamics the potential function cannot decrease and, more importantly, that the Nash equilibria are the local maxima of the potential. It is also possible to prove that potential maxima are in pure-strategy Nash equilibria, result that is generally true for games of strategic complements \cite{bramoulle2001complementarity}. A mixed strategy profile is a probability vector $\vec{\pi} \in [0,1]^N$ in which $\pi_i$ indicates the probability with which agent $i$ plays action $1$. The generalization of the potential function to mixed strategies is 
\begin{equation}\label{potentialfunctionMixed}
V(\vec{\pi};\vec{\theta})  =   \frac{1}{2}\sum_{i, j}a_{ij} \pi_i \pi_j  - \sum_i  \theta_i k_i \pi_i,
\end{equation}
where $A=\{a_{ij}\in \{0,1\}\}_{i,j=1}^N$ is the symmetric adjacency matrix.
The first-order maximization condition 
\begin{equation}
\frac{\partial V}{\partial \pi_i}  =  \sum_{j} a_{ij} \pi_j  - \theta_i k_i  = 0, \quad \forall i \in \mathcal{V},
\end{equation}
 is possible for both pure and mixed equilibria. The Hessian matrix is just given by $\frac{\partial^2 V}{\partial \pi_i\partial \pi_j}  =   a_{ij}$ that is not negative (nor positive) semi-definite because, in the absence of self-loops on the graph, it is surely ${\rm Tr}[A]=0$. It follows that mixed-strategy Nash equilibria cannot be global maxima of the potential $V$, which instead have to stay on the corners of the hypercube.

Some important results connect  stochastic stability with the concept of risk dominance.  For instance, in a $2\times 2$ game, only the risk-dominant equilibrium is stochastically stable \cite{young1993evolution}. This result was generalized by Maruta \cite{maruta1997relationship} to  $n\times n$ symmetric games (two players with a finite set of $n$ actions): if a globally risk dominant equilibrium exists it is stochastically stable. Recently, Opolot  extended the relationship between $p$-dominance and stochastic stability to coordination games defined on networks \cite{opolot2018relationship}. 

It is worth noticing that multi-player games with discrete actions defined on network structures were also subject of intense investigation in the computer science community. Algorithmic game theory \cite{nisan2007algorithmic,roughgarden2010computing} studies the computational aspects of Nash equilibria and other solution concepts. The computational problem of defining and finding  Nash equilibria for games in which agents interact locally on a graph was introduced by La Mura \cite{la2000game} and Kearns et al. \cite{kearns2013graphical}. An efficient (i.e. polynomial in the size of the game, but the latter can be exponential in the number of strategies or players) algorithm to find all, generally mixed-strategy, approximate Nash equilibria (called $\epsilon$-Nash equilibria) on trees was originally proposed in \cite{kearns2013graphical}, then belief propagation and other heuristics for constraint-satisfaction problems were proposed for loopy graphs \cite{ortiz2003nash,vickrey2002multi,soni2007constraint}. Computational approaches to other solution concepts, such as correlated equilibria  \cite{kakade2003correlated} and Bayes-Nash equilibria \cite{soni2007constraint}, were also considered. Focusing on pure Nash equilibria, an important result is the exact mapping of any graphical game onto a Markov random field by Daskalakis and Papadimitriou \cite{daskalakis2006computing}, that makes possible to tackle exact and approximate computations of pure Nash equilibria with the algorithmic techniques usually employed to find the maximum a posteriori configurations of the associated Markov random field model.

\section{Related literature on dynamical equilibrium selection in coordination games}\label{app-equilibrium-selection}

Historically, the problem of dynamical equilibrium selection in coordination games is solved by introducing bounded rationality. 
Harsanyi and Selten \cite{harsanyi1988general} proposed the concept of risk dominance as a refinement of Nash equilibria in $2\times 2$ games in order to move the attention from the efficiency of the equilibrium to its robustness against mistakes, because of the higher payoff such strategies provide in case of miscoordination compared to the payoff-dominant ones.  As discussed in App.~\ref{app-game}, risk dominance has no straightforward and unique generalization to multi-player games, where it is usually replaced by the concept of stochastic stability, i.e. the long run equilibria of the game are those which are visited with positive probability in the limit of vanishingly small noise \cite{foster1990stochastic,blume1993statistical,kandori1993learning,young1993evolution}. The long-run behaviour of a wide class of evolutionary and learning dynamics in the presence of non-trivial  (fixed or changing) interaction structure has been considerably investigated over the last three decades \cite{ellison1993learning,ellison2000basins,blume1995statistical,robson1996efficient,anderlini1996path,berninghaus1996conventions,eshel1998altruists,lee2000noisy,sandholm2001almost,alos2007partial,montanari2010spread,young2011dynamics,kreindler2013fast,azomahou2014convergence,opolot2016contagion}. Most results are obtained for specific dynamical rules and on systems with no disorder, such as regular lattices. For instance, noisy best response converges to the risk-dominant equilibrium when played by agents organized on a circle with nearest-neighbour interactions \cite{ellison1993learning}, whereas in the same setup some imitation-based learning dynamics was shown to preferentially select the payoff-dominant equilibrium \cite{robson1996efficient,eshel1998altruists,alos2007partial}. 
On $d$-dimensional lattices and general networks, different types of equilibria can be achieved depending on the details of the dynamics \cite{blume1995statistical,berninghaus1996conventions,morris2000contagion,kosfeld2002stochastic} and two or more stochastically stable states can coexist in the same system \cite{anderlini1996path,jackson2002formation}. Moreover, the convergence to the stochastically stable state is much faster when agents interact locally (e.g. low-dimensional lattices) than for systems with global and random interactions (e.g. random graphs), where the convergence time can even be exponential in the number of players \cite{montanari2010spread,young2011dynamics}. 

Recent works have considered the effects of topological network properties, such as degree heterogeneity, clustering and community structure on the dynamical equilibrium selection in coordination games. By means of numerical simulations, it was recently shown that higher density of edges, clustering and community structure promote convergence to the efficient outcomes under noisy best response \cite{tomassini2010evolution,antonioni2013coordination,buskens2016effects}, whereas the existence of bottlenecks or other topological traps might prevent convergence to one of the equilibria in the case of imitation dynamics \cite{roca2010topological}.
Finally, coordination to the efficient outcome is favoured by degree heterogeneity \cite{lopez2006contagion,cimini2015dynamics,konno2015coordination,mazzoli2017equilibria}, though heterogeneity mixes positive effects due to higher centralization with negative effects due to higher segmentation of the network \cite{buskens2016effects}.

Some of these theoretical results were confirmed in laboratory experiments, where we can identify a core of findings that appear robust across the different experimental setups. We focus again on  the case in which a distinction between payoff-dominant and risk-dominant equilibria is possible. First, the number of players may affect equilibrium selection, because players in large groups tend to be more risk-averse \cite{van1990tacit}.  The efficient equilibrium is more preferably obtained in complete network structures or  when clustering is large \cite{keser1998coordination,my1999global,berninghaus2002conventions,cassar2007coordination}. On the other hand, local interactions between players arranged on a regular low-dimensional structure (e.g. lattice) favor convergence to the risk-dominant equilibrium \cite{keser1998coordination,my1999global,berninghaus2002conventions}. This effect decreases increasing the dimensionality of the lattice \cite{berninghaus2002conventions,rosenkranz2008network}. Convergence to the efficient equilibrium is faster in networks with small world properties and clustering \cite{cassar2007coordination}.
Moreover, the equilibrium selection is also affected by the duration of the repeated game, as the coordination to the efficient equilibrium usually increases with the number of rounds of the myopic play \cite{van1990tacit,berninghaus1998time}. In this respect, recent experiments argue that in most cases the efficient equilibrium is  played with low effects due to the network structure \cite{cassar2007coordination,frey2012equilibrium,charness2014experimental}.
Finally, degree heterogeneity seems to promote coordination with a crucial role played by well-connected individuals which trigger the adoption cascades \cite{rosenkranz2008network,kearns2009behavioral}. Finally, coordination emergence is more difficult when the network exhibits community structure and cliquishness \cite{roca2010topological,judd2010behavioral,antonioni2013coordination}. For a review of the experimental results on coordination games, see e.g. \cite{kosfeld2004economic,choi2016networks}.

\section{Properties of the pivotal equilibria on random graphs with uniform thresholds}\label{app-pivotal}

\begin{figure}[tb]
\begin{center}
\includegraphics[width=0.6\columnwidth]{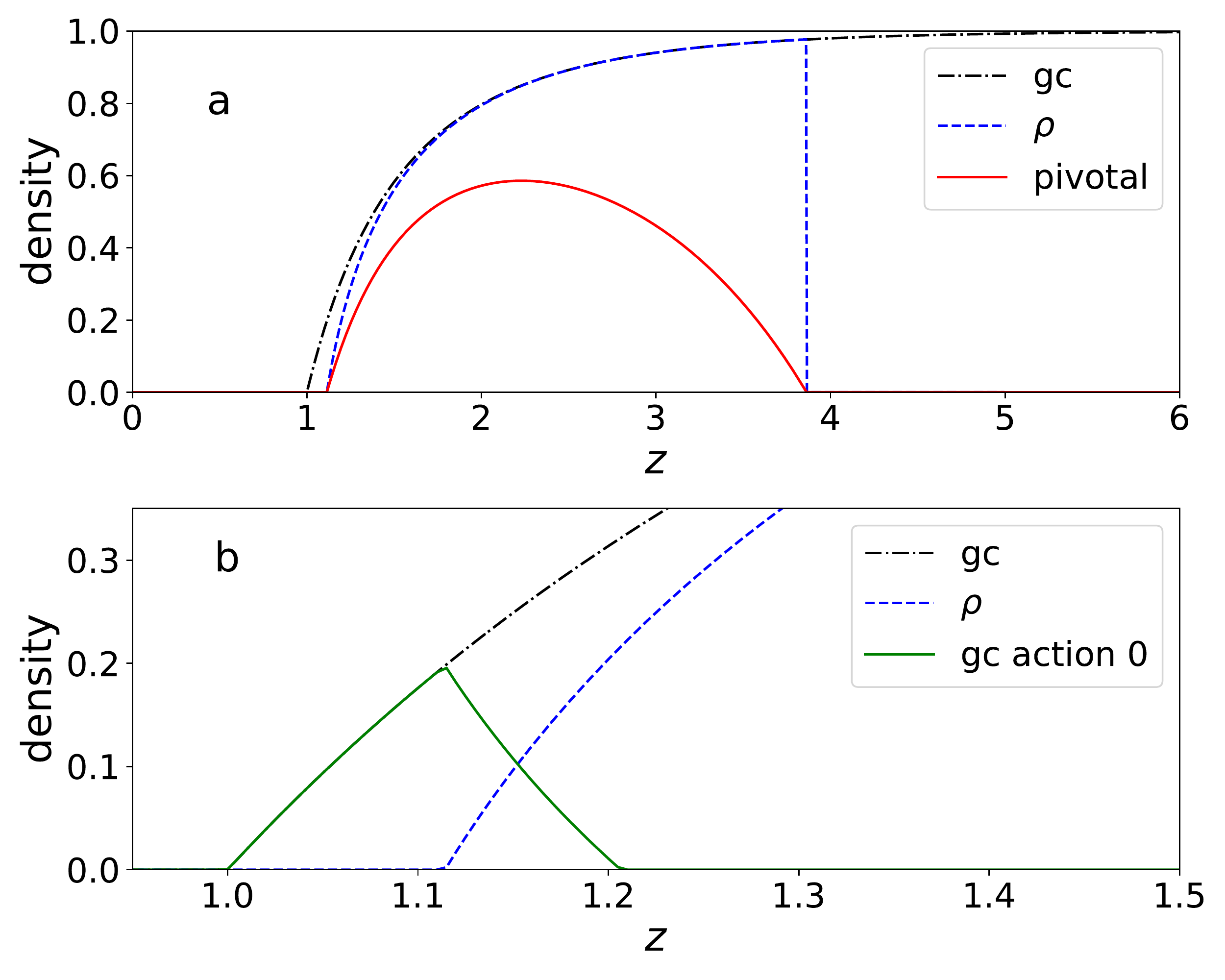}
    \caption{Properties of pivotal equilibria on Erd\H{o}s-R\'enyi (ER) random graph of average degree $z$ and uniform thresholds ($\mu=0.23$, $\sigma=0$): (a) size of the largest connected component (black, dash-dotted line), density $\rho$ of nodes playing action $1$ (blue dashed line), density of pivotal nodes (red solid line);  (b) size of the largest connected component (black, dash-dotted line), density $\rho$ of nodes playing action $1$ (blue dashed line), size of the largest connected component of agents playing action $0$  (green solid line). }\label{fig:ERlelarge}
\end{center}
\end{figure}

In the case of uniform thresholds ($\sigma=0$) one can easily define the concept of {\em pivotal equilibrium}. Following Lelarge \cite{lelarge2012diffusion}, we call  pivotal agents  the nodes belonging to the largest connected component of the induced subgraph in which only nodes of degree strictly less than $\mu^{-1}$ are retained. The pivotal equilibria can be obtained from the trivial minimal equilibrium $\vec{0}$ by just switching two neighbouring pivotal agents. On  random regular  graphs of degree $K$, the set of pivotal agents is, respectively, empty for $\mu>\mu_{\rm c,m}=1/K$ or the whole (giant connected component of the) graph for $\mu\leq \mu_{\rm c,m}$. On general random graphs, instead, coexistence of the two actions is possible in a pivotal equilibrium. For Erd\H{o}s-R\'enyi (ER) random graphs with average degree $z$, there is an interval  $[z_{\rm m}(\mu),z_{\rm M}(\mu)]$ in which the minimal equilibrium at $\vec{0}$  becomes unstable in favour of pivotal  equilibria with a finite density of agents playing action $1$, as shown in Fig.~\ref{fig:ERlelarge}a. Agents playing action $0$  are usually connected in small clusters and only in some specific regime two giant connected components can do actually exist  \cite{lelarge2012diffusion}, result that can be straightforwardly obtained in a non-rigorous way as follows.  
In the usual tree-like approximation, if $h$ is the solution of \eqref{bp-lo}, the probability $\xi_0$ that an edge chosen randomly leads to the giant component formed by agents with action $0$ is obtained solving the self-consistent equation
\begin{equation}\label{bp-xi}
\xi_0 =  \sum_{k} \frac{k p_k}{\avg{k}} \sum_{l=0}^{k-1} \sum_{m=1}^{k-1-l} M_{k-1,l,m}\left(h,\xi_0,1-h-\xi_0\right) F_{0}\left(\frac{l}{k}\right) 
\end{equation}
with the multinomial expression $M_{k,l,m}(x,y,z)=\binom{k-1}{l}  \binom{k-1-l}{m}x^l y^m z^{k-l-m}$. Then the  probability $\hat{\xi}_0$ that a randomly chosen node belongs to the giant component of action $0$ is 
\begin{equation}\label{bp-hatxi}
\hat{\xi}_0 =  \sum_{k} p_k \sum_{l=0}^{k} \sum_{m=1}^{k-l} M_{k,l,m}\left(h,\xi_0,1-h-\xi_0\right) F_{0}\left(l/k\right).
\end{equation}
Based on these results, Fig.~\ref{fig:ERlelarge}b shows the coexistence of two giant components for $\mu=0.23$ and $\sigma=0$ as function of the average degree $z$ of Erd\H{o}s-R\'enyi (ER) random graphs. It is remarkable that this phenomenon takes place in a very limited interval of values of $z$, outside of which the agents playing action $0$ in the pivotal equilibria are mostly organised in many small clusters rather than in a large connected component.
\begin{figure}[tb]
\begin{center}
\includegraphics[width=0.6\columnwidth]{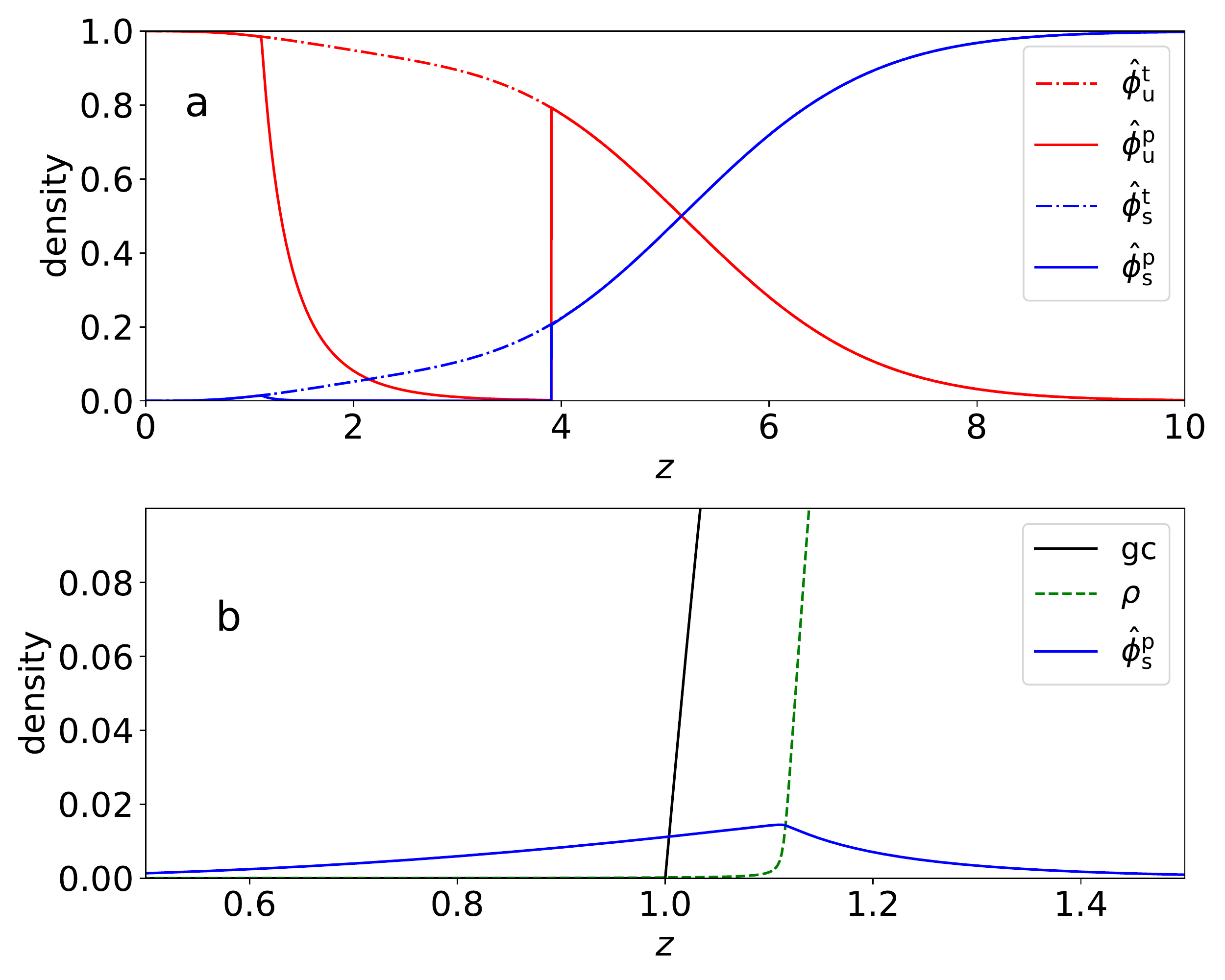}
    \caption{Properties of pivotal equilibria on Erd\H{o}s-R\'enyi (ER) random graphs of average degree $z$ (after removal of isolated nodes) and uniform thresholds ($\mu=0.23$, $\sigma=0$): (a) densities of stable (blue lines) and unstable (red lines) not-isolated nodes playing action $0$ in the trivial $\vec{0}$ equilibrium (dashed lines for $\hat{\phi}_{\rm s}^{\rm t}$, $\hat{\phi}_{\rm u}^{\rm t}$) and in the pivotal one (solid lines for $\hat{\phi}_{\rm s}^{\rm p}$, $\hat{\phi}_{\rm u}^{\rm p}$); (b) Highlight of the behaviour of $\hat{\phi}_{\rm s}^{\rm p}$  (blue solid line) in the neighbourhood of the transition and comparison with the rescaled size of the largest connected component (black solid line) and with the density of agents playing action $1$ (green dashed line). }\label{fig:ERunstable}
\end{center}
\end{figure}

Another interesting quantity is the probability $\phi$ that a randomly chosen edge emerging from any node $j$ with $x_j=0$ leads to a node $i$ that also plays action $x_i=0$ and this choice does not change if $x_j$ is turned to $1$. On a tree-like graph, it satisfies the self-consistent equation 
\begin{equation}\label{bp-phi}
\phi =  \sum_{k} \frac{k p_k}{\avg{k}} \sum_{l=0}^{k-1} \binom{k-1}{l} h^l (1-h)^{k-1-l} F_{0}\left(\frac{l+1}{k}\right) 
\end{equation}
where $h$ satisfies \eqref{bp-lo}. This quantity can be used to compute  
the probability $\hat{\phi}_{\rm u}$ that a node plays action $0$ but is unstable to perturbations, given by 
\begin{equation}\label{bp-qu}
 \hat{\phi}_{\rm u}  =   \sum_{k} p_k \sum_{l=0}^{k} \sum_{m=0}^{k-l} M_{k,l,m}(h,1-h-\phi,\phi)  \left\{ F_{1}\left(\frac{l+m}{k}\right) - F_{1}\left(\frac{l}{k}\right)\right\}.
\end{equation}
Finally, the probability  that a node plays action $0$ and it does not trigger any cascade when turned to $1$ but, necessarily, switches back to $0$ is $\hat{\phi}_{\rm s}=1-\rho- \hat{\phi}_{\rm u}$.
Figure~\ref{fig:ERunstable} displays these quantities  for  Erd\H{o}s-R\'enyi (ER) random graphs as function of the average degree $z$ for $\mu=0.23$ and $\sigma=0$ (the behaviour is qualitatively similar for other threshold values). For $z <z_{\rm m}$, all nodes play action $0$, but they are mostly unstable to small perturbations. Apart from isolated nodes, that are always trivially stable to perturbations (and are removed in Fig.~\ref{fig:ERunstable}), some stable $0$s appear around the continuous transition at $z_{\rm m}$. Strikingly, for  $z>z_{\rm M}$, a large fraction of nodes are unstable to perturbations even though this fraction decreases with $z$. It means that other Nash equilibria with agents playing action 1 are expected to exist in addition to the trivial one. In addition to the minimal and maximal equilibria, a large number of other equilibria is in fact shown to exist in Sec.~\ref{subsec:cavityGRG}.

\section{Spectrum of equilibria in the fully-connected limit}\label{app-fullyconnected}

A particularly simple limit of the model under study is obtained when the underlying graph is fully-connected: every agent interacts with all the others, even though they are endowed with different threshold values $\{\theta_i\}_{i=1}^N$ taken from a common distribution $f(\theta)$. For $N\to \infty$, the local fields $h_i = \frac{1}{N-1}\sum_{j\neq i} x_j$ concentrate around the average value $\rho_{\rm fc}$, which is the solution of the non-linear self-consistent equation $\rho_{\rm fc} = F_1(\rho_{\rm fc})$, obtained averaging the best-response relation \eqref{best-response} over the disorder. As function  of the parameters of $f(\theta)$, the solution is in general not unique, with well-known bi-stability phenomena \cite{sethna2004random,bouchaud2013crises,lucas2013multistable}.  It is natural to ask about entropic effects, that is how many different Nash equilibria exist for the same value of the average density, for which we can use a simple combinatorial approach (see also \cite{rosinberg2008stable}). In a fully-connected system of size $N$, the average number of Nash equilibria with exactly $X$ agents playing action $1$ is given by 
\begin{align}
\mathcal{N}_{NE}(X) = \binom{N}{X} F_1\left( \frac{X-1}{N}\right)^X \left[1- F_1\left(\frac{X}{N} \right) \right]^{N-X}.
\end{align}
In the large $N$ limit, called $\rho= X/N$, we have
\begin{equation}\label{NumberNEmf}
\mathcal{N}_{NE}(\rho) = \frac{1}{\sqrt{2\pi N \rho(1-\rho)}} e^{-N \Psi(\rho) - F_1'(\rho)\frac{\rho}{F_1(\rho)}},
\end{equation} 
with the large deviation function $\Psi(\rho) = \rho \log\frac{\rho}{F_1(\rho)} +(1-\rho)\log\frac{(1-\rho)}{1-F_1(\rho)}$ and  $F_1'(\rho_{\rm fc})=\frac{d F_1}{d \rho}|_{\rho = \rho_{\rm fc}}$. The average number $\mathcal{N}_{NE} = \sum_X \mathcal{N}_{NE}(X)$ is computed using Laplace's method and noticing that, at the saddle point, the mean-field self-consistent condition $\rho_{\rm fc} = F_1(\rho_{\rm fc})$ holds. 
For $N = \infty$, the multistability can only occur exactly at $\rho_{\rm fc}$, and the corresponding average number of equilibria is obtained performing the gaussian integral around the saddle point, i.e.  
\begin{equation}\label{eqNneMF}
\mathcal{N}_{NE} = \frac{e^{-F_1'(\rho_{\rm fc})}}{\left| 1 - F_1'(\rho_{\rm fc})\right|},
\end{equation}
which diverges at $F_1'(\rho_{\rm fc})=1$.\footnote{It must be noticed that rather than the average number, the most relevant quantity to be computed is the typical number. The calculation can be extended with little more effort to compute the full distribution of the number of Nash equilibria, that turns out to be exponential, with most probabile value being $0$ or $1$ depending on $\sigma$ \cite{rosinberg2008stable}.}
For finite $N$,  the probability of Nash equilibria with density $\rho$ is Gaussian around the typical density $\rho_{\rm fc}$ with variance $\Delta_{NE}^2 = \frac{\rho_{\rm fc}(1 - \rho_{\rm fc})}{N(1-F_1'(\rho_{\rm fc}) )^2}$. 
In fact, the interval of density values for which Nash equilibria exist shrinks to zero for $N\to \infty$ because, in a fully-connected graph, density fluctuations also vanish in the same limit.  
Nevertheless, the term $1/(1-F_1'(\rho_{\rm fc}))$ has an interesting interpretation in terms of branching processes. For large $N$, the probability that one agent moves from action $0$ to action $1$ as a consequence of a previous change of another action is roughly given by 
\begin{equation}
F_1\left(\frac{X+1}{N}\right) - F_1\left(\frac{X}{N}\right) \approx  \frac{1}{N}F_1'(\rho_{\rm fc}),
\end{equation}
it follows that the average number of agents that revise their action roughly grows in time as a branching process with branching ratio $F_1'(\rho_{\rm fc})$ and average final size inversely proportional to $1-F_1'(\rho_{\rm fc})$.  

On finitely-connected graphs, local fields $\{h_i\}_{i=1}^N$ keep fluctuating even in the thermodynamic limit, suggesting the existence of non-trivial entropic effects over a finite region of density values. A simple argument for homogeneous random graphs \cite{lucas2013multistable} can be used to demonstrate that this should be the case for any finite value of the average degree $z$. For $z\gg 1$, the probability that a  node is induced to flip from $0$ to $1$ because of the previous flip of a neighbouring node is approximately $F_1'(\rho_{\rm fc})/z$, where we approximate the density with that of the fully connected model. Then, since a node $i$ has approximately $z$ neighbours, the probability with which  two neighbouring nodes can mutually sustain each other in either $0$ or $1$ state is  $z  (F_1'(\rho_{\rm fc})/z)^2 = F_1'(\rho_{\rm fc})^2/z$. If we assume that the $Nz/2$ pairs of nodes in the graph can be flipped independently, then a rough calculation suggests a number of possible equilibria $\mathcal{N}_{NE} \sim 2^{N F_1'(\rho_{\rm fc})^2/(2z)}$. In fact, in the presence of weak correlations, different Nash equilibria are connected by avalanches of strategy revision. 
Suppose that we focus on a region of the graph around node $i$ in which nodes play action $0$ (the argument works in the opposite direction as well). If node $i$ is flipped to 1 together with a neighbour $j$, then the remaining $z-1$ neighbours of $j$ can be induced to turn to $1$ each with probability $F_1'(\rho_{\rm fc})/z$. The revision process propagates as an avalanche with final average size
\begin{equation}
 \frac{F_1'(\rho_{\rm fc})^2}{z} \sum_{\ell=0}^{\infty} \left(\frac{F_1'(\rho_{\rm fc})(z-1)}{z}\right)^\ell \approx \frac{ F_1'(\rho_{\rm fc})^2}{z\left(1 - F_1'(\rho_{\rm fc})\right)}.
\end{equation}
There are possibly $O(N)$ of such finite avalanche processes, meaning that we should expect a density spectrum of Nash equilibria of finite width $\sim O(1/z)$ even in the thermodynamic limit.
This can be correct only at very low values of $F_1'(\rho_{\rm fc})/z$ ( e.g. at very large $z$), where it is reasonable to expect that pair flips do not trigger avalanches of rearrangements throughout the whole system.

\section{Derivation of distributional BP equations from the Replica Method}\label{sec:replica}
The  distributional BP equations for ensembles of random graphs, equivalent to \eqref{bp-instancePregular} in the case of random regular graphs, can be also obtained by means of an application of the Replica Method originally developed to study metastable states in finitely-connected graphs \cite{dean2000metastable,lefevre2001metastable,berg2001metastable,detcheverry2005metastable,rosinberg2008stable,rosinberg2009t}.
The starting point for the application of the replica method is the average replicated  partition function, representing the $n$-th moment of the overall statistical weight of the Nash equilibria averaged over the distribution of quench disorder (i.e. threshold values and random graph structure). The variable $x_i^a \in \{0,1\}$ (with $i=1,\dots,N$ and $a=1,\dots, n$) represents the possible actions of agents $i$ in the copy $a$ of the system. The best-response constraints, together with the constraint on the degrees of the nodes and on the overall number of agents playing action $1$, are included by means of Kronecker delta functions. It follows that the average replicated partition function reads 
\begin{equation}
 \left\langle \left\langle \mathcal{Z}_{NE}(\mathbf{X})^n  \right\rangle \right\rangle = \left\langle \left\langle \text{Tr} \prod_{a=1}^n \left[ \prod_{i=1}^{N} \delta_{\rm K}\left(x_i^a,\Theta\left[\sum_{j\neq i} a_{ij} x_j^a-\theta_i K\right]\right) \delta_{\rm K}\left(X^a,\sum_i x_i^a\right) \right]\delta\left(\sum_{j\neq i} a_{ij}, K\right) \right\rangle_A \right\rangle_f
\end{equation} 
where the averages are performed over the graph ensemble specified by the adjacency matrix $A=\{a_{ij}\}$ with $a_{ij}\in\{0,1\}$ and over the threshold distribution $f(\theta)$. Before performing the averages we focus on the replicated partition function that can be rewritten as 
\begin{subequations}
\begin{align}
   [\mathcal{Z}_{NE}(\mathbf{X};\{A,\vec{\theta}\})]^n =  &  \prod_{a=1}^n\left[ \sum_{\vec{x}^a} \prod_{i=1}^{N} \int_{-\infty}^{+\infty} dy_i^a \delta(y_i^a - \sum_{j\neq i} a_{ij} x_j^a+\theta_i K)\delta_{\rm K}\left(x_i^a,\Theta(y_i^a)\right) \delta_{\rm K}\left(X^a,\sum_i x_i^a\right)\right] \\ 
  =  & \prod_{a=1}^n \int_{0}^{2\pi \rm{i}} d\hat{X}^a \left[ \sum_{\vec{x}^a} \prod_{i=1}^{N} \int_{-\infty}^{+\infty} dy_i^a   \int_{-\mathrm{i}\infty}^{+\mathrm{i}\infty} d\hat{y}_i^a     e^{-\hat{y}_i^a \left(y_i^a -\sum_{j} a_{ij} x_j^a+ \theta_i K\right)} \delta_{\rm K}\left(x_i^a,\Theta(y_i^a)\right)  e^{-\hat{X}^a\left(X^a - \sum_i x_i^a\right)}\right]
\end{align}
\end{subequations}
where we used the integral representation of Dirac delta function $\delta(x)=\int_{-\mathrm{i}\infty}^{+\mathrm{i}\infty}  \frac{d\hat{x}}{2\pi \rm i} e^{-\hat{x} x}$ and of the Kronecker delta $\delta_{\rm K}(X)=\int_0^{2\pi \rm{i}} \frac{d\hat{X}}{2\pi \rm i} e^{-\hat{X} X}$, in which the normalisation factors $1/(2\pi \rm{i})$ are absorbed in the integration differential elements for convenience.
Isolating the terms containing the dependence on the adjacency matrix $A$ and then averaging over the distribution $P(A) = \prod_{i<j} P(a_{ij})$ with $P(a_{ij}) = \left(1-\frac{z}{N}\right)\delta_{\rm K}(a_{ij}) + \frac{z}{N}\delta_{\rm K}(a_{ij}-1)$, that corresponds to Erd\H{o}s-R\`enyi random graphs of average degree $z$, we obtain
\begin{subequations}  
\begin{align}
\nonumber & \left\langle   \prod_{a=1}^{n} \prod_{i=1}^{N}  e^{-\hat{y}_i^a \left(y_i^a -\sum_{j} a_{ij} x_j^a+ \theta_i K\right)}  e^{-\hat{X}^a\left(X^a - \sum_i x_i^a\right)} \delta_{\rm K}\left(\sum_{j\neq i} a_{ij}, K\right) \right\rangle_A\\
 & \quad =   \left\langle   \prod_{a=1}^{n} \prod_{i=1}^{N} \int_0^{2\pi \rm{i}} d\lambda_i e^{-\hat{y}_i^a \left(y_i^a -\sum_{j} a_{ij} x_j^a+ \theta_i K\right)} e^{-\lambda_i\left(\sum_{j} a_{ij}- K\right)} e^{-\hat{X}^a\left(X^a - \sum_i x_i^a\right)} \right\rangle_A  \\
 & \quad =      \prod_{a=1}^{n} \prod_{i=1}^{N}   \int_0^{2\pi \rm{i}} d\lambda_i  e^{-\mathbf{\hat{X}}\cdot \mathbf{X}}  e^{ \sum_{i}(\mathbf{\hat{X}}\cdot \mathbf{x}_i +K \lambda_i - \mathbf{\hat{y}}_i \cdot \mathbf{y}_i - \mathbf{\hat{y}}_i \cdot \mathbf{1} \theta_i K )} \prod_{i<j} \left(1-\frac{z}{N}+\frac{z}{N} e^{-\lambda_i-\lambda_j +\mathbf{\hat{y}_i}\cdot \mathbf{x_j} + \mathbf{\hat{y}_j}\cdot \mathbf{x_i} }\right)  \\
 & \quad=  \prod_{a=1}^{n} \prod_{i=1}^{N}  \int_0^{2\pi \rm{i}} d\lambda_i   e^{-\mathbf{\hat{X}}\cdot \mathbf{X}}  e^{ \sum_{i}(\mathbf{\hat{X}}\cdot \mathbf{x}_i +K \lambda_i - \mathbf{\hat{y}}_i \cdot \mathbf{y}_i - \mathbf{\hat{y}}_i \cdot \mathbf{1} \theta_i K )}   \exp\left\{ -\frac{z N}{2} + \frac{z}{2N} \sum_{ij} e^{-\lambda_i-\lambda_j +\mathbf{\hat{y}}_i \cdot \mathbf{x}_j + \mathbf{\hat{y}}_j \cdot \mathbf{x}_i }\right\}. 
\end{align}
\end{subequations}
Putting all terms together and introducing the functional order parameter $\phi(\boldsymbol{\sigma},\boldsymbol{ \tau}) = \frac{1}{N}\sum_i \delta_{\rm K}(\mathbf{x}_i,\boldsymbol{\sigma}) e^{-\lambda_i + \mathbf{\hat{y}}_i \cdot \boldsymbol{\tau}}$ for $\boldsymbol{\sigma},\boldsymbol{\tau} \in\{ 0,1\}^n$,  the average replicated partition function becomes
\begin{align}
\left\langle \left\langle \mathcal{Z}_{NE}(\mathbf{X})^n  \right\rangle \right\rangle & = \int \prod_{\boldsymbol{\sigma},\boldsymbol{\tau}} D\hat{\phi} (\boldsymbol{\sigma}, \boldsymbol{\tau} ) D \phi(\boldsymbol{\sigma}, \boldsymbol{\tau} )  e^{- N \sum_{\boldsymbol{\sigma},\boldsymbol{\tau}} \hat{\phi}(\boldsymbol{\sigma},\boldsymbol{\tau}) \phi(\boldsymbol{\sigma},\boldsymbol{\tau}) }  \int d\mathbf{\hat{X}} e^{-\mathbf{\hat{X}}\cdot \mathbf{X}}  \exp\left[\frac{z N}{2} \left(\sum_{\boldsymbol{\sigma},\boldsymbol{\tau}} \phi(\boldsymbol{\sigma},\boldsymbol{\tau}) \phi(\boldsymbol{\tau},\boldsymbol{\sigma})  -1 \right)  \right]  \\
\nonumber &\quad \times   \left\langle \sum_{\vec{\mathbf{x}}}  \left[\int  d\vec{\mathbf{y}}  d\vec{\hat{\mathbf{y}}} d\vec{\lambda} \prod_{a=1}^n \delta_{\rm K}\left(\vec{x}^a,\Theta\left[\vec{y}^a\right]\right)   \right]  e^{ \sum_i(\mathbf{\hat{X}}\cdot \mathbf{x}_i + \sum_{\boldsymbol{\sigma},\boldsymbol{\tau}}\hat{\phi}(\boldsymbol{\sigma},\boldsymbol{\tau}) \delta_{\rm K}(\mathbf{x}_i,\boldsymbol{\sigma}) e^{-\lambda_i+\mathbf{\hat{y}}_i\cdot \boldsymbol{\tau}}  +K \lambda_i - \mathbf{\hat{y}}_i \cdot \mathbf{y}_i - \mathbf{\hat{y}}_i \cdot \mathbf{1} \theta_i K )} \right\rangle_f. 
\end{align}
Taking the saddle-point with respect to $\phi(\boldsymbol{\sigma},\boldsymbol{ \tau})$ gives  $\hat{\phi}(\boldsymbol{\sigma},\boldsymbol{ \tau})=z \phi(\boldsymbol{\tau},\boldsymbol{ \sigma})$, hence again we have  
\begin{subequations}
\begin{align}
\nonumber \left\langle \left\langle \mathcal{Z}_{NE}(\mathbf{X})^n  \right\rangle \right\rangle & 
 =  \int d\mathbf{\hat{X}} e^{-\mathbf{\hat{X}}\cdot \mathbf{X}} \int \prod_{\boldsymbol{\sigma},\boldsymbol{\tau}} D \phi(\boldsymbol{\sigma}, \boldsymbol{\tau} )    \exp\left[-\frac{z N}{2} \left(\sum_{\boldsymbol{\sigma},\boldsymbol{\tau}} \phi(\boldsymbol{\sigma},\boldsymbol{\tau}) \phi(\boldsymbol{\tau},\boldsymbol{\sigma})  +1 \right)  \right]  \\
 &\quad \times   \left\langle \sum_{\vec{\mathbf{x}}}  \left[\int  d\vec{\mathbf{y}}  d\vec{\hat{\mathbf{y}}} d\vec{\lambda} \prod_{a=1}^n \delta_{\rm K}\left(\vec{x}^a,\Theta\left[\vec{y}^a\right]\right)   \right]  e^{ \sum_i(\mathbf{\hat{X}}\cdot \mathbf{x}_i +z  \sum_{\boldsymbol{\sigma},\boldsymbol{\tau}} \phi(\boldsymbol{\tau},\boldsymbol{\sigma}) \delta_{\rm K}(\mathbf{x}_i,\boldsymbol{\sigma}) e^{-\lambda_i+\mathbf{\hat{y}}_i\cdot \boldsymbol{\tau}}  +K \lambda_i - \mathbf{\hat{y}}_i \cdot \mathbf{y}_i - \mathbf{\hat{y}}_i \cdot \mathbf{1} \theta_i K )} \right\rangle_f \\
\nonumber &=  \int d\mathbf{\hat{X}} e^{-\mathbf{\hat{X}}\cdot \mathbf{X}}  \int \prod_{\boldsymbol{\sigma},\boldsymbol{\tau}} D \phi(\boldsymbol{\sigma}, \boldsymbol{\tau} )    \exp\left[-\frac{z N}{2} \left(\sum_{\boldsymbol{\sigma},\boldsymbol{\tau}} \phi(\boldsymbol{\sigma},\boldsymbol{\tau}) \phi(\boldsymbol{\tau},\boldsymbol{\sigma})  +1 \right)  \right]  \\
 &\quad \times  \left[\sum_{\mathbf{x}} \int f(\theta) d\theta \int  d\mathbf{y}
 d\hat{\mathbf{y}} d\lambda \prod_{a=1}^n\delta_{\rm K}\left(x^a,\Theta\left[y^a\right]\right)  e^{\hat{\mathbf{X}}\cdot  \mathbf{x} + z \sum_{\boldsymbol{\tau}} \phi(\tau, \mathbf{x}) e^{-\lambda +\hat{\mathbf{y}}\cdot \boldsymbol{\tau}}  +K \lambda - \hat{\mathbf{y}}\cdot \mathbf{y} - \hat{\mathbf{y}}\cdot \mathbf{1} \theta K }\right]^N \\
\nonumber & =   \int d\mathbf{\hat{X}} e^{-\mathbf{\hat{X}}\cdot \mathbf{X}}  \int \prod_{\boldsymbol{\sigma},\boldsymbol{\tau}} D \phi(\boldsymbol{\sigma}, \boldsymbol{\tau} )    \exp\left[-\frac{z N}{2} \left(\sum_{\boldsymbol{\sigma},\boldsymbol{\tau}} \phi(\boldsymbol{\sigma},\boldsymbol{\tau}) \phi(\boldsymbol{\tau},\boldsymbol{\sigma})  +1 \right)  \right]  \\
 & \quad \times \left[\frac{z^K}{K!} \sum_{\mathbf{x}} \int f(\theta) d\theta \int  d\mathbf{y} d\hat{\mathbf{y}} \prod_{a=1}^n\delta_{\rm K}\left(x^a,\Theta\left[y^a\right]\right) e^{\hat{\mathbf{X}}\cdot \mathbf{x} - \hat{\mathbf{y}}\cdot  \mathbf{y} - \hat{\mathbf{y}}\cdot \mathbf{1}\theta K }  \sum_{\boldsymbol{\tau}_1,\dots,\boldsymbol{\tau}_K} \phi(\boldsymbol{\tau}_1, \mathbf{x}) \cdots \phi(\boldsymbol{\tau}_K, \mathbf{x}) e^{\hat{\mathbf{y}}\cdot \sum_{\ell=1}^K \boldsymbol{\tau}_{\ell}} \right]^N   \\
\nonumber & =   \int d\mathbf{\hat{X}} e^{-\mathbf{\hat{X}}\cdot \mathbf{X}} \int \prod_{\boldsymbol{\sigma},\boldsymbol{\tau}} D \phi(\boldsymbol{\sigma}, \boldsymbol{\tau} )   \exp\left[-\frac{z N}{2} \left(\sum_{\boldsymbol{\sigma},\boldsymbol{\tau}} \phi(\boldsymbol{\sigma},\boldsymbol{\tau}) \phi(\boldsymbol{\tau},\boldsymbol{\sigma})  +1 \right)  \right]   \\
 &\quad\times   \left[\frac{z^K}{K!} \sum_{\mathbf{x}}  \sum_{\boldsymbol{\tau}_1,\dots,\boldsymbol{\tau}_K} \phi(\boldsymbol{\tau}_1, \mathbf{x}) \cdots \phi(\boldsymbol{\tau}_K, \mathbf{x})  e^{\hat{\mathbf{X}}\cdot \mathbf{x}} \int f(\theta) d\theta \prod_{a=1}^n \delta_{\rm K}\left(x^a,\Theta\left[ \sum_{\ell=1}^K \tau_{\ell}^a - \theta K\right]\right)   \right]^N  \\
 & =   \int d\mathbf{\hat{X}} e^{-\mathbf{\hat{X}}\cdot \mathbf{X}} \int \prod_{\boldsymbol{\sigma},\boldsymbol{\tau}} D \phi(\boldsymbol{\sigma}, \boldsymbol{\tau} )   e^{-\frac{z N}{2} \left(\sum_{\boldsymbol{\sigma},\boldsymbol{\tau}} \phi(\boldsymbol{\sigma},\boldsymbol{\tau}) \phi(\boldsymbol{\tau},\boldsymbol{\sigma})  +1 \right)  + N \Gamma[\phi,\hat{\mathbf{X}}] + N K \log{z} - N \log{K!}}   \label{Zreplicated}
\end{align}
\end{subequations}
where  we first used the identity $\frac{1}{2\pi \rm i} \int_0^{2\pi \rm i}d\lambda e^{\lambda z+\alpha e^{-\lambda}}=\frac{\alpha^z}{z!}$ and then we defined  
\begin{equation}
\Gamma[\phi,\hat{\mathbf{X}}] =  \log\left\{ \sum_{\boldsymbol{\sigma},\boldsymbol{\tau}_1,\dots,\boldsymbol{\tau}_K} \phi(\boldsymbol{\tau}_1,\boldsymbol{\sigma}) \cdots \phi(\boldsymbol{\tau}_K, \boldsymbol{\sigma})  e^{\hat{\mathbf{X}}\cdot \boldsymbol{\sigma}} \int f(\theta) d\theta \prod_{a=1}^n \delta_{\rm K}\left(\sigma^a,\Theta\left[ \sum_{\ell=1}^K \tau_{\ell}^a - \theta K\right]\right) \right\}. \label{gammaReplica}
\end{equation}
Further defining 
\begin{equation}
F_{\bs{\sigma}}\left(\bs{\tau} + \sum_{\ell=1}^{K-1}\bs{\tau}_\ell \right)=  \int f(\theta) d\theta \prod_{a=1}^n \delta_{\rm K}\left(\sigma^a,\Theta\left[\tau^a + \sum_{\ell=1}^{K-1} \tau_{\ell}^a - \theta K\right]\right)
\end{equation}
at the saddle-point,  the functional order parameter $\phi(\bs{\tau},\bs{\sigma})$ has to satisfy the self-consistent functional equation
\begin{equation}
\phi(\bs{\sigma},\bs{ \tau}) = \frac{K}{z} \frac{\sum_{\bs{\tau}_1,\dots,\bs{\tau}_{K-1}} \phi(\bs{\tau}_1,\bs{ \sigma}) \cdots \phi(\bs{\tau}_{K-1}, \bs{\sigma}) e^{\hat{\mathbf{X}} \cdot \bs{\sigma} }  F_{\bs{\sigma}}\left( \bs{\tau}+\sum_{\ell=1}^{K-1} \bs{\tau}_{\ell} \right) }{\sum_{\bs{\sigma'},\bs{\tau}_1,\dots,\bs{\tau}_{K}} \phi(\bs{\tau}_1,\bs{ \sigma'}) \cdots \phi(\bs{\tau}_{K}, \bs{\sigma'}) e^{\hat{\mathbf{X}} \cdot \bs{\sigma'} }  F_{\bs{\sigma'}}\left( \sum_{\ell=1}^{K} \bs{\tau}_{\ell} \right) }
\end{equation}
which gives the normalization condition $\sum_{\bs{\sigma},\bs{\tau}}\phi(\bs{\sigma},\bs{ \tau}) \phi(\bs{\tau},\bs{\sigma}) = K/z$.  It is convenient to rescale the order parameters $\psi(\bs{\sigma},\bs{\tau}) = \sqrt{\frac{z}{K}}\phi(\bs{\sigma},\bs{\tau})$ so that the self-consistent equation becomes independent of $z$,
\begin{equation}
\psi(\bs{\sigma},\bs{ \tau}) =  \frac{\sum_{\bs{\tau}_1,\dots,\bs{\tau}_{K-1}} \psi(\bs{\tau}_1,\bs{ \sigma}) \cdots \psi(\bs{\tau}_{K-1}, \bs{\sigma}) e^{\hat{\mathbf{X}} \cdot \bs{\sigma} }  F_{\bs{\sigma}}\left( \bs{\tau}+\sum_{\ell=1}^{K-1} \bs{\tau}_{\ell} \right) }{\sum_{\bs{\sigma'},\bs{\tau}_1,\dots,\bs{\tau}_{K}} \psi(\bs{\tau}_1,\bs{ \sigma'}) \cdots \psi(\bs{\tau}_{K}, \bs{\sigma'}) e^{\hat{\mathbf{X}} \cdot \bs{\sigma'} }  F_{\bs{\sigma'}}\left( \sum_{\ell=1}^{K} \bs{\tau}_{\ell} \right) }.\label{self-consistentPSI}
\end{equation}
In terms of the new variables the  the average replicated partition function in the constrained Erd\H{o}s-R\`enyi ensemble becomes
\begin{align}
\left\langle \left\langle \mathcal{Z}_{NE}(\mathbf{X})^n  \right\rangle \right\rangle & =   \int d\mathbf{\hat{X}} e^{-\mathbf{\hat{X}}\cdot \mathbf{X}} \int \prod_{\boldsymbol{\sigma},\boldsymbol{\tau}} D \psi(\boldsymbol{\sigma}, \boldsymbol{\tau} )   e^{-\frac{K N}{2} \left(\sum_{\boldsymbol{\sigma},\boldsymbol{\tau}} \psi(\boldsymbol{\sigma},\boldsymbol{\tau}) \psi(\boldsymbol{\tau},\boldsymbol{\sigma})  +\frac{z}{K} \right)  + N \Gamma[\psi,\hat{\mathbf{X}}]  + \frac{N K}{2} \log{K} +  \frac{N K}{2} \log{z}  - N \log{K!}}   \label{Zreplicated2}
\end{align}
The random regular graphs are a subset of the random graph ensemble considered, with entropy $s_{RRG}(K,z)=\frac{K}{2}(\log{K}+\log{z}-1)-\log{K!}-\frac{z}{2}$ \cite{dean2000metastable}. Subtracting this term to the expression \eqref{Zreplicated2} we find the replicated partition function of the Nash equilibria on random regular graphs,
\begin{align}
\left\langle \left\langle \mathcal{Z}^{RRG}_{NE}(\mathbf{X})^n  \right\rangle \right\rangle & =   \int d\mathbf{\hat{X}} e^{-\mathbf{\hat{X}}\cdot \mathbf{X}} \int \prod_{\boldsymbol{\sigma},\boldsymbol{\tau}} D \psi(\boldsymbol{\sigma}, \boldsymbol{\tau} )   e^{-\frac{K N}{2} \left(\sum_{\boldsymbol{\sigma},\boldsymbol{\tau}} \psi(\boldsymbol{\sigma},\boldsymbol{\tau}) \psi(\boldsymbol{\tau},\boldsymbol{\sigma}) - 1 \right)   + N \Gamma[\psi,\hat{\mathbf{X}}]   }   \label{ZreplicatedRRG}
\end{align}
We fix the density of agents playing action $1$ in all replicas of the system to assume the same value $\rho= X/N$, i.e. $X^a=X, \forall  a=1,\dots, n$. From \eqref{Zreplicated} it is given by the saddle-point with respect to $X^a, \forall a$,
\begin{subequations}
\begin{align}
 \rho & = \frac{\partial \Gamma[\psi,\hat{\mathbf{X}}] }{\partial \hat{X}^a}\\
 & = \frac{ \sum_{\boldsymbol{\sigma},\boldsymbol{\tau}_1,\dots,\boldsymbol{\tau}_K} \psi(\boldsymbol{\tau}_1,\boldsymbol{\sigma}) \cdots \psi(\boldsymbol{\tau}_K, \boldsymbol{\sigma})  e^{\hat{\mathbf{X}}\cdot \boldsymbol{\sigma}} \sigma^a  F_{\bs{\sigma}}\left(\sum_{\ell=1}^{K} \bs{\tau}_{\ell} \right)}{\sum_{\boldsymbol{\sigma'},\boldsymbol{\tau}_1',\dots,\boldsymbol{\tau}_K'} \psi(\boldsymbol{\tau}_1',\boldsymbol{\sigma'}) \cdots \psi(\boldsymbol{\tau}_K', \boldsymbol{\sigma'})  e^{\hat{\mathbf{X}}\cdot \boldsymbol{\sigma'}}F_{\bs{\sigma}'}\left(\sum_{\ell=1}^{K} \bs{\tau}_{\ell}' \right)}\\
& =  \frac{ \sum_{\boldsymbol{\sigma},\boldsymbol{\tau}} \sigma^a \psi(\boldsymbol{\sigma},\boldsymbol{\tau}) \psi(\boldsymbol{\tau},\boldsymbol{\sigma})}{ \sum_{\boldsymbol{\sigma}',\boldsymbol{\tau}'} \psi(\boldsymbol{\sigma}',\boldsymbol{\tau}') \psi(\boldsymbol{\tau}',\boldsymbol{\sigma}')}.
\end{align}
\end{subequations}
In the next subsections both annealed and quenched expressions for the entropy of Nash equilibria are obtained.

\subsection{Annealed Calculation}\label{app:annealed}
The annealed calculation of the entropy of Nash equilibria, consisting in averaging the partition function over the disorder instead of its logarithm, provides an upper bound to the exact quenched entropy. 
Setting $n=1$, the average partition function is given by 
\begin{align}
\left\langle \left\langle \mathcal{Z}^{RRG}_{NE}(X)  \right\rangle \right\rangle & =   \int d\hat{X} e^{-N \hat{X} X} \int \prod_{\sigma,\tau} D \psi(\sigma, \tau )   e^{-\frac{K N}{2} \left(\sum_{\sigma,\tau} \psi(\sigma,\tau) \psi(\tau,\sigma) - 1 \right)   + N \log\left\{ \sum_{\sigma,\tau_1,\dots,\tau_K} \psi(\tau_1, \sigma) \cdots \psi(\tau_K, \sigma) e^{\hat{X} \sigma} F_ \sigma \left[ \frac{\sum_{\ell=1}^K \tau_{\ell}}{K}\right]  \right\}  }.   \label{ZreplicatedRRG}
\end{align}
Introducing the density $\rho= X/N$ and calling $\epsilon =\hat{X}$, the annealed entropy of Nash equilibria is
\begin{subequations}
\begin{align}
 s^{\rm ann}_{NE}(\rho) & =\max_{\epsilon,\psi}\left\{ -\epsilon \rho  -\frac{K}{2} \left(\sum_{\sigma,\tau} \psi(\sigma,\tau) \psi(\tau,\sigma) - 1 \right) + \log\left( \sum_{\sigma,\tau_1,\dots,\tau_K} e^{\epsilon \sigma} F_ \sigma \left( \frac{\sum_\ell \tau_\ell}{K}\right)  \prod_{\ell=1}^{K}\psi(\tau_\ell, \sigma) \right) \right\}\\
 & = \max_{\epsilon}\left\{ -\epsilon \rho  + \log\left( \sum_{\sigma,\tau_1,\dots,\tau_K} e^{\epsilon \sigma} F_ \sigma \left( \frac{\sum_\ell \tau_\ell}{K}\right)  \prod_{\ell=1}^{K}\psi(\tau_\ell, \sigma) \right) \right\} \\
& =  \max_{\epsilon}\left\{ -\epsilon \rho - f_{NE}^{\rm ann}(\epsilon)\right\} 
\end{align}
\end{subequations}
where the functional order parameter satisfies the saddle-point equations 
\begin{equation}\label{SaddlePointAnn}
\psi(\sigma, \tau) = \frac{\sum_{\tau_1,\dots,\tau_{K-1}} \psi(\tau_1, \sigma) \cdots \psi(\tau_{K-1}, \sigma) e^{\epsilon \sigma}   F_ \sigma\left[ \frac{\tau+\sum_{\ell=1}^{K-1} \tau_{\ell}}{K}\right] }{\sum_{\sigma',\tau_1,\dots,\tau_K} \psi(\tau_1, \sigma') \cdots \psi(\tau_K, \sigma') e^{\epsilon \sigma'}  F_ {\sigma'}\left[ \frac{\sum_{\ell=1}^K \tau_{\ell}}{K}\right] }
\end{equation}
with $F_ {1}\left[\theta\right] = \int_0^{\theta}  f(\theta') d\theta' = 1-F_0\left[\theta \right]$, while the average ``annealed" density $\rho=\rho_{\rm ann}$ of agents playing action $1$ is fixed  by the Legendre transform
\begin{align}\label{eq:rho_ann}
\rho_{\rm ann}  = -\frac{\partial f_{NE}^{\rm ann}(\epsilon)}{\partial \epsilon} =  \frac{ \sum_{{\sigma},{\tau}} \sigma \psi({\sigma},{\tau}) \psi({\tau},{\sigma})}{ \sum_{{\sigma},{\tau}} \psi({\sigma},{\tau}) \psi({\tau},{\sigma})} = \frac{\psi(1,1)^2+\psi(1,0)\psi(0,1)}{\psi(0,0)^2 + \psi(1,1)^2+2 \psi(1,0)\psi(0,1)}.
\end{align}
 
The generalization of the derivation to uncorrelated random graphs with distribution $p_k$ is straightforward. In the absence of on-site disorder (i.e. uniform thresholds), Eqs.~\eqref{SaddlePointAnn} coincide with the BP equations \eqref{bp-instance} for ensembles of random regular graphs (see also \eqref{eq-BPregular}) and the annealed entropy is  correct  \cite{mori2011connection}. 
On the contrary, for general distributions $f(\theta)$, Eqs.~\eqref{SaddlePointAnn} are just an annealed approximation of the (distributional) BP equations, providing an upper bound to the actual number of Nash equilibria.

The annealed approximation is a good starting point to derive standard mean-field equations. Neglecting local correlations implies that the order parameter does not depend on the second variable, i.e. $\psi(\sigma,\tau)\approx \psi_{\rm mf}(\sigma)$ with the normalization condition $\sum_\sigma \psi_{\rm mf}(\sigma)= 1$. Using this approximation in \eqref{eq:rho_ann} together with \eqref{SaddlePointAnn}, we obtain 
\begin{subequations}
\begin{align}
 \frac{\sum_{\tau,\sigma} \sigma\psi(\sigma,\tau)\psi(\tau,\sigma)}{\sum_{\tau,\sigma}\psi(\sigma,\tau)\psi(\tau,\sigma)}
& = \frac{1}{\sum_{\tau,\sigma}\psi(\sigma,\tau)\psi(\tau,\sigma)}\frac{\sum_{\sigma,\tau_1,\dots,\tau_K} \psi(\tau_1,\sigma)\cdots \psi(\tau_K,\sigma) \sigma F_{\sigma}\left[ \frac{\sum_{\ell=1}^K \tau_{\ell}}{K}\right]}{\sum_{\sigma,\tau_1,\dots,\tau_K} \psi(\tau_1,\sigma)\cdots \psi(\tau_K,\sigma) F_{\sigma}\left[ \frac{\sum_{\ell=1}^K \tau_{\ell}}{K}\right]}\\
  & \approx \frac{1}{\sum_{\sigma}\psi_{\rm mf}(\sigma)\sum_{\tau} \psi_{\rm mf}(\tau)}\frac{\sum_{\tau_1,\dots,\tau_K} \psi_{\rm mf}(\tau_1)\cdots \psi_{\rm mf}(\tau_K) F_{1}\left[ \frac{\sum_{\ell=1}^K \tau_{\ell}}{K}\right]}{\sum_{\sigma} \sum_{\tau_1,\dots,\tau_K} \psi_{\rm mf}(\tau_1)\cdots \psi_{\rm mf}(\tau_K) F_{\sigma}\left[ \frac{\sum_{\ell=1}^K \tau_{\ell}}{K}\right]}\\
  & = \frac{\sum_{\tau_1,\dots,\tau_K} \psi_{\rm mf}(\tau_1)\cdots \psi_{\rm mf}(\tau_K) F_{1}\left[ \frac{\sum_{\ell=1}^K \tau_{\ell}}{K}\right]}{\sum_{\tau_1,\dots,\tau_K} \psi_{\rm mf}(\tau_1)\cdots \psi_{\rm mf}(\tau_K) \sum_\sigma F_{\sigma}\left[ \frac{\sum_{\ell=1}^K \tau_{\ell}}{K}\right]}\\
& =  \frac{\sum_{\tau_1,\dots,\tau_K} \psi_{\rm mf}(\tau_1)\cdots \psi_{\rm mf}(\tau_K) F_{1}\left[ \frac{\sum_{\ell=1}^K \tau_{\ell}}{K}\right]}{\sum_{\tau_1,\dots,\tau_K} \psi_{\rm mf}(\tau_1)\cdots \psi_{\rm mf}(\tau_K) } \\
& = \sum_{\tau_1,\dots,\tau_K} \psi_{\rm mf}(\tau_1)\cdots \psi_{\rm mf}(\tau_K) F_{1}\left[ \frac{\sum_{\ell=1}^K \tau_{\ell}}{K}\right],
\end{align}
\end{subequations}
which corresponds to the mean-field equation
\begin{align}
 \psi_{\rm mf}(\sigma) = \sum_{\tau_1,\dots,\tau_K} \psi_{\rm mf}(\tau_1)\cdots \psi_{\rm mf}(\tau_K) F_{\sigma}\left[ \frac{\sum_{\ell=1}^K \tau_{\ell}}{K}\right].
\end{align}
Defining the mean-field density as $\rho_{\rm mf} = \psi_{\rm mf}(1)$ and generalizing the expression for uncorrelated random graphs with degree distribution $p_k$, one gets the well-known self-consistent mean-field equations  \cite{galeotti2010network,anand2013network,anand2013epidemics,cimini2015dynamics,porter2016dynamical}, 
\begin{align}\label{MFpk_rho}
\rho_{\rm mf} =  \Phi_{\rm mf}[\rho_{\rm mf}] =  \sum_k p_k \sum_{\ell}\binom{k}{\ell}\rho_{\rm mf}^{\ell}\left(1- \rho_{\rm mf}\right)^{k-\ell}F_{1}\left(\frac{\ell}{k}\right) .
\end{align}

\subsection{Quenched Calculation}\label{app:quenched}
In order to compute the quenched free energy, it is now necessary to take the limit $n\to 0$, i.e. to evaluate 
\begin{equation}
\langle \langle \log \mathcal{Z}_{NE}^{RRG}(\rho) \rangle \rangle = \lim_{n\to 0} \frac{\left\langle \left\langle \mathcal{Z}^{RRG}_{NE}(\mathbf{X})^n  \right\rangle \right\rangle - 1}{n}.
\end{equation}

Under the replica-symmetric ansatz, the functional order parameter $\psi(\bs{\sigma},\bs{\tau})$ should be a function only of the permutation-invariant quantities $\sum_a \sigma^a$, $\sum_a \tau^a$ and $\sum_a \sigma^a \tau^a$. For later convenience we use a linear combination of these quantities, defining $r = \sum_a \sigma^a \tau^a$, $s=\sum_a \sigma^a-  \sum_a  \sigma^a \tau^a $, $t=\sum_a \tau^a-  \sum_a \sigma^a \tau^a $. One is then tempted to introduce the  Fourier transform  
\begin{align}
\psi(\bs{\sigma},\bs{\tau}) \equiv \psi(r,s,t) = \int du dv dw  \hat{\psi}(u,v,w) e^{r u+ s v + t w}
\end{align}
whose inverse transform is defined as 
\begin{align}
\hat{\psi}(u,v,w) = \int \frac{dr}{2\pi} \frac{ds}{2\pi} \frac{dt}{2\pi}  \psi({\rm i}r,{\rm i} s,{\rm i} t) e^{-{\rm i} r u- {\rm i} s v - {\rm i}t w}.
\end{align}
Introducing this expression in the self-consistent functional equation \eqref{self-consistentPSI} and setting $\hat{\mathbf{X}}=\epsilon \mathbf{1}$, it becomes  
\begin{subequations}
\begin{align}
\nonumber \psi(r,s,t)  & \propto \int f(\theta) d\theta  e^{\epsilon \sum_a \sigma^a} \int \prod_{\ell=1}^{K-1} \left[ du_\ell dv_\ell dw_\ell \hat{\psi}(u_\ell,v_\ell, w_\ell)\right] \\
 & \quad \times \sum_{\bs{\tau}_1,\dots, \bs{\tau}_{K-1} }  e^{ \sum_{\ell=1}^{K-1}\left\{ u_\ell \sum_a \sigma^a  \tau_\ell^a +   v_\ell \left( \sum_a \sigma^a - \sum_a \sigma^a \tau_\ell^a\right)+ w_\ell \left( \sum_a \tau^a_\ell - \sum_a \sigma^a \tau_\ell^a \right)   \right\} }  \prod_a \delta\left(\sigma^a,\Theta\left[\tau^a + \sum_{\ell=1}^{K-1} \tau_{\ell}^a - \theta K\right]\right)\\
& \nonumber =  \int  \prod_{\ell=1}^{K-1}\left[ du_\ell dv_\ell dw_\ell \hat{\psi}(u_\ell,v_\ell, w_\ell)\right]  \int f(\theta) d\theta \\
 & \quad \times \prod_{a=1}^n \left\{ e^{\epsilon  \sigma^a}  \sum_{\bs{\tau}_1,\dots, \bs{\tau}_{K-1} }  e^{ \sum_{\ell=1}^{K-1} \left\{ u_\ell  \sigma^a  \tau_\ell^a +   v_\ell \left(  \sigma^a -  \sigma^a \tau_\ell^a\right)+ w_\ell \left( \tau^a_\ell -  \sigma^a \tau_\ell^a \right)   \right\}}  \delta\left(\sigma^a,\Theta\left[\tau^a + \sum_{\ell=1}^{K-1} \tau_{\ell}^a - \theta K\right]\right)\right\}\\
 & = \int  \prod_{\ell=1}^{K-1}\left[ du_\ell dv_\ell dw_\ell \hat{\psi}(u_\ell,v_\ell, w_\ell)\right] \mathbb{E}_\theta \left[ \exp \left\{ \sum_{a=1}^n \log g\left(\sigma^a,\tau^a  \right) \right\}\right]\\
 & = \int  \prod_{\ell=1}^{K-1}\left[ du_\ell dv_\ell dw_\ell \hat{\psi}(u_\ell,v_\ell, w_\ell)\right] \mathbb{E}_\theta \left[ \exp \left\{\sum_{\sigma, \tau}\left( \sum_{a=1}^n \delta_{\rm K}(\sigma,\sigma^a)\delta_{\rm K}(\tau,\tau^a)\right) \log g\left(\sigma,\tau \right) \right\}\right]
\end{align}
\end{subequations}
where  
\begin{subequations} \label{eq-BPg}
\begin{align}
g\left(\sigma,\tau \right) & \equiv  g\left(\sigma,\tau | \{ u_\ell, v_\ell, w_\ell \}_{\ell =1}^{K-1},\theta, \epsilon \right)\\
 &  = e^{\epsilon \sigma} \sum_{\tau_1, \dots, \tau_{K-1}} \prod_{\ell =1}^{K-1} e^{  u_\ell  \sigma  \tau_\ell +   v_\ell \left(  \sigma -  \sigma \tau_\ell\right)+ w_\ell \left( \tau_\ell -  \sigma \tau_\ell \right)    } \delta\left(\sigma,\Theta\left[\tau + \sum_{\ell=1}^{K-1} \tau_{\ell} - \theta K\right]\right) . 
\end{align}
\end{subequations}
Using the identity
\begin{align}
\sum_{a=1}^n \delta_{\rm K}(\sigma,\sigma^a)\delta_{\rm K}(\tau,\tau^a) = 
\begin{cases}
 n - \sum_a \sigma^a - \sum_a \tau^a + \sum_a \sigma^a \tau^a  & \qquad   \sigma=\tau=0 \\
  \sum_a \tau^a - \sum_a \sigma^a \tau^a  & \qquad   \sigma=0, \tau=1 \\
 \sum_a \sigma^a -  \sum_a \sigma^a \tau^a   & \qquad   \sigma=1, \tau=0 \\
  \sum_a \sigma^a \tau^a  & \qquad   \sigma=\tau=1 
\end{cases}
\end{align}
one finally obtains, after taking the $n\to 0$ limit 
\begin{subequations}
\begin{align}
  \psi(r,s,t) & \propto  \int  \prod_{\ell=1}^{K-1}\left[ du_\ell dv_\ell dw_\ell \hat{\psi}(u_\ell,v_\ell, w_\ell)\right] \mathbb{E}_\theta \left[  e^{ \left( - s -  t - r \right) \log g\left(0,0 \right) +   t \log  g\left(0,1  \right) + s  \log g\left(1,0  \right) + r \log g\left(1,1  \right) } \right] \\
  & \propto \int  \prod_{\ell=1}^{K-1}\left[ du_\ell dv_\ell dw_\ell \hat{\psi}(u_\ell,v_\ell, w_\ell)\right] \mathbb{E}_\theta \left[   \left(\frac{g\left(0,1 \right)}{g\left(0,0 \right) }\right)^t  \left(\frac{g\left(1,0  \right)}{g\left(0,0  \right) }\right)^s  \left(\frac{g\left(1,1  \right)}{g\left(0,0\right) }\right)^r \right]. 
\end{align}
\end{subequations}
Inverting the Fourier transform, 
\begin{subequations} \label{psiRS}
\begin{align}
  \hat{\psi}(u,v,w) & 
\propto    \int  \prod_{\ell=1}^{K-1}\left[ du_\ell dv_\ell dw_\ell \hat{\psi}(u_\ell,v_\ell, w_\ell)\right]   \int d\theta  f(\theta)  \int  \frac{dr}{2\pi} e^{-{\rm i} r \left(u -  \log{\frac{g\left(1,1  \right)}{g\left(0,0  \right)}} \right)} \int  \frac{ds}{2\pi} e^{-{\rm i} s \left(v -  \log{\frac{g\left(1,0  \right)}{g\left(0,0 \right)}} \right)}  \int  \frac{dt}{2\pi} e^{-{\rm i} t \left(w -  \log{\frac{g\left(0,1  \right)}{g\left(0,0  \right)}} \right)}     \\
  & \propto \int  \prod_{\ell=1}^{K-1}\left[ du_\ell dv_\ell dw_\ell \hat{\psi}(u_\ell,v_\ell, w_\ell)\right] \mathbb{E}_\theta \left[  \delta\left(u -  \log{\frac{g\left(1,1  \right)}{g\left(0,0  \right)}}  \right)\delta\left(v -  \log{\frac{g\left(1,0  \right)}{g\left(0,0  \right)}}  \right) \delta\left(w -  \log{\frac{g\left(0,1  \right)}{g\left(0,0  \right)}}  \right) \right]. 
\end{align}
\end{subequations}
In order to obtain the free energy in the RS ansatz, we rewrite \eqref{gammaReplica} in the $n\to 0$ limit as 
\begin{subequations}
\begin{align}
 e^{N\Gamma[\psi,\epsilon]} & = \sum_{\boldsymbol{\sigma},\boldsymbol{\tau}_1,\dots,\boldsymbol{\tau}_K} \psi(\boldsymbol{\tau}_1,\boldsymbol{\sigma}) \cdots \psi(\boldsymbol{\tau}_K, \boldsymbol{\sigma})  e^{\hat{\mathbf{X}}\cdot \boldsymbol{\sigma}} \int f(\theta) d\theta \prod_{a=1}^n \delta_{\rm K}\left(\sigma^a,\Theta\left[ \sum_{\ell=1}^K \tau_{\ell}^a - \theta K\right]\right) \\
 & =  \int f(\theta) d\theta \int \prod_{\ell=1}^{K-1} du_\ell dv_\ell dw_\ell \hat{\psi}(u_\ell,v_\ell, w_\ell)  \left[ \sum_{\sigma, \tau_1,\dots, \tau_{K} }  e^{\epsilon  \sigma +  \sum_{\ell=1}^{K}\left( u_\ell  \sigma  \tau_\ell +   v_\ell \left(  \sigma - \sigma \tau_\ell\right)+ w_\ell \left(  \tau_\ell -  \sigma \tau_\ell \right)   \right) }   \delta\left(\sigma,\Theta\left[  \sum_{\ell=1}^{K} \tau_{\ell} - \theta K\right]\right) \right]^n \\
& \overset{n\to 0}{\approx}  \nonumber  1 +  n \int f(\theta) d\theta  \int \prod_{\ell=1}^{K-1} \left[ du_\ell dv_\ell dw_\ell \hat{\psi}(u_\ell,v_\ell, w_\ell)\right] \\
& \qquad \times \log\left\{ \sum_{\sigma, \tau_1,\dots, \tau_{K} }  e^{\epsilon  \sigma +  \sum_{\ell=1}^{K}\left( u_\ell  \sigma  \tau_\ell +   v_\ell \left(  \sigma - \sigma \tau_\ell\right)+ w_\ell \left(  \tau_\ell -  \sigma \tau_\ell \right)   \right) }   \delta\left(\sigma,\Theta\left[  \sum_{\ell=1}^{K} \tau_{\ell} - \theta K\right]\right) \right\} 
\end{align}
\end{subequations}
while the other term of the replicated partition function  \eqref{ZreplicatedRRG} gives  
\begin{subequations}
\begin{align}
\nonumber & e^{-\frac{K N}{2} \left(\sum_{\boldsymbol{\sigma},\boldsymbol{\tau}} \psi(\boldsymbol{\sigma},\boldsymbol{\tau}) \psi(\boldsymbol{\tau},\boldsymbol{\sigma}) - 1 \right) } \\
 & \quad = \exp\left\{\frac{KN}{2} \left[1 - \int \prod_{\ell=1}^{2} \left[ du_\ell dv_\ell dw_\ell \hat{\psi}(u_\ell,v_\ell, w_\ell)\right]  \sum_{\bs{\sigma},\bs{\tau}} e^{(v_1+w_2)\left(\sum_a \sigma^a - \sum_a \sigma^a\tau^a\right) + (v_2+w_1)\left(\sum_a \tau^a - \sum_a \sigma^a\tau^a\right) + (u_1+u_2)\sum_a \sigma^a\tau_a} \right]\right\}\\
 &  \quad =   \exp\left\{\frac{KN}{2} \left[1 - \int \prod_{\ell=1}^{2} \left[ du_\ell dv_\ell dw_\ell \hat{\psi}(u_\ell,v_\ell, w_\ell)\right]  \left( \sum_{\sigma, \tau} e^{(v_1+w_2)\left( \sigma -  \sigma\tau\right) + (v_2+w_1)\left( \tau - \sigma\tau\right) + (u_1+u_2) \sigma\tau } \right)^n \right] \right\}\\
 & \quad  \overset{n\to 0}{\to}  \exp\left\{\frac{KN}{2} \left[1 - \int \prod_{\ell=1}^{2} \left[ du_\ell dv_\ell dw_\ell \hat{\psi}(u_\ell,v_\ell, w_\ell)\right]  \left\{ 1 + n \log\left( \sum_{\sigma, \tau} e^{(v_1+w_2)\left( \sigma -  \sigma\tau\right) + (v_2+w_1)\left( \tau - \sigma\tau\right) + (u_1+u_2) \sigma\tau } \right) \right] \right\} \right\}\\
 & \quad  = \exp\left\{-\frac{KNn}{2}  \int \prod_{\ell=1}^{2} \left[ du_\ell dv_\ell dw_\ell \hat{\psi}(u_\ell,v_\ell, w_\ell)\right]   \log\left( \sum_{\sigma, \tau} e^{(v_1+w_2)\left( \sigma -  \sigma\tau\right) + (v_2+w_1)\left( \tau - \sigma\tau\right) + (u_1+u_2) \sigma\tau } \right)  \right\}\\
& \quad  \approx  1 - \frac{KNn}{2}  \int \prod_{\ell=1}^{2} \left[ du_\ell dv_\ell dw_\ell \hat{\psi}(u_\ell,v_\ell, w_\ell)\right]   \log\left( \sum_{\sigma, \tau} e^{(v_1+w_2)\left( \sigma -  \sigma\tau\right) + (v_2+w_1)\left( \tau - \sigma\tau\right) + (u_1+u_2) \sigma\tau } \right). 
\end{align}
\end{subequations}
Putting all together we finally find the replica symmetric (free) entropy of Nash equilibria as 
\begin{subequations}
\begin{align}
 s_{NE}(\rho) & =  \frac{1}{N}\lim_{n\to 0} \frac{\left\langle \left\langle \mathcal{Z}^{RRG}_{NE}(\mathbf{X})^n  \right\rangle \right\rangle - 1}{n} \\ 
 & =  - \rho\epsilon - \frac{K}{2} \int \prod_{\ell=1}^{2} \left[ du_\ell dv_\ell dw_\ell \hat{\psi}(u_\ell,v_\ell, w_\ell)\right]   \log\left( \sum_{\sigma, \tau} e^{(v_1+w_2)\left( \sigma -  \sigma\tau\right) + (v_2+w_1)\left( \tau - \sigma\tau\right) + (u_1+u_2) \sigma\tau } \right)  \\
\nonumber &  \quad +  \int \prod_{\ell=1}^{K} \left[ du_\ell dv_\ell dw_\ell \hat{\psi}(u_\ell,v_\ell, w_\ell)\right]  \mathbb{E}_{\theta} \left[  \log\left\{
 \sum_{\sigma, \tau_1,\dots, \tau_{K} }  e^{\epsilon  \sigma +  \sum_{\ell=1}^{K}\left( u_\ell  \sigma  \tau_\ell +   v_\ell \left(  \sigma - \sigma \tau_\ell\right)+ w_\ell \left(  \tau_\ell -  \sigma \tau_\ell \right)   \right) }   \delta\left(\sigma,\Theta\left[  \sum_{\ell=1}^{K} \tau_{\ell} - \theta K\right]\right) \right\}\right]  \label{fRRGrs}\\
  & =  - \rho \epsilon - f_{NE}(\epsilon)
\end{align}
\end{subequations}
with $\rho(\epsilon)=-\frac{\partial f_{NE}(\epsilon)}{\partial \epsilon}$ and the functions $\hat{\psi}$ given by \eqref{psiRS}.
Equations \eqref{psiRS} are equivalent to the distributional BP equations \eqref{bp-instancePregular} for random regular graph. The variable $u$, $v$, and $w$ are cavity fields, which can be obtained from the four components of $\vec{\eta}$ with a simple change of variable. In fact, from BP equations \eqref{bp-instance} and \eqref{eq-BPg} it is trivial to check that the quantities $\tilde{\eta}(x,y)=\eta(x,y)/\eta(0,0)$ and $\tilde{g}\left(\sigma,\tau \right) = g\left(\sigma,\tau  \right)/g\left(0,0 \right)$ satisfy the same set of equations on random regular graphs.

\section{Bayes-Nash equilibria}\label{app-BayesNash}
In the game $\left(\mathcal{G},\{0,1\}^{|\mathcal{V}|},\{u_i\}_{i\in \mathcal{V}},\{\theta_i\}_{i\in \mathcal{V}}\right)$, we assumed that all players perfectly know the structure of the game, that is the interaction graph, all individual thresholds and utility functions. Note that, even if the dynamic processes described in Sec.~\ref{sec:dynamics} could be performed by the agents without necessarily assuming that they know all details about the structure of the system, the assumption of complete information is necessary to define the very concept of Nash equilibrium. In most real cases, however, the agents do not know the whole structure of the graph, requiring a different formulation in terms of {\em games of incomplete information}. According to Galeotti et al. \cite{galeotti2010network}, the uncertainty about the structure of the game can be expressed in terms of two only payoff-relevant parameters, called {\em types}: individual thresholds $\{\theta_i\}_{i\in \mathcal{V}}$ and degrees $\{k_i\}_{i\in \mathcal{V}}$. Each agent $i$  knows exactly her own threshold $\theta_i$ and the number of neighbours $k_i$, whereas has only some belief on the distribution of types $(\theta, k)$ across the network. We assume that agents'  types are i.i.d. random variables drawn from the degree distribution $p_k$ and the threshold distribution $f(\theta)$.  Let us define a {\em Bayesian game} in which agents have some probabilistic priors $f(\theta)$ and $p_k$ about the game structure and play according to their types, that is the pure strategy $x_i \in\{0,1\}$ of agent $i$ is function of her types $(k_i,\theta_i)$. Since the degree and thresholds of the neighbours are unknown to agent $i$ and drawn from a common distribution, agent $i$ will use such information to form a belief on the probability with which neighbours play an action. Let us call $\hat{\rho}_{\rm bn}$ the probability that a neighbour of a randomly chosen node plays action $1$, it satisfies the self-consistent equation 
$\hat{\rho}_{\rm bn} = \hat{\Phi}_{\rm bn}[\hat{\rho}_{\rm bn}]$, with
\begin{align}\label{BNrho1}
\hat{\Phi}_{\rm bn}[\rho] = \sum_k \frac{k p_k}{\langle k\rangle} \sum_{\ell}\binom{k}{\ell}\rho^{\ell}\left(1- \rho\right)^{k-\ell}F_{1}(\ell/k).
\end{align}
The quantity $\hat{\rho}_{\rm bn}$ identifies a symmetric {\em Bayes-Nash equilibrium} \cite{galeotti2010network,jackson2015games}, in which  agent $i$ with types $(k_i,\theta_i)$  plays action $x_i$ if and only if 
\begin{subequations}
\begin{align}
 x_i & = \arg\max_{x} \mathbb{E}\left[u_i(x,\vec{x}_{\partial i})\right] \\
 & = \arg\max_{x} \left[x \sum_{m=0}^{k_i} B_{k_i,m}(\hat{\rho}_{\rm bn}) m + (1-x) \theta_i k_i\right] = \begin{cases} 1 & \text{if } \hat{\rho}_{\rm bn}\geq \theta_i \\  0 & \text{if } \hat{\rho}_{\rm bn}<\theta_i \end{cases}.
\end{align}
\end{subequations}
Finally, using \eqref{MFpk_rho}, the probability  $\rho_{\rm bn}$ that a randomly chosen agent plays action $1$ is given by 
\begin{align}\label{BNrho}
\rho_{\rm bn} = \Phi_{\rm mf}[\hat{\rho}_{\rm bn}].
\end{align}
In the present context, the definition of Bayes-Nash equilibrium resembles the mean-field approximation discussed in App.~\ref{app:annealed} and provides to the latter a clear game-theoretical justification. 
In fact, the properties of Bayes-Nash equilibria in the present model (and in similar coordination models as well) are not representative of the typical properties of the multitude of pure Nash equilibria of the underlying coordination game of complete information (see Fig. \ref{fig:BRcomparison} for a comparison between numerical results in the case of random regular graphs). There are other possible ways of defining a game of incomplete information starting from the tuple $\left(\mathcal{G},\{0,1\}^{|\mathcal{V}|},\{u_i\}_{i\in \mathcal{V}},\{\theta_i\}_{i\in \mathcal{V}}\right)$, for instance assuming that the graph is common knowledge but the thresholds are private information. Although of certain interest, this subject goes beyond the purpose of the present work.

\section{Analysis of the BP equations for  random regular graphs and uniform thresholds}\label{app-BPregular}
The  BP equations \eqref{bp-instance} derived by means of the cavity approach developed in Sec.\ref{subsec:cavity} can be considerably simplified in the case of  random regular graphs of degree $K$ and uniform thresholds of value $\Theta=K\theta = K \mu$. At the ensemble level, in the replica symmetric hypothesis, the cavity marginal (or message) $\eta_{ij}(x_i,x_j)$ can be written as a single four-dimensional vector $\vec{\eta}= \left(\eta(0,0),\eta(0,1),\eta(1,0),\eta(1,1)\right)$, whose components satisfy the BP equations 
\begin{subequations}
\begin{align}
\eta(0,0)& =\frac{1}{Z_{\rm c}} \sum_{\ell<\Theta}\binom{K-1}{\ell}\eta(1,0)^{\ell}\eta(0,0)^{K-1-\ell} \\
\eta(0,1)& =\frac{1}{Z_{\rm c}} \sum_{\ell<\Theta-1}\binom{K-1}{\ell}\eta(1,0)^{\ell}\eta(0,0)^{K-1-\ell} \\
\eta(1,0)& = \frac{e^{\epsilon}}{Z_{\rm c}} \sum_{\ell\geq \Theta}\binom{K-1}{\ell}\eta(1,1)^{\ell}\eta(0,1)^{K-1-\ell} \\
\eta(1,1)& =  \frac{e^{\epsilon}}{Z_{\rm c}} \sum_{\ell\geq \Theta-1}\binom{K-1}{\ell}\eta(1,1)^{\ell}\eta(0,1)^{K-1-\ell}. 
\end{align} \label{eq-BPregular}
\end{subequations} 
Eqs.~\ref{eq-BPregular} can be solved by iteration and the fixed points as function of the parameter $\epsilon$ can be used to compute the properties of Nash equilibria with a density $\rho(\epsilon)$ of agents playing action $1$.

\begin{figure}[tb]
\begin{center}
\includegraphics[width=0.6\columnwidth]{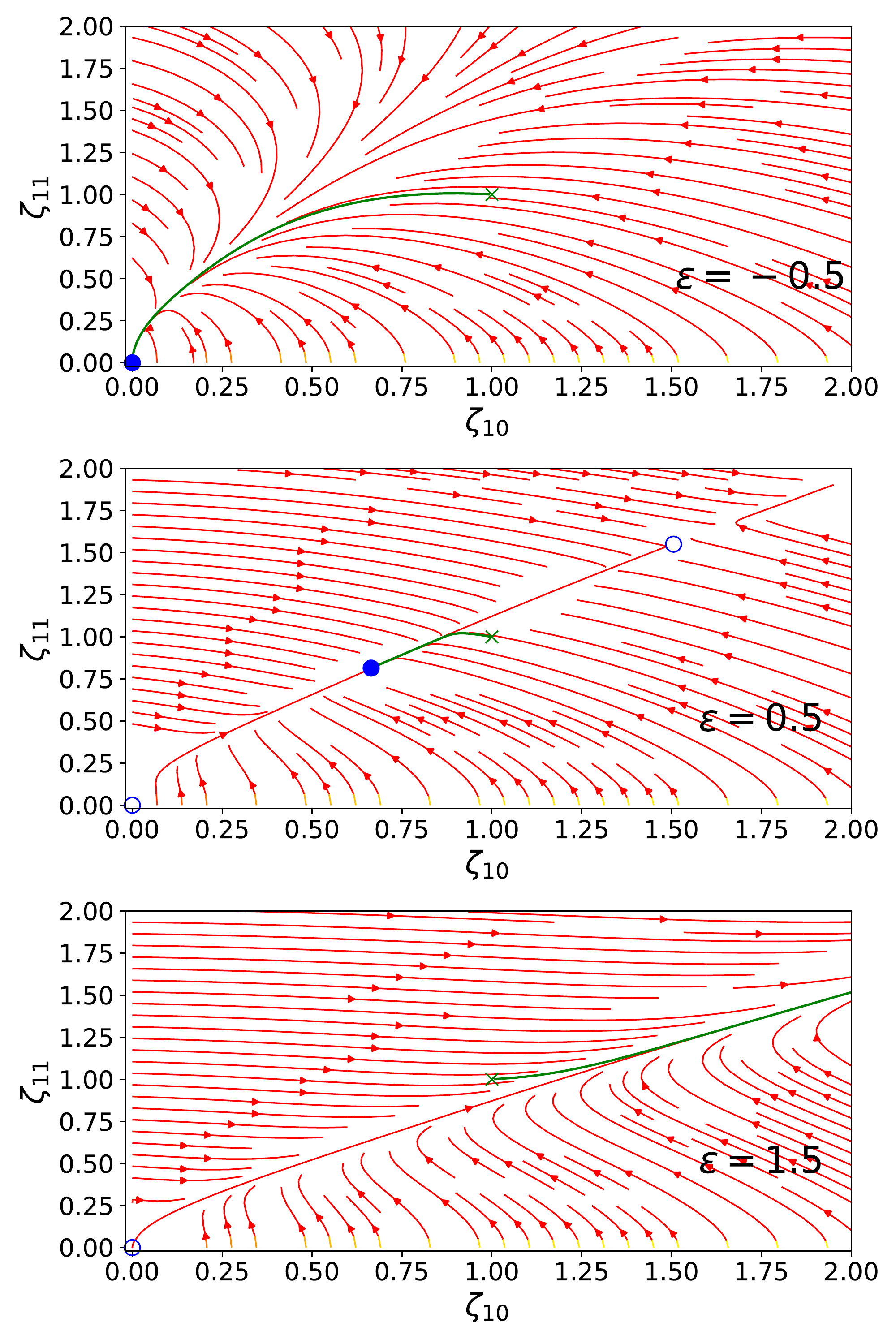}
\caption{Random regular graphs with $K=4$, $\Theta=2$: Stable (solid blue circle) and unstable (empty blue circle) fixed points of Eqs.~\ref{fig:bifurcBPk4t2} for $\epsilon=-0.5,0.5,1.5$. The flow lines of the dynamical system used to solve iteratively these equations are also displayed.}\label{fig:bifurcBPk4t2}
\end{center}
\end{figure}

In order to understand the results discussed in Sec.~\ref{subsec-regularRG} we can start to further simplify these equations because only three independent variables are required. We define the difference of cavity marginals 
 \begin{equation}
 \delta\eta_0   = \eta(0,0) - \eta(0,1) = \frac{1}{Z_c} \binom{K-1}{\Theta-1}\eta(1,0)^{\Theta-1} \eta(0,0)^{K-\Theta}
 \end{equation}
 and divide both sides of all equations by $\delta\eta_0$, 
 \begin{subequations}
\begin{align}
\frac{\eta(1,1)}{\delta\eta_0} & =  e^{\epsilon} \sum_{m\geq \Theta-1} C_{k,m}^{\Theta} \frac{\eta(1,1)^m \eta(0,1)^{K-1-m}}{\eta(1,0)^{\Theta-1} \eta(0,0)^{K-\Theta}},\\
\frac{\eta(1,0)}{\delta\eta_0} & = e^{\epsilon} \sum_{m\geq \Theta} C_{k,m}^{\Theta} \frac{\eta(1,1)^m\eta(0,1)^{K-1-m}}{\eta(1,0)^{\Theta-1} \eta(0,0)^{K-\Theta}},\\
\frac{\eta(0,1)}{\delta\eta_0} & =  \sum_{m< \Theta-1} C_{k,m}^{\Theta}\frac{\eta(1,0)^m \eta(0,0)^{K-1-m}}{\eta(1,0)^{\Theta-1} \eta(0,0)^{K-\Theta}},\\
\frac{\eta(0,0)}{\delta\eta_0} & = \sum_{m < \Theta} C_{k,m}^{\Theta}\frac{\eta(1,0)^m \eta(0,0)^{K-1-m}}{\eta(1,0)^{\Theta-1} \eta(0,0)^{K-\Theta}},
\end{align}
\end{subequations}
where we defined the combinatorial term
\begin{equation}
C_{k,m}^{\Theta} = \frac{\binom{K-1}{m}}{\binom{K-1}{\Theta-1} }= \frac{\frac{K-1!}{m! K-1-m!}}{\frac{K-1!}{\Theta-1!K-\Theta!}} = \frac{\Theta-1!K-\Theta!}{m!K-1-m!}.
\end{equation}
Now dividing numerator and denominator of both sides by $\eta(0,0)$ we get a set of new variables $\zeta_{ab}=\eta(a,b)/\eta(0,0)$, defined implicitly by 
\begin{subequations}
\begin{align}
\frac{\zeta_{11}}{1-\zeta_{01}} & =  e^{\epsilon} \zeta_{10}^{1-\Theta} \sum_{m\geq \Theta-1} C_{k,m}^{\Theta} \zeta_{11}^m \zeta_{01}^{K-1-m},\\
\frac{\zeta_{10}}{1-\zeta_{01}} & = e^{\epsilon} \zeta_{10}^{1-\Theta} \sum_{m\geq \Theta} C_{k,m}^{\Theta} \zeta_{11}^m\zeta_{01}^{K-1-m},\\
\frac{\zeta_{01}}{1-\zeta_{01}} & = \zeta_{10}^{1-\Theta} \sum_{m< \Theta-1} C_{k,m}^{\Theta}\zeta_{10}^m,\\
\frac{1}{1-\zeta_{01}} & = \zeta_{10}^{1-\Theta} \sum_{m < \Theta} C_{k,m}^{\Theta}\zeta_{10}^m.
\end{align}
\end{subequations}
Reordering the terms, we find  
\begin{subequations}\label{eqforzetaBP}
\begin{align}
\zeta_{01} & = 1- \frac{\zeta_{10}^{\Theta-1}}{\sum_{m < \Theta} C_{k,m}^{\Theta}\zeta_{10}^m} \equiv \mathcal{F}(\zeta_{10}),\\
\zeta_{11} &= \zeta_{10} \left\{1 + \frac{\zeta_{11}^{\Theta-1} \mathcal{F}(\zeta_{10})^{K-\Theta} }{\sum_{m\geq \Theta} C_{k,m}^{\Theta} \zeta_{11}^m \mathcal{F}(\zeta_{10})^{K-1-m}}\right\},\\
\zeta_{10} & = e^{\epsilon} \mathcal{F}(\zeta_{10}) \frac{\sum_{m\geq \Theta} C_{k,m}^{\Theta} \zeta_{11}^m \mathcal{F}(\zeta_{10})^{K-1-m}}{\sum_{m < \Theta-1 } C_{k,m}^{\Theta} \zeta_{10}^m }.
\end{align}
\end{subequations}
The first equation is a function of the others, therefore the original system is reduced to a two-dimensional one for $\zeta_{11}$ and $\zeta_{10}$. For the sake of simplicity we focus on the cases $K=3,4$ and $\Theta=2$ already discussed in Sec.~\ref{subsec-regularRG}, but the general analysis gives qualitatively similar results.
For $K=3$ e $\Theta=2$, \eqref{eqforzetaBP} become
\begin{subequations}
\begin{align}
\zeta_{11} & =  \zeta_{10} +  \frac{2\zeta_{10} }{\zeta_{11}(1+2\zeta_{10})}  \\
\zeta_{10} & = \frac{e^{\epsilon} \zeta_{11}^2}{1+2\zeta_{10}}.
\end{align}
\end{subequations}
For  $K=4$ and $\Theta=2$, instead we have
\begin{subequations}\label{eq-BPk4t3app}
\begin{align}
\zeta_{11}& =  \zeta_{10} \left\{\frac{2 \mathcal{F}(\zeta_{10})+3 \zeta_{11}}{\mathcal{F}(\zeta_{10}) +3\zeta_{11}}\right\} \\
\zeta_{10} & = e^{\epsilon} \mathcal{F}(\zeta_{10})^4 \frac{ \mathcal{F}(\zeta_{10}) +3 \zeta_{11} }{3\mathcal{F}(\zeta_{10})}
\end{align}
\end{subequations}
with $\mathcal{F}(\zeta_{10})  = (3  + \zeta_{10}-\zeta_{10}^2)/(3 + \zeta_{10})$.
The structure of the fixed points of Eqs.~\eqref{eq-BPk4t3app}  is reported, as function of $\epsilon$, in Fig.~\ref{fig:bifurcBPk4t2}. The flow lines describe the convergence properties of the BP equations  in the $(\zeta_{10},\zeta_{11})$ plane when solved by iteration.  
 The trivial Nash equilibria $\vec{0}$ and $\vec{1}$ always exist and correspond respectively to the fixed point at $(0,0)$ and the runaway solution $(+\infty,+\infty)$. Varying $\epsilon$ other fixed points can appear, representing Nash equilibria with an intermediate density of agents playing action $1$ (see e.g. Fig.~\ref{fig:rhoentBPk4t2}a). For sufficiently large negative values of $\epsilon$, the $(0,0)$ fixed point, associated with the trivial Nash equilibrium $\vec{0}$, is stable. At  $\epsilon\approx -1.13$, for $K=4$ and $\Theta=2$, a new stable fixed point emerges from  $(0,0)$, that instead becomes unstable by means of a transcritical bifurcation. There is an additional unstable fixed point, which annihilates with the stable one at  $\epsilon\approx 0.596$, in a saddle-node bifurcation. Beyond this value for $\epsilon$, the runaway solution (i.e. trivial Nash equilibrium at $\rho=1$) is the only attractive one. A similar qualitative behaviour, characterized by a transcritical bifurcation followed by a saddle-node bifurcation, is observed for the other values of $K$ and $\Theta$, even for the apparently different case of  $K=3$, $\Theta=2$ in which $\rho(\epsilon)$ goes continuously from 0 to 1 as function of $\epsilon$. 

\begin{figure}[tb]
\begin{center}
\includegraphics[width=0.6\columnwidth]{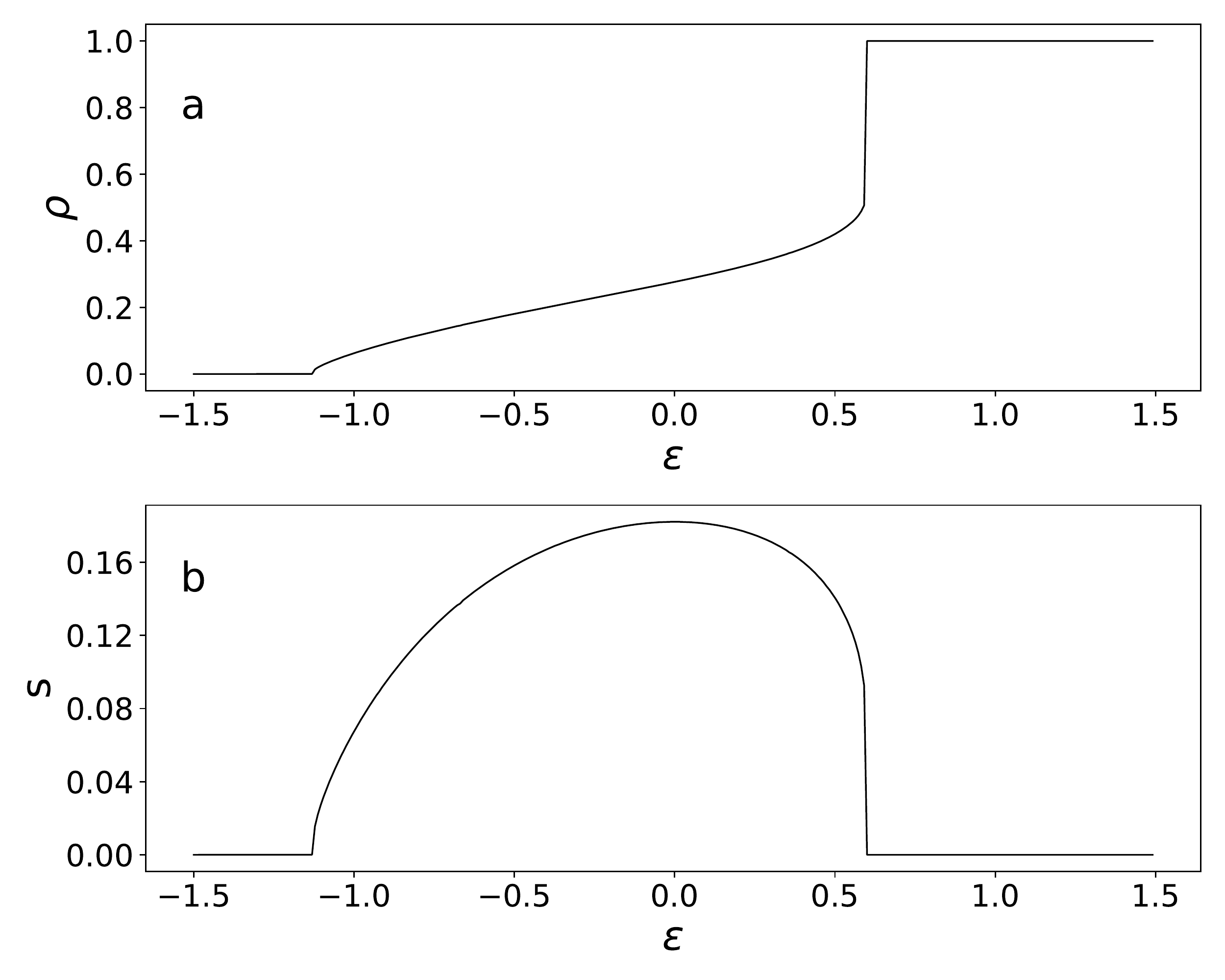}
\caption{Density $\rho$ of agents playing action $1$ in the Nash equilibria and their entropy (b)  as function of the parameter $\epsilon$ for random regular graphs of degree $K=4$ and uniform threshold $\Theta=2$.}\label{fig:rhoentBPk4t2}
\end{center}
\end{figure}

\section{Approximate Master Equation for best-response dynamics}\label{app-AME}
The approximate master equation (AME) \cite{gleeson2011high,gleeson2013binary} is a method to approximate the (Markov) dynamics of binary state models on infinitely large random graphs (random graph ensembles). Similar approximations, based on keeping as minimal structures the star-shaped clusters composed of a central node and all her neighbours, can be also derived within the framework of cluster-variational methods \cite{pelizzola2017variational} and lumping techniques \cite{kiss2017mathematics}. Consider the approximate probability marginal $q^{x}_{k,\theta}(m,t)$ representing the probability that a node of degree $k$ and threshold $\theta$ plays action $x$ and $m$ of her neighbours play action $1$ at time $t$ (we can assume $\Delta t = 1/N$ and rescale time by $N$ to get a continuous-time evolution in the large $N$ limit). The AME equations for the continuous-time limit of the asynchronous best-response dynamics  defined in \eqref{brDyn} are 
\begin{subequations}\label{AMEequations}
\begin{align}
\nonumber \frac{d}{dt} q^0_{k,\theta}(m,t)  = & - \left\{ \Theta(m-\theta k) + (k-m)\beta^0(t) + m \gamma^0(t) \right\} q^0_{k,\theta}(m,t)   + (k-m+1)\beta^0 q^0_{k,\theta}(m-1,t) \\
 & + (m+1)\gamma^0 q^0_{k,\theta}(m+1,t) + \left(1-\Theta(m-\theta k)\right) q^1_{k,\theta}(m,t),\\
\nonumber \frac{d}{dt} q^1_{k,\theta}(m,t)  = &    \left\{ \Theta(m-\theta k)-1 -(k-m)\beta^1  - m \gamma^1 \right\} q^1_{k,\theta}(m,t)    + (k-m+1)\beta^1 q^1_{k,\theta}(m-1,t) \\
 &+ (m+1)\gamma^1 q^1_{k,\theta}(m+1,t) + \Theta(m-\theta k) q^0_{k,\theta}(m,t)  
\end{align}
\end{subequations}
with the closure conditions (see \cite{gleeson2011high,gleeson2013binary} for a general definition)
\begin{subequations}\label{ratesAME}
\begin{align}
\beta^0 & = \frac{\int d\theta f(\theta) \sum_k p_k \sum_{m=\lceil \theta k\rceil}^k (k-m) q^0_{k,\theta}(m) }{\int d\theta f(\theta)\sum_k p_k \sum_{m=0}^k (k-m) q^0_{k,\theta}(m) },\\
\gamma^0 & = \frac{\int d\theta f(\theta) \sum_k p_k \sum_{m=0}^{\lfloor \theta k\rfloor} (k-m) q^1_{k,\theta}(m) }{\int d\theta f(\theta) \sum_k p_k \sum_{m=0}^k (k-m) q^1_{k,\theta}(m) },\\
\beta^1 & = \frac{\int d\theta f(\theta) \sum_k p_k \sum_{m=\lceil \theta k\rceil}^k m q^0_{k,\theta}(m) }{\int d\theta f(\theta)\sum_k p_k \sum_{m=0}^k m q^0_{k,\theta}(m) },\\
\gamma^1 & =  \frac{\int d\theta f(\theta) \sum_k p_k \sum_{m=0}^{\lfloor \theta k\rfloor} m  q^1_{k,\theta}(m) }{\int d\theta f(\theta)\sum_k p_k \sum_{m=0}^k m q^1_{k,\theta}(m) }.
\end{align}
\end{subequations}
The r.h.s. of  \eqref{ratesAME} provide some approximate estimate of the average rates $\beta^{x},\gamma^{x}$ with which a neighbour of node playing action $x$ switches from  action $0$ to action $1$ and from $1$ to $0$ respectively.   
If the individual states at time $t=0$ are i.i.d. random variables, the approximate probability marginals assume factorized initial conditions, 
\begin{subequations}
\begin{align}
q^0_{k,\theta}(m,0) & = \left(1-\rho_0\right) \binom{k}{m} \rho_0^m (1-\rho_0)^{k-m}, \\
q^1_{k,\theta}(m,0) & = \rho_0 \binom{k}{m} \rho_0^m (1-\rho_0)^{k-m}.
\end{align}
\end{subequations}
Averaging over the distributions $f(\theta)$ and $p_k$ at every time step represents a potential computational bottleneck in the integration of \eqref{AMEequations}. For a faster implementation, averages can be performed by sampling over a population of $M$ representative individuals, in which each individual $i$ endowed with a pair of values $\{k_i, \theta_i\}$ drawn from the corresponding distributions. 

\section{Marginal equilibria and improved mean-field method for best response}\label{app-marginal}

The best-response process is an individually greedy process, in which agents change action only when this is strictly favourable. It is thus not particularly surprising that, when starting from a completely random strategy profile (i.e. with density $\rho=0.5$), the reorganization process of the action profile by best-response rapidly gets stuck into some Nash equilibrium far from the Pareto optimum (maximal equilibrium). It means that the equilibria reached by best-response are, in general, very ``superficial'' ones, that is they are equilibria in which  actions are on average supported by a minimal number of neighbours playing accordingly. In the following, we define a particular subset of Nash equilibria of the one-shot coordination game, which we call {\em marginal equilibria}, that are meant to reproduce, at least approximately, such properties. Suppose that an agent has the opportunity to revise her action only once and this process takes place in a random sequential way from a given random initial condition. Since the revision is done only once, we can apply the principle of deferred decisions and average over the realisations of the random variables $\{\theta_i\}$ at the time when revision occurs. This is a sort of generalisation to random initial conditions of the approach discussed in Sec.~\ref{sec:monotoneBR} for best-response dynamics from extremal configurations (i.e. $\vec{0}$ and $\vec{1}$). 
We define $\chi_{ij}(x|\ast)$ to be the (cavity) marginal probability that $i$ plays action $x$ after revision even in the absence of $j$ (the symbol $\ast$ indicates that $x_j$ can be any of the possible actions). This is possible if a sufficiently large number of neighbours $k \in \partial i\sm j$ play actions $x_k=x_i$ after revision. Otherwise, agent $i$ should choose an action conditionally to $j$ taking the same choice. The corresponding probability $\chi_{ij}(\ast)$ has the same expression for both actions. The above reasoning does not care about neighboring nodes $i$ and $j$ playing different actions before revision, because in such case neither of the two is directly sensitive to a revision of the strategy of the other. 
On a tree, these  probabilities can be determined by recursion
\begin{subequations}\label{MEchi1}
\begin{align}
 \chi_{ij}(x|\ast) & = \frac{1}{Z^{\chi}_{ij}}  \sum_{\vec{x}_{\partial i\sm j}}  F_{x}\left[ \frac{(1-x)+\sum_{k\in \partial i\setminus j} x_k}{ |\partial i|}\right]   \prod_{k \in \partial i\sm j} \left[\chi_{ki}(x_k|\ast) + \chi_{ki}(\ast)\right]  \\
 \chi_{ij}(\ast) & =\frac{1}{Z^{\chi}_{ij}}  \sum_{\vec{x}_{\partial i\sm j}} \left\{F_1\left[ \frac{1+ \sum_{k\in \partial i\setminus j} x_k}{ |\partial i|}\right]- F_1\left[ \frac{ \sum_{k\in \partial i \setminus j} x_k}{ |\partial i|}\right]\right\}  \prod_{k \in \partial i\sm j} \left[\chi_{ki}(x_k|\ast) + \chi_{ki}(\ast)\right],
\end{align}
\end{subequations}
with the normalization  $Z^{\chi}_{ij} = \chi_{ij}(0|\ast)+\chi_{ij}(1|\ast) + 2\chi_{ij}(\ast)$.
Defining $\chi_{ij}(x) = \chi_{ij}(x|\ast) + \chi_{ij}(\ast)$, the equations \eqref{MEchi1} become 
\begin{subequations}
\begin{align}
 \nonumber \chi_{ij}(1)  & = \frac{1}{Z^{\chi}_{ij}}  \sum_{\vec{x}_{\partial i\sm j}}  \left\{ F_{1}\left[ \frac{\sum_{k\in \partial i\setminus j} x_k}{ |\partial i|}\right]  + F_1\left[ \frac{1+ \sum_{k\in \partial i\setminus j} x_k}{ |\partial i|}\right]- F_1\left[ \frac{ \sum_{k\in \partial i \setminus j} x_k}{ |\partial i|}\right] \right\} \prod_{k \in \partial i\sm j} \left[\chi_{ki}(x_k|\ast) + \chi_{ki}(\ast)\right]  \\
 & =\frac{1}{Z^{\chi}_{ij}}  \sum_{\vec{x}_{\partial i\sm j}}  F_1\left[ \frac{1+ \sum_{k\in \partial i\setminus j} x_k}{ |\partial i|}\right]   \prod_{k \in \partial i\sm j} \left[\chi_{ki}(x_k|\ast) + \chi_{ki}(\ast)\right],\\
\nonumber \chi_{ij}(0)  & = \frac{1}{Z^{\chi}_{ij}}  \sum_{\vec{x}_{\partial i\sm j}}  \left\{ F_{0}\left[ \frac{1+\sum_{k\in \partial i\setminus j} x_k}{ |\partial i|}\right]  + F_1\left[ \frac{1+ \sum_{k\in \partial i\setminus j} x_k}{ |\partial i|}\right]- F_1\left[ \frac{ \sum_{k\in \partial i \setminus j} x_k}{ |\partial i|}\right] \right\} \prod_{k \in \partial i\sm j} \left[\chi_{ki}(x_k|\ast) + \chi_{ki}(\ast)\right]\\
\nonumber  & =\frac{1}{Z^{\chi}_{ij}}  \sum_{\vec{x}_{\partial i\sm j}} \left\{ 1 - F_1\left[ \frac{ \sum_{k\in \partial i \setminus j} x_k}{ |\partial i|}\right] \right\}  \prod_{k \in \partial i\sm j} \left[\chi_{ki}(x_k|\ast) + \chi_{ki}(\ast)\right]\\
 &  =\frac{1}{Z^{\chi}_{ij}}  \sum_{\vec{x}_{\partial i\sm j}}  F_0\left[ \frac{ \sum_{k\in \partial i \setminus j} x_k}{ |\partial i|}\right] \prod_{k \in \partial i\sm j} \left[\chi_{ki}(x_k|\ast) + \chi_{ki}(\ast)\right]
\end{align}
\end{subequations}
that can be reduced to
\begin{equation}\label{MEchi2}
\chi_{ij}(x)  =\frac{1}{Z_{ij}^\chi }   \sum_{\vec{x}_{\partial i\sm j}}  F_{x}\left[ \frac{x+\sum_{k\in \partial i\setminus j} x_k}{ |\partial i|}\right] \prod_{k \in \partial i\sm j} \chi_{ki}(x_k),
\end{equation}
with the normalization condition $Z_{ij}^\chi  = \chi_{ij}(0)+ \chi_{ij}(1)$. From the fixed-point solution of \eqref{MEchi2}, the probability that node $i$ plays action $x$ in the marginal equilibrium can be straightforwardly computed as 
 \begin{equation}
\chi_i(x) \propto  \sum_{\vec{x}_{\partial i}}  F_{x}\left[ \frac{\sum_{k\in \partial i} x_k}{ |\partial i|}\right] \prod_{k \in \partial i} \chi_{ki}(x_k).
 \end{equation}
At the level of random graphs ensembles, the notion of marginal equilibrium defines an improved mean-field (IMF) approximation in which it is possible to partially account for the conditional dependence of agents' choices on the choice of some neighbours. The average density $\rho_{\rm Imf}$ of agents playing action $1$ in marginal equilibria on random graphs ensembles with degree distribution $p_k$ is given by 
\begin{equation}\label{rho_IMF}
\rho_{\rm Imf}=\sum_{k} p_k \sum_{m=0}^{k}\binom{k}{m}{\hat{\rho}_{\rm Imf}(1)}^{m}{\hat{\rho}_{\rm Imf}(0)}^{k-m}F_1 \left[ \frac{m}{k}\right]
\end{equation}
where the quantities $\hat{\rho}_{\rm Imf}(x)$ are the mean-field probabilities that a randomly chosen neighbour plays action $x$ conditional on the (possibly wrong) belief that the central node also plays $x$. They satisfy the self-consistent mean-field equations
\begin{subequations}\label{rho_IMF2}
\begin{align}
\hat{\rho}_{\rm Imf}(1) &\propto \sum_{k} \frac{k p_k}{\langle k\rangle} \sum_{m=0}^{k-1}\binom{k-1}{m}{\hat{\rho}_{\rm Imf}(1)}^{m}{\hat{\rho}_{\rm Imf}(0)}^{k-1-m} F_1 \left[ \frac{1+m}{k}\right]\\
\hat{\rho}_{\rm Imf}(0) & \propto \sum_{k} \frac{k p_k}{\langle k\rangle} \sum_{m=0}^{k-1}\binom{k-1}{m}{\hat{\rho}_{\rm Imf}(1)}^{m}{\hat{\rho}_{\rm Imf}(0)}^{k-1-m} F_0 \left[ \frac{m}{k}\right],
\end{align}
\end{subequations}
obtained from \eqref{MEchi2}.
In a marginal equilibrium, defined by \eqref{MEchi1} and \eqref{MEchi2}, the agents play optimistically, conditioning their actions on the neighbours only if strictly necessary. This mechanism generates states that are minimally constrained with respect, for instance, to Bayes-Nash equilibria described by \eqref{BNrho1} and \eqref{BNrho}, although they could possibly present stronger correlations. Because of that, it is not surprising that their properties are close to those of the subset of Nash equilibria reached by best-response dynamics from uniformly random initial conditions, in particular much closer than for Bayes-Nash equilibria (see Figure \ref{fig:BRcomparison}).

\section{Max-Sum equations for stochastically stable states}\label{app-stoch}
For a potential game, the local maxima of the potential function are in one-to-one relation with the Nash equilibria \cite{monderer1996potential}. Moreover, we restrict to the binary configurations defined by pure strategy profiles because we have seen (in App.~\ref{app-game}) that mixed strategy equilibria are not global maxima of the potential function in \eqref{potential0}. We define an auxiliary optimization problem on the binary configurations $\vec{x}\in \{0,1\}^N$ associated with the maximization of $V(\vec{x};\vec{\theta})$.  Consider the graphical model defined on a locally tree-like graph by the joint probability distribution $\mathcal{P}(\vec{x};\vec{\theta}) = e^{\beta V(\vec{x};\vec{\theta})}/Z_V[\vec{\theta}]$ with partition function $Z_V[\vec{\theta}] = \sum_{\vec{x}} e^{\beta V(\vec{x};\vec{\theta})}$,  where $\beta$ is a fictitious temperature.
By marginalising around a node $j$ and introducing the cavity marginals $P_{ij}(x_i)$ we get
\begin{equation}
P_i(x_j) \propto \sum_{\{x_i\}_{i \in \partial j} } e^{-\beta  k_j \theta_j x_j + \beta \sum_{i\in \partial j} x_j x_i}  \prod_{i\in\partial j} P_{ij}(x_i)
\end{equation}
in which 
\begin{equation}
P_{ij}(x_i) \propto \sum_{\{x_k\}_{k \in \partial i\setminus j} }  e^{-\beta  k_i \theta_i x_i}  \prod_{k\in\partial i\sm j} e^{\beta x_i x_k} P_{ki}(x_k).
\end{equation}
Defining $h_{ij} = \lim_{\beta \to \infty} \frac{1}{\beta}\frac{P_{ij}(1)}{P_{ij}(0)}$ and $y_{ij} = \min\left\{1,\max\{0,h_{ij}+1\}\right\}$, we obtain the {\em max-sum} equations
\begin{equation}
\label{eq:ms-cavity}
 y_{ij}  = \Psi \left( \sum_{k\in\partial i\sm j} y_{ki} ; \theta_i \right)=   \min\left\{1,\max\left\{0, 1 -k_i\theta_i  +  \sum_{k\in\partial i\sm j} y_{ki}\right\}\right\}\\
\end{equation} 
that can be solved by iteration until a fixed-point is found. At the fixed point, the action $x_i$ of agent $i$ is determined by the logarithm of the ratio between the probability to play $1$ and that of $0$, i.e. 
\begin{equation}\label{eq:ms-total}
x_i = \Theta\left[ \lim_{\beta \to \infty} \frac{1}{\beta}\frac{P_{i}(1)}{P_{i}(0)} \right] =  \Theta\left[-k_i \theta_i + \sum_{k\in\partial i} y_{ki}\right], \quad \forall i \in \mathcal{V}.
\end{equation} 
For a given instance of threshold values $\vec{\theta}$, Eqs.~\eqref{eq:ms-cavity}-\eqref{eq:ms-total} completely define the solution of the optimization problem, i.e. the maximum of the potential function, that is in general unique if the thresholds are real numbers. If we are interested in some average quantity computed over the distribution $f(\theta)$ of threshold values, the max-sum equations \eqref{eq:ms-cavity} become the first (or internal) level of a more complex two-level BP approach developed for studying stochastic optimization problems \cite{altarelli2011stochasticPRL,altarelli2011stochasticJSTAT,altarelli2014containing,altarelli2015statics}. Max-sum messages $y_{ij}$ are promoted to distributions $Y_{ij}(y)$ representing the probability of message $y_{ij}=y$. On random regular graphs, at the ensemble level, the two-level formalism simplifies because of the  homogeneity of the structure, reducing to the solution of the self-consistent distributional equation 
\begin{equation}\label{eq:ms-cavityP} 
Y(y) = \int   \id \left[ y = \Psi \left( \sum_{k=1}^{K-1} y_{k} ; \theta \right) \right] f(\theta) d\theta \prod_{k=1}^{K-1} Y(y_k) dy_k,
\end{equation}
which  can be obtained numerically, for instance by means of a population dynamics approach. Finally, the probability of an agent playing action $x=1$ in the maximum of the potential (i.e. in the stochastically stable state) is obtained as 
\begin{equation}\label{eq:ms-totalP} 
{\rm Prob}\left[x=1\right] = \int   \Theta\left[- K \theta + \sum_{k=1}^{K} y_{k}\right] f(\theta) d\theta \prod_{k=1}^{K} Y(y_k) dy_k.
\end{equation}

\section{Computation of the typical potential $V_{\rm typ}$ using the BP equations with potential bias}\label{BPVtyp}

Consider the partition function in \eqref{pf} with the modified energy function \eqref{eq-HV},  the single instance BP equations are
\begin{equation}\label{bp-instanceV}
 \eta_{ij}(x_i,x_j)  =\frac{1}{Z_{\rm c}}  \sum_{\vec{x}_{\partial i\sm j}}\id\left[x_i = b_i(\vec{x}_{\partial i}; \theta_i) \right] e^{\left(\epsilon  - \nu\theta_i k_i \right) x_i + \nu x_i x_j  }    \prod_{k \in \partial i\sm j} \eta_{ki}(x_k,x_i).
\end{equation}
With this definition of the BP marginals, the  expression of the free-energy $f(\epsilon, \nu;\vec{\theta})$ has to be modified in order to prevent double counting of the pairwise interaction terms of the energy, i.e. 
\begin{equation}\label{bp-fepsnu}
 Nf(\epsilon,\nu;\vec{\theta})  =   \sum_{i\in V}f_i(\epsilon,\nu;\vec{\theta}) - \frac{1}{2}\sum_{(i,j)\in E} f_{ij}(\nu;\vec{\theta})  
\end{equation}
with 
\begin{subequations}
\begin{align}
f_i(\epsilon,\nu;\vec{\theta}) & = -\ln \left\{\sum_{\vec{x}_{\partial i \cup i}} \id\left[x_i = b_i(\vec{x}_{\partial i}; \theta_i) \right]  e^{\left(\epsilon - \nu\theta_i k_i \right) x_i} \prod_{k \in \partial i} \eta_{ki}(x_k,x_i)  \right\},\\
f_{ij}(\nu;\vec{\theta}) & = - \ln\left\{ \sum_{ x_i,x_j} e^{\nu x_i x_j  } \eta_{ij}(x_i,x_j ) \eta_{ji}(x_j,x_i)\right\}.
\end{align}
\end{subequations}
By computing the value of the potential $V(\vec{x};\vec{\theta})$ from the solutions of the above BP equations, we obtain its average value over the set of typical pure Nash equilibria at some given density $\rho$. This typical potential is given by 
\begin{equation}\label{Vtyp}
 V_{\rm typ}(\epsilon, \nu;\vec{\theta})   = \frac{1}{Z_{V}} \sum_i \left\{\sum_{\vec{x}_{\partial i}} \id\left[x_i = b_i(\vec{x}_{\partial i}; \theta_i) \right] \left(-\theta_i k_i x_i +\frac{1}{2} x_i x_j \right)  e^{\left(\epsilon - \nu\theta_i k_i \right) x_i} \prod_{k \in \partial i} \eta_{ki}(x_k,x_i)\right\}, 
\end{equation}
with normalisation 
\begin{equation}
Z_{V}=  \sum_i \left\{\sum_{\vec{x}_{\partial i}} \id\left[x_i = b_i(\vec{x}_{\partial i}; \theta_i) \right]  e^{\left(\epsilon - \nu\theta_i k_i \right) x_i} \prod_{k \in \partial i} \eta_{ki}(x_k,x_i)\right\}.
\end{equation}

\section{Computational analysis of the internal fixed points of the Sato-Crutchfield equations}\label{app-MSsato}
Putting together \eqref{RLrule1}-\eqref{RLrule3} we obtain 
\begin{subequations}
\begin{align}\label{RLruleA}
\pi_i(t+1) & = \frac{\pi_i(t)^{1-\alpha} e^{\beta  \mathbb{U}_i\left[ 1 | \vec{\pi}_{\partial i}(t)\right]}}{\pi_i(t)^{1-\alpha} e^{  \beta  \mathbb{U}_i\left[ 1 | \vec{\pi}_{\partial i}(t)\right]} + \left(1-\pi_i(t)\right)^{1-\alpha} e^{  \beta  \mathbb{U}_i\left[ 0 | \vec{\pi}_{\partial i}(t)\right]} } \\
& = \frac{\pi_i(t)^{1-\alpha} e^{\beta \sum_{j \in \partial i} \pi_j }}{ \pi_i(t)^{1-\alpha} e^{  \beta   \sum_{j \in \partial i} \pi_j } + \left(1-\pi_i(t)\right)^{1-\alpha} e^{  \beta  k_i\theta_i} }  .
\end{align}
\end{subequations}
In order to take the continuous-time limit we have to replace time increment by $\Delta t \to 0$, and rescale both $\alpha \to \alpha \Delta t$ and $\beta \to \beta \Delta t$, obtaining 
\begin{align}
\pi_i(t)+\Delta t \frac{d}{dt}\pi_i(t) & \approx  \frac{\pi_i(t) \left[1 + \Delta t \left(\beta \sum_{j \in \partial i} \pi_j -\alpha \log{\pi_i(t)} \right)\right]}{\pi_i(t) \left[1 + \Delta t \left(\beta \sum_{j \in \partial i} \pi_j(t) -\alpha \log{\pi_i(t)} \right)\right] + \left( 1 - \pi_i(t)\right)\left[1 + \Delta t \left( \beta \Delta t k_i\theta_i -\alpha \log{\left(1-\pi_i(t)\right)}\right) \right]} \\
& \approx \pi_i(t) + \Delta t \left\{ \pi_i(t)\left(1-\pi_i(t)\right) \left[ \left(\beta \sum_{j \in \partial i} \pi_j(t) -\alpha \log{\pi_i(t)} \right) - \left( \beta k_i\theta_i -\alpha \log{\left(1-\pi_i(t)\right)}\right)\right]\right\}   
\end{align}
and finally we obtain the Sato-Crutchfield equations \eqref{SCeq}
\begin{equation}
\frac{d}{dt}\pi_i(t) = \beta \pi_i(t)\left(1-\pi_i(t)\right)\left[   \sum_{j \in \partial i} \pi_j(t)  - k_i\theta_i  -\lambda \log{\frac{\pi_i(t)}{1-\pi_i(t)}}\right].
\end{equation}
Since for $\lambda >0$ the entropic term pushes strategies towards internal fixed points, we focus on the latter, analyzing the possible solutions of the system of self-consistent equations 
\begin{equation}\label{eq-relaxedBR}
\pi_i = {\rm S}(\vec{\pi}_{\partial i}) = \frac{1}{1+e^{-\frac{1}{\lambda}\left(\sum_{j \in \partial i} \pi_j  - k_i\theta_i\right)}},
\end{equation}
with $\pi_i \in [0,1]$ $\forall i$. This system can be seen as a relaxed version of the best-response relations (which are recovered in the limit $\lambda \to 0$ where the mixed strategies converge to pure actions).
Equations \eqref{eq-relaxedBR} can be seen as a set of constraints between continuous variables with support $[0,1]$, therefore it is tempting to employ a belief propagation approach to extract information about the statistical properties of the corresponding solutions. Following Ref.~\cite{altarelli2014containing}, on a tree we can write the distributional BP equation 
\begin{equation}
P_{ij}(\pi_i,\pi_j) \propto \int\prod_{k\in\partial i\setminus j} d\pi_k  \delta\left[\pi_i- {\rm S}(\vec{\pi}_{\partial i}) )\right] e^{-\epsilon \pi_i} \prod_{k \in \partial i\setminus j} P_{ki}(\pi_k,\pi_i)
\end{equation}
in which $P_{ij}(\pi_i,\pi_j)$ is a cavity marginal over neighbouring variables and the term $e^{-\epsilon \pi_i}$ is included to bias the overall measure. The total marginal $P_i(\pi)$, representing the probability that the mixed strategy $\pi_i$ of agent $i$ is equal to $\pi$ in the (possibly biased) set of solutions of \eqref{eq-relaxedBR}, is obtained as 
\begin{equation}
P_{i}(\pi)  \propto \int\prod_{k\in\partial i } d\pi_k  \delta\left[\pi- {\rm S}(\vec{\pi}_{\partial i}) )\right] e^{-\epsilon \pi} \prod_{k \in \partial i} P_{ki}(\pi_k,\pi_i).
\end{equation}
The equations can be solved numerically, for instance discretizing the support $[0,1]^2$ using a grid of bins.   

It is obvious that even in the presence of the artificial bias for $\epsilon\neq 0$, the associated measure over the set of internal fixed points is not the one obtained dynamically by means of the SC equations, let alone the one determined by the original stochastic learning rule. Some further insight into the process of dynamical selection of the fixed points comes from noticing that the function 
\begin{align}
 \nonumber G(\vec{\pi}; \vec{\theta})  &=\sum_{(i,j)} \pi_i \pi_j -  \sum_i k_i \theta_i \pi_i +\lambda \sum_i s(\pi_i)  \\
  & =   \sum_{(i,j)} \pi_i \pi_j -  \sum_i k_i \theta_i \pi_i  - \lambda \sum_i \left[\pi_i \log{\pi_i} + (1-\pi_i)\log{(1-\pi_i)}\right],
\end{align}
defined on individual mixed strategies $\pi_i \in [0,1]$, is a Lyapunov function for the Sato-Crutchfield equations \eqref{SCeq}. Indeed, 
\begin{align}
\nonumber \frac{d G(\vec{\pi}; \vec{\theta}) }{dt} & = \nabla G \cdot \frac{d \vec{\pi}}{dt} \\
& = \sum_{i = 1}^{N} \frac{\partial G}{\partial \pi_i} \frac{d\pi_i}{dt} = \beta  \sum_i \left\{ \pi_i (1-\pi_i) \left[\sum_{j\in\partial i} \pi_j - k_i \theta_i + \lambda \frac{d s(\pi_i)}{d\pi_i}\right]^2\right\} \geq 0. 
\end{align}
The function $G(\vec{\pi}; \vec{\theta})$ always grows along the dynamics, acting as a generalized potential, and stops only at the fixed points of the Sato-Crutchfield equations. It is reasonable to expect that, for  $\lambda >0$, the dynamics preferentially converges to the (internal) global maxima of the function $G$. As already done for stochastic stability in App.~\ref{app-stoch}, the latter can be identified employing the zero-temperature cavity method to study an auxiliary optimization problem. Consider the graphical model defined on a locally tree-like graph by the joint probability distribution $\mathcal{P}(\vec{\pi};\vec{\theta}) = e^{\Gamma G(\vec{\pi};\vec{\theta})}/Z_G[\vec{\theta}]$ with partition function $Z_G[\vec{\theta}] = \int_{\vec{\pi}} e^{\Gamma G(\vec{\pi};\vec{\theta})}$, the marginal probability of variable $\pi_j$ can be written as 
\begin{equation}
Q_{j}(\pi_j) \propto \sum_{\{\pi_i\}_{i \in \partial j} } e^{\Gamma  \left(\sum_{i\in \partial j} \pi_j \pi_i  - k_j \theta_j \pi_j + \lambda s(\pi_j)\right)} \prod_{i\in\partial j} Q_{ij}(\pi_i|\theta_i)
\end{equation}
in which the cavity marginals $Q_{ij}(\pi_i)$ satisfy the equations
\begin{equation}
Q_{ij}(\pi_i) \propto   e^{-\Gamma  \left(k_i \theta_i \pi_i - \lambda s(\pi_i)\right)}  \prod_{k\in\partial i\sm j} \int_{0}^{1} d\pi_ke^{\Gamma \pi_i \pi_k} Q_{ki}(\pi_k).
\end{equation}
Then introducing $\hat{Q}_{ij}(\pi_j) = \int_{0}^{1} d\pi_i e^{\Gamma \pi_i \pi_j}Q_{ij}(\pi_i)$, we find 
\begin{equation}
\hat{Q}_{ij}(\pi_j) \propto  \int_{0}^{1} d\pi_i  e^{-\Gamma  \left(k_i \theta_i \pi_i - \lambda s(\pi_i) - \pi_i \pi_j\right)}  \prod_{k\in\partial i\sm j} \hat{Q}_{ki}(\pi_i),
\end{equation}
and taking the $\Gamma \to +\infty$ limit we obtain the max-sum equations
\begin{equation}\label{eq:maxsumSC}
q_{ij}(\pi_j) =  \max_{\hat{\pi} \in [0,1]} \left\{ \lambda s(\hat{\pi}) + \pi_j \hat{\pi} - k_i \theta_i \hat{\pi} +  \sum_{k\in\partial i\sm j} q_{ki}(\hat{\pi})\right\} - C
\end{equation}
with $q_{ij}(\pi_j)=\lim_{\Gamma \to \infty} \frac{1}{\Gamma}\log \hat{Q}_{ij}(\pi_j)$ and $C=\max_{\pi} q_{ij}(\pi)$.
In practice, in order to produce the results presented in Fig.~\ref{fig:rlK4theory}, the Eqs. \eqref{eq:maxsumSC} were solved numerically on single graph instances, or at the level of random graph ensembles, by discretising the interval $[0,1]$ and treating max-sum messages as histograms.

\bibliographystyle{aipauth4-1}
\bibliography{netcoord}

\end{document}